\documentclass[11pt,a4paper]{article}
\pdfoutput=1
\usepackage{jheppub}
\usepackage{amsfonts,amssymb,amsmath}
\usepackage{mathrsfs}
\usepackage{bbm}
\usepackage{youngtab}
\usepackage{rotfloat}
\usepackage{rotating}
\usepackage{tikz}
\usepackage{lscape}
\usepackage[12pt]{moresize}

\def\a{\alpha}

\newcommand{\bC}{\ensuremath{\mathbb{C}}}

\newcommand{\bR}{\ensuremath{\mathbb{R}}}

\newcommand{\bZ}{\ensuremath{\mathbb{Z}}}

\newcommand{\scA}{\ensuremath{\mathscr{A}}}

\newcommand{\scF}{\ensuremath{\mathscr{F}}}

\newcommand{\scH}{\ensuremath{\mathscr{H}}}

\newcommand{\scM}{\ensuremath{\mathscr{M}}}

\newcommand{\scP}{\ensuremath{\mathscr{P}}}

\newcommand{\frakg}{\ensuremath{\mathfrak{g}}}

\newcommand{\fraksl}{\ensuremath{\mathfrak{sl}}}
\newcommand{\frakso}{\ensuremath{\mathfrak{so}}}
\newcommand{\fraksp}{\ensuremath{\mathfrak{sp}}}

\newcommand{\cH}{\mathcal{H}}

\newcommand{\cL}{\mathcal{L}}

\newcommand{\cN}{\mathcal{N}}
\newcommand{\cO}{\mathcal{O}}

\newcommand{\cW}{\mathcal{W}}

\newcommand{\Tr}{\mbox{Tr}}

\newcommand{\Li}{{\rm Li}}

\newcommand{\half}{\frac{1}{2}}

\def\duuu{d_{\rlap{\raise 1.2pt\hbox{$\scriptstyle\rightarrow$}}\lower 1.2pt
\hbox{$\scriptstyle\leftarrow$}}}

\def\bea{\begin{eqnarray}}
\def\eea{\end{eqnarray}}
\def\be{\begin{equation}}
\def\ee{\end{equation}}
\def\ba{\begin{align}}
\def\ea{\end{align}}

\definecolor{forestgreen}{rgb}{0.133,0.545,0.133}
\definecolor{purple}{rgb}{0.62745098,0.125490196,0.941176471}

\newcommand{\bem}{\begin{pmatrix}}
\newcommand{\eem}{\end{pmatrix}}

\def\={\;  = \;}
\def\+{\, + \,}

\def\wt{\widetilde}
\def\wh{\widehat}
\def\bar{\overline}

\def\rt2{\sqrt{2}}

\newcommand{\HOMFLY}{{\rm HOMFLY}}
\newcommand{\Kauffman}{{\rm Kauffman}}
\newcommand{\univ}{{\rm univ}}
\newcommand{\thin}{{\rm thin}}

\preprint{NIKHEF-2013-025}

\title{Colored Kauffman Homology and Super-A-polynomials}

\author[1]{Satoshi Nawata,}
\author[2]{P. Ramadevi,}
\author[2]{and Zodinmawia}
\affiliation[1]{NIKHEF theory group,\\ Science Park 105,
1098 XG Amsterdam, The Netherlands \vspace{.2cm}}

\affiliation[2]{Department of Physics, Indian Institute of Technology Bombay,\\
 Mumbai, India, 400076\vspace{.2cm}}

\emailAdd{s.nawata@nikhef.nl}
\emailAdd{ramadevi@phy.iitb.ac.in}
\emailAdd{zodin@phy.iitb.ac.in}

\abstract{We study the structural properties of colored Kauffman homologies of knots. Quadruple-gradings play an essential role in revealing the differential structure of colored Kauffman homology. Using the differential structure, the Kauffman homologies carrying the symmetric tensor products of the vector representation for the trefoil and the figure-eight are determined. In addition, making use of relations from representation theory, we also obtain the HOMFLY homologies colored by rectangular Young tableaux with two rows for these knots. Furthermore, the notion of super-$A$-polynomials is extended in order to encompass two-parameter deformations of $PSL(2,\bC)$ character varieties.}


\begin{document}
\Yboxdim4pt

\maketitle


\section{Introduction}\label{sec:intro}
The celebrated paper \cite{Witten:1988hf} by Witten showed that Chern-Simons theory provides a natural framework
for the study of 3-manifolds and knot invariants. The paper \cite{Witten:1988hf} shed new light on the study of low-dimensional topology and has resulted in rigorous formulations of numerous quantum knot invariants.  In particular, the formulation of knot invariants by the expectation values of Wilson loops in Chern-Simons theory has given rise to  colored $\frakg$ quantum invariants \cite{Reshetikhin:1990}, generalizing the Jones polynomials \cite{Jones:1985dw}. Interestingly, the explicit evaluations of colored quantum invariants have posed the question: ``Why are all colored $\frakg$ quantum invariants polynomials of $q$ with integer coefficients?''

This simple question has led to one of the most dramatic developments in knot theory initiated by Khovanov \cite{Khovanov:2000}, categorifications of 
quantum knot invariants.  In the ground-breaking work \cite{Khovanov:2000}, Khovanov constructed the bi-graded homology which is itself a knot invariant and its $q$-graded Euler characteristics is the Jones polynomial $J(K;q)$ of a knot $K$. This construction immediately ignited  research towards the categorifications of colored  $\frakg$ quantum invariants, which has resulted in colored $\fraksl(2)$ homology \cite{Webster:2010,Cooper:2010,Frenkel:2010} and uncolored $\fraksl(N)$ homology \cite{Khovanov:2004} thus far. In a similar way, the triply-graded homology has been defined in \cite{Khovanov:2005} for the categorification of an uncolored HOMFLY polynomial $P_{\yng(1)}(K;a,q)$ which is a two-variable polynomial invariant \cite{Freyd:1985dx,przytycki1987conway} associated to  $SU(N)$ Chern-Simons theory. 

The construction of the uncolored HOMFLY homology in \cite{Khovanov:2005} was also influenced by the physical insight from topological string \cite{Gukov:2004hz}, where the uncolored HOMFLY homology is identified with the space of BPS states. Moreover, it is predicted in \cite{Dunfield:2005si} that HOMFLY homology is gifted with the differential $d_N$ which provides the relation to the $\fraksl(N)$ homology. The viewpoint from topological string theory also predicts the existence of colored HOMFLY homology  with rich colored differential structure \cite{Gukov:2011ry}. To elucidate the structural features of the HOMFLY homology colored by rectangular Young diagrams, the authors of \cite{Gorsky:2013jxa} have taken into account two homological degrees, $t_r$- and  $t_c$-degrees, on equal footing, which led to quadruple-gradings: $(a,q,t_r,t_c)$-gradings. Furthermore,  the introduction of an auxiliary grading $Q$  written in terms of a linear combination of $q$, $t_r$ and $t_c$ leads to an elegant appearance of the structural features. 

As  two-variable polynomial invariants of yet another kind, the colored Kauffman polynomials $F_R(K;\lambda,q)$ \cite{Kauffman:1987} can be obtained from Chern-Simons theory with $SO(N)/Sp(N)$ gauge groups.  The explicit evaluations of colored Kauffman  polynomials  have been demonstrated only for torus knots \cite{Stevan:2010jh}  within Chern-Simons theory. However, for non-torus knots, colored Kauffman polynomials are not available so far. 

Chern-Simons theory with $SO(N)/Sp(N)$ gauge groups on $S^3$ can be realized in the A-model on $T^*S^3$ with an anti-holomorphic involution \cite{Sinha:2000ap}. Therefore,  colored Kauffman polynomials can be connected to M2-M5 bound states via the geometric transition. In this paper, we will also carry out the non-trivial check for the integrality conjecture associated with the M2-M5 bound states proposed by Mari\~no \cite{Marino:2009mw} for the figure-eight. Furthermore, it is argued in  \cite{Gukov:2005qp} that the space of the M2-M5 bound states in the presence of an orientifold can be identified with the Kauffman homology. 

Thus, the physical picture suggests the existence of knot homology associated with $\frakso/\fraksp$ Lie algebras and any representations. As for categorifications of Kauffman polynomials, the uncolored case has been explored in \cite{Gukov:2005qp}. The goal of this paper is to clarify the structural properties of colored Kauffman homology as done for the colored HOMFLY homology \cite{Gukov:2011ry,Gorsky:2013jxa}. When we introduce quadruple-gradings in an appropriate manner, the structural features of Kauffman homology colored by the symmetric tensor product ($[r]$-color) and the anti-symmetric tensor product ($[1^r]$-color)  of the vector representation appear to be also evident. Surprisingly, the colored Kauffman homology and the colored HOMFLY homology hold several common properties: (i) the existences of $d_N$ differentials,  colored differentials and  universal colored differentials, (ii) mirror/transposition symmetry, (iii) refined exponential growth property for a certain class of knots. To the contrary, the difference lies in the fact that the colored Kauffman homology does not have the self-symmetry although the colored HOMFLY homology does. 

The most interesting fact is that, for every knot, the $[r]$-colored Kauffman homology  contains the $[r]$-colored HOMFLY homology. More precisely, in the $[r]$-colored Kauffman homology, there exist differentials of two kinds whose homologies are isomorphic to  the $[r]$-colored HOMFLY homology. The one kind is the universal differential found in \cite{Gukov:2005qp}. The quadruple-gradings allow us to see the other kind, called diagonal differentials. 

Furthermore, there is a similarity in the structure between $[r]$-colored Kauffman homology and  $[r,r]$-colored HOMFLY homology. In particular, one can find the counterparts in $[r]$-colored Kauffman homology of all the differentials present in $[r,r]$-colored  HOMFLY homology. Besides, the dimensions of both the homologies are the same for thin knots. All these properties are discussed in \S\ref{sec:quad-differential} in greater detail.

It is worth mentioning that refined Chern-Simons theory based on $SU(N)$ gauge group \cite{Aganagic:2011sg} is another way to approach the HOMFLY homology for torus knots. In fact, in the case of  the symmetric representations, refined Chern-Simons invariants coincide with the Poincar\'e polynomials of the HOMFLY homology for torus knots  \cite{Aganagic:2011sg,DuninBarkowski:2011yx,Fuji:2012pm,Gorsky:2013jna} though the reason is not fully understood yet. On the other hand, for $SO(N)/Sp(N)$ gauge groups, the refined Chern-Simons invariants  are different from the Poincar\'e polynomials of the Kauffman homology \cite{Cherednik:2011nr,Aganagic:2012au}. 

In this paper, we shall provide the Poincar\'e polynomials of the $[r]$-colored Kauffman homology and the $[r,r]$-colored HOMFLY homology for the trefoil and the figure-eight, which are obtained using
the  structural properties  and  representation theory. The results facilitate the study of  the large color behaviors of the $[r]$-colored Kauffman homology, which provide two-parameter deformations of  $PSL(2,\bC)$ character varieties, called super-$A$-polynomials of $SO$-type. For $[r]$-colored HOMFLY homology, the analogous computations have been performed in \cite{Fuji:2012pm,Fuji:2012nx,Fuji:2012pi,Nawata:2012pg}. Remarkably, via the 3d/3d correspondence, the super-$A$-polynomial admits a physical interpretation as the defining equation for the moduli space of supersymmetric vacua of the 3d $\cN = 2$ theory associated to a knot
 \cite{Fuji:2012nx,Fuji:2012pi}. 

Finally, the authors would like to remark to mathematicians that no statements except the results in \S\ref{sec:polynomial-inv} have been proven in this paper so that precise mathematical formulation is waiting to be given.

The organization of the paper is as follows. In \S \ref{sec:polynomial-inv}, we 
briefly discuss the polynomial invariants, mainly focusing on Kauffman polynomials of torus knots. This section paves the way for the other sections.  In \S \ref{sec:string-theory}, the appearance of these 
polynomial invariants  in the context of string theory is summarized. This approach has been extended to identify the knot 
homology  with the space of BPS states. We also discuss the mirror geometry in the B-model in the presence of an orientifold.
\S \ref{sec:quad-differential} is devoted to describe the structure of the colored Kauffman homology. We give a detailed list of the properties of the quadruply-graded  colored Kauffman homology.  In addition, we  explicitly show the degrees of all colored differentials and re-gradings in the corresponding colored isomorphisms. Furthermore, the isomorphisms of knot homologies are provided from the viewpoint of representation theory.
In \S \ref{sec:homologi}, the structural features are used to obtain the Poincar\'e polynomials of  the $[r]$-colored Kauffman homology for the trefoil and the figure-eight. We also present $[r,r]$-colored HOMFLY homology for these knots making use of the relations from representation theory. To determine these homological invariants, the refined exponential growth property plays an essential role. \S\ref{sec:super-a-polynomials} contains  computations for  the super $A$-polynomials where $PSL(2,\bC)$ character varieties of the knot complements are obtained by taking the appropriate limit.  In \S \ref{3d}, the relation to the 3d/3d correspondence is briefly discussed. Finally, we conclude
with comments and open-problems in \S\ref{sec:discussions}.
 Appendix \ref{conv} provides a list of the conventions
and notations in this paper. 
In Appendix \ref{sec:integrality}, the non-trivial check for the integrality conjecture proposed by Mari\~no for the figure-eight has been carried out by using the invariants obtained in Appendix \ref{sec:homologi}. The figures for colored Kauffman homology of the trefoil  that are too big for the main text are placed in Appendix \ref{sec:tab-fig}. 


\section{Polynomial invariants}\label{sec:polynomial-inv}
\subsection{Skein relations}\label{sec:skein}  
The unreduced HOMFLY polynomial,
$\bar P(K;a,q)$ is a two-variable polynomial invariant of any oriented knot $K$ in $S^3$, defined
using the following skein relations of oriented planar diagrams
\bea\label{HOMFLY-skein}
&&\bar P\left({\raisebox{-.2cm}{\includegraphics[width=.6cm]{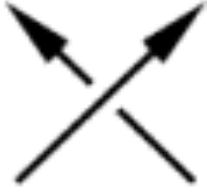}}}\right)
- ~\bar P\left({\raisebox{-.2cm}{\includegraphics[width=.6cm]{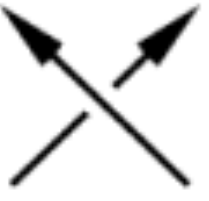}}}\right)
=
(q-q^{-1}) \bar P\left({\raisebox{-.2cm}{\includegraphics[width=.6cm]{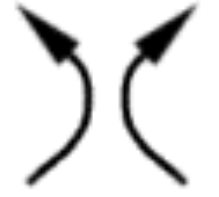}}}\right)\,,\cr
&&\bar P\left({\raisebox{-.2cm}{\includegraphics[width=1cm]{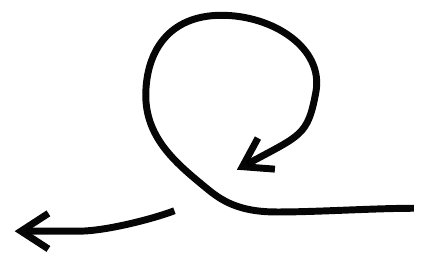}}}\right)=a~\bar P\left({\raisebox{-.05cm}{\includegraphics[width=.8cm]{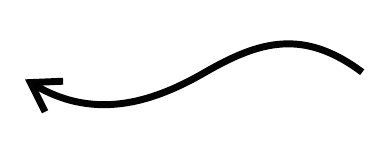}}}\right)\,,\quad\quad  \bar P\left({\raisebox{-.2cm}{\includegraphics[width=1cm]{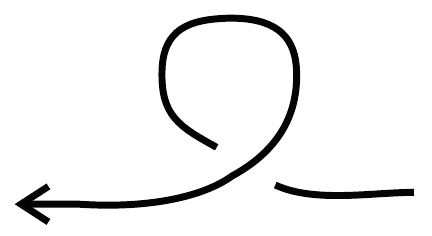}}}\right)=a^{-1}~\bar P\left({\raisebox{-.05cm}{\includegraphics[width=.8cm]{line}}}\right)\,,
\eea
with normalization such that the invariant for the unknot is given by
\be
\bar P ({\raisebox{-.1cm}{\includegraphics[width=.4cm]{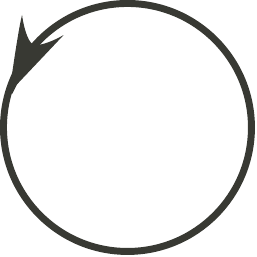}}}) = {a - a^{-1} \over q -
q^{-1} }~. \ee

The unreduced Kauffman polynomial, $\bar F (K;\lambda,q)$ is another two-variable invariant polynomial  of  unoriented knots 
defined via the skein relations 
\bea \label{Kauffman-skein}
 &&\overline F \left({\raisebox{-.2cm}{\includegraphics[width=.6cm]{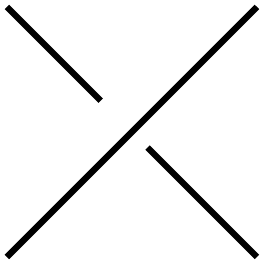}}}\right) -\overline F
\left({\raisebox{-.2cm}{\includegraphics[width=.6cm]{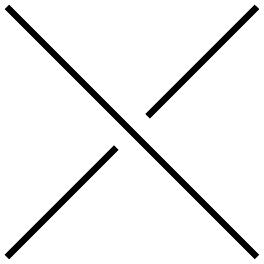}}}\right) =
(q-q^{-1}) \left[\overline F \left({\raisebox{-.2cm}{\includegraphics[width=.6cm]{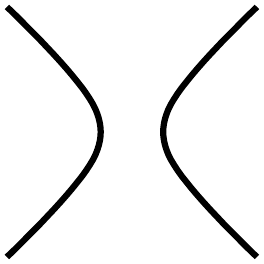}}}\right) -\overline F
\left({\raisebox{-.2cm}{\includegraphics[width=.6cm]{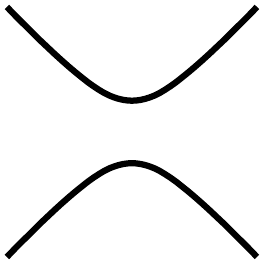}}}\right)\right]\,,\cr
&&\overline F \left({\raisebox{-.1cm}{\includegraphics[width=1cm]{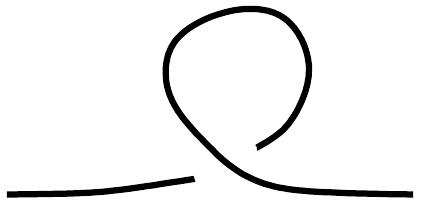}}}\right) = \lambda \overline F ({\includegraphics[width=.7cm]{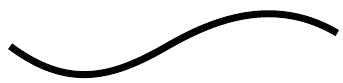}})\,,
\quad\quad
\overline F\left({\raisebox{-.1cm}{\includegraphics[width=1cm]{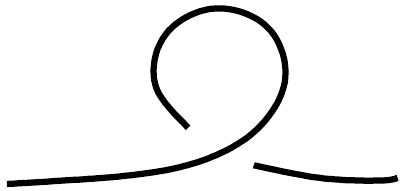}}}\right) = \lambda^{-1}\overline F ({\includegraphics[width=.7cm]{line2}})\,,
\eea
with normalization
\bea \overline F ({\raisebox{-.1cm}{\includegraphics[width=.4cm]{unknot}}}) = {\lambda - \lambda^{-1} \over q -
q^{-1} } + 1. \eea

The reduced versions of the HOMFLY and Kauffman polynomials,
$P(K)$ and $F(K)$, can be obtained by dividing the corresponding unreduced polynomials by the unknot factor
\bea
\bar P (K) = \bar P ({\raisebox{-.1cm}{\includegraphics[width=.4cm]{unknot}}}) P(K)~, \quad\quad
\bar F (K) = \bar F ({\raisebox{-.1cm}{\includegraphics[width=.4cm]{unknot}}}) F(K) ~,
\eea
so that  $P({\raisebox{-.1cm}{\includegraphics[width=.4cm]{unknot}}})=F
({\raisebox{-.1cm}{\includegraphics[width=.4cm]{unknot}}})=1$.

\subsection{Chern-Simons theory and  polynomial invariants}\label{sec:CS}
Chern-Simons gauge theory based on any compact semi-simple group ${ G}$ provides a 
natural  framework  for the study of knots and links \cite{Witten:1988hf}. 
The Chern-Simons action $S$  is 
\begin{equation}
S = {k \over 4 \pi} \int_{S^3} \Tr\left (A \wedge dA + {2 \over 3} A \wedge
A \wedge A \right)  
\end{equation}
where $A$ is the $\frakg$-valued gauge connection  and $k$ is the
coupling constant. A natural metric-independent observable in Chern-Simons theory is a Wilson loop operator $W_{R}(K)~=~\Tr_{R} ~\mathcal{U}_{ K}$ along a knot $K$ carrying  the 
representation $R$ of $\frakg$ where  $\mathcal{U}_{ K}=P\left[\exp \oint_{{ K}} A\right]$. It was heuristically outlined in \cite{Witten:1988hf} that the expectation value of the Wilson loop 
operator
\begin{equation}
\overline J^{\frakg}_{R}(K;q):= 
\langle W_{R}(K) \rangle= {\int[{\cal D}A]e^{iS}
 W_{R}(K) \over \int[{\cal D} A]e^{iS}}~, 
 \label {linki} 
\end{equation}
beomes a quantum invariant of the knot $K$. Here, the quantum parameter $q$ is expressed by 
\bea
q=\exp\left(\tfrac{\pi i }{k+h^{\vee}}\right)~,
\eea
where $h^{\vee}$ represents the dual Coxeter number of the gauge group. The constructions of \cite{Witten:1988hf} soon led to a rigorous formulation of the quantum invariants by the representation theory of quantum groups $U_q(\frakg)$ \cite{Reshetikhin:1990}. It turns out that the quantum invariant $\overline J^{\frakg}_{R}(K;q)$ becomes a polynomial with respect to $q$ for any representation $R$ of $\frakg$. 

The evaluations of quantum invariants can be also carried out by using the relation between  Chern-Simons theory and the Wess-Zumino-Novikov-Witten conformal field theory \cite{Kaul:1991vt,Kaul:1992rs,RamaDevi:1992dh,Kaul:1993hb}. In fact, the skein relations in \S\ref{sec:skein} can be obtained by braiding operations on four point conformal blocks. On one hand, in the context of $SU(N)$ Chern-Simons theory, braiding operations on the four point conformal block with the fundamental representation provides the HOMFLY skein relation \eqref{HOMFLY-skein}  by substituting $a$ for $q^N$ \cite{Witten:1988hf}. On the other hand, by placing the defining representation on the conformal block, a similar method gives the Kauffman skein relation   \eqref{Kauffman-skein}  in $SO(N)/Sp(N)$ Chern-Simons theory where  $\lambda=q^{N-1}$ for  $SO(N)$ or $\lambda=q^{N+1}$ for  $Sp(N)$  \cite{Yamagishi:1989im,Yamagishi:1990,Horne:1989ue}.

Assigning higher rank representations $R$ on a knot, the $\fraksl(N)$ quantum invariant $\overline J^{\fraksl(N)}_{R}(K;q)$  and  the $\frakso(N)/\fraksp(N)$ quantum invariant $\overline J^{\frakso(N)/\fraksp(N)}_{R}(K;q)$ turn into the colored HOMFLY invariant and 
the colored Kauffman invariant, respectively, with the same changes of variables.
It is appropriate  to mention  that a  representation $R$  of  $\fraksl(N)$
placed on a component of an oriented link is changed to the conjugate 
representation $\overline R$ when the orientation of that component is reversed. The change of relative orientations in a link will affect colored HOMFLY invariants of the link.
On the other hand,  all the representations of   $\frakso(N)$ and $\fraksp(N)$  are real $R={\overline R}$, implying that colored Kauffman invariants are suitable for the study of unoriented knots and links.  In Chern-Simons theory,  the quantum invariant of the unknot  proves to be equal to the quantum dimension of the representation $R$  
living on the unknot:
\begin{equation}\label{unknot}
\overline J^\frakg_{R}({\raisebox{-.1cm}{\includegraphics[width=.4cm]{unknot}}})=\dim_q^\frakg R ~.
\end{equation}
Note that the quantum dimension of the representation $R$ of $\frakg$ with
highest weight $\Lambda_R$ is given by 
\begin{equation}\label{unknot2}
\dim^\frakg_q R=\prod_{\alpha>0}\frac{[\alpha\cdot (\varrho+\Lambda_R)]_q}{[\alpha\cdot\varrho]_q}~,
\end{equation}
where $\alpha>0$ are positive roots of  $\frakg$  and $\varrho $ is the Weyl vector. 
The square bracket refers to the quantum number, defined by
\begin{equation}
[x]_q={q^{x}-q^{-x} \over q- q^{-1} }~.
\end{equation}
Normalizing the unknot invariant, the reduced colored HOMFLY and Kauffman invariants of a knot are in polynomial form with respect to two variables.

As for explicit computations of invariants, the colored Kauffman polynomials of torus knots can be implemented for any representation $R$, making use of the modular $SL(2,\bZ)$ transformations \cite{Stevan:2010jh}. To date, however, no calculations have been performed for  colored Kauffman polynomials of non-torus knots  by any method. In this paper, we will report some progress for  the colored Kauffman polynomials of the figure-eight.

\subsection{Framed unknot}
It follows from \eqref{unknot} and \eqref{unknot2} that the colored Kauffman polynomial for the unknot carrying rank-$r$ symmetric representation is
\be\label{Kauffman-unknot}
\overline F_{[r]}({\raisebox{-.1cm}{\includegraphics[width=.4cm]{unknot}}};\lambda,q)=\frac{q^r(\lambda^2;q^2)_r}{\lambda^r(q^2;q^2)_r}+\frac{q^{r-1}(\lambda^2;q^2)_{r-1}}{\lambda^{r-1}(q^2;q^2)_{r-1}}~,
\ee
where we denote the $q$-Pochhammer symbols by $(z;q^2)_{k}=\prod_{j=0}^{k-1} (1-zq^{2j})$. In fact,  Chern-Simons invariants are framing dependent \cite{Witten:1988hf}. One can incorporate the framing dependence by the modular $T$ transformation whose eigenvalue is $q^{\kappa_R}$. Here, the quadratic Casimir $\kappa_R$ of the representation $R$ specified by the Young diagram $[r_1,r_2,\cdots,r_{N-1}]$ is given by
\bea
\kappa_R=N|R|+\sum_{j} (r_j^2-2jr_j)~,
\eea
where $|R|$ is the total number of boxes in the Young diagram. Hence, the $[r]$-colored Kauffman polynomial for the framed unknot ${\raisebox{-.1cm}{\includegraphics[width=.4cm]{unknot}}}_f$ with framing $f$ is
\be
\overline F_{[r]}({\raisebox{-.1cm}{\includegraphics[width=.4cm]{unknot}}}_f;\lambda,q)=(-1)^{f r}\lambda^{f r}q^{fr(r-1)}F_{[r]}({\raisebox{-.1cm}{\includegraphics[width=.4cm]{unknot}}};\lambda,q)~.
\ee

Let us calculate both the classical and the quantum $\lambda$-deformed $A$-polynomial of $SO$-type for the framed unknot ${\raisebox{-.1cm}{\includegraphics[width=.4cm]{unknot}}}_f$ with framing $f$. For a detailed explanation of the $A$-polynomials, we refer the reader to \S\ref{sec:super-a-polynomials}. The large color limit
\be
q = e^{\hbar} \to 1 \,, \qquad \lambda = \text{fixed} \,, \qquad t = \text{fixed} \,, \qquad x = q^{2r} = \text{fixed} \ ,
\label{reflimit1}
\ee
of  the $[r]$-colored Kauffman polynomial for the framed unknot  $\overline F_{[r]}({\raisebox{-.1cm}{\includegraphics[width=.4cm]{unknot}}}_f;\lambda,q)$ is of form
\be
\overline F_{[r]}({\raisebox{-.1cm}{\includegraphics[width=.4cm]{unknot}}}_f;\lambda,q)\sim e^{\frac{1}{2\hbar}\widetilde\cW^{SO}({\raisebox{-.1cm}{\includegraphics[width=.3cm]{unknot}}}_f;x,\lambda)+{\cal O}(\hbar)}~. 
\ee
Making use of the asymptotic of $q$-Pochhammer symbol 
\begin{eqnarray}\label{q-Poch-asymptotic}
(z;q^2)_{k}\overset{{\hbar \to 0}}{\sim} e^{\frac{1}{2\hbar}\left({\rm Li}_2(z)-{\rm Li}_2(zq^{2k})\right)}~, \,
\end{eqnarray}
we find that
\bea\label{W-unknot}
\widetilde\cW^{SO}({\raisebox{-.1cm}{\includegraphics[width=.4cm]{unknot}}}_f;x;\lambda)&=& i\pi f\log{x}+f\log{x}\log{\lambda}+f\log^2{x}-\log{x}\log{\lambda}\cr
&&+\Li_2(\lambda^2)-\Li_2(\lambda^2 x)-\frac{\pi^2}{6}+\Li_2(x)~.
\eea
The zero locus of the classical $\lambda$-deformed $A$-polynomial of $SO$-type is given by the equation
\be
y=\exp\left({x\tfrac{\partial{\widetilde\cW^{SO}({\raisebox{-.1cm}{\includegraphics[width=.3cm]{unknot}}}_f;x;\lambda)}}{\partial x}}\right)~.
\ee
Using the expression \eqref{W-unknot}, we obtain
\be\label{A-poly-unknot}
A^{SO}({\raisebox{-.1cm}{\includegraphics[width=.4cm]{unknot}}}_f;x, y;\lambda)=(-1)^f x^{2 f} \lambda ^f \left(1-x \lambda ^2\right)+y (-\lambda +x \lambda )~. 
\ee
Hence, the $\lambda$-deformed $A$-polynomial of $SO$-type can be identified with  the $a$-deformed $A$-polynomial of $SU$-type with $\lambda=a$ for the framed unknot.

To find the quantum version, we find the recursion relation for $\overline F_{[r]}({\raisebox{-.1cm}{\includegraphics[width=.4cm]{unknot}}}_f;\lambda,q)$
\bea
&&q^{2+2 f} \left(-1+q^{2 r}\right) \left(-1+q^{2+2 r}\right) \lambda ^2 \bar{F}_{[r+2]}-(-1)^f q^{1+2 f (1+r)} \left(1+q^2\right) \left(-1+q^{2 r}\right) \lambda ^{1+f} \cr
&&\times\left(-q^2+q^{2 r} \lambda ^2\right) \bar{F}_{[r+1]}+q^{4 f r} \lambda ^{2 f} \left(q^6-q^{2+2 r} \lambda ^2-q^{4+2 r} \lambda ^2+q^{4 r} \lambda ^4\right) \bar{F}_{[r]}=0~.
\eea 
 From this recursion relation, we can read off the quantum  $\lambda$-deformed $A$-polynomial of $SO$-type for the framed unknot
\bea
\wh{A}^{SO}({\raisebox{-.1cm}{\includegraphics[width=.4cm]{unknot}}}_f;\hat{x}, \hat{y};\lambda,q)&=&x^{4 f} \lambda ^{2 f} \left(q^6-q^2 x^2 \lambda ^2-q^4 x^2 \lambda ^2+x^4 \lambda ^4\right)\cr
&&-(-1)^f q^{-3}\left(1+q^2\right) x^{2 f} \left(q^2-x^2\right) \lambda ^{1+f} \left(q^4-x^2 \lambda ^2\right) \hat{y}\cr
&&+q^{-4+2 f} \left(q^6-q^2 x^2-q^4 x^2+x^4\right) \lambda ^2 \hat{y}^2.
\eea
Taking $q=1$ limit, it reduces to the classical $\lambda$-deformed $A$-polynomial of $SO$-type \eqref{A-poly-unknot}
\be
\wh{A}^{SO}({\raisebox{-.1cm}{\includegraphics[width=.4cm]{unknot}}}_f; x,y;\lambda,q=1)= \left((-1)^f x^{2 f} \lambda ^f \left(1-x \lambda ^2\right)+y (-\lambda +x \lambda )\right)^2.
\ee 
 Particularly,  the classical $\lambda$-deformed $A$-polynomial for the unknot with zero framing is expressed by
 \be
A^{SO}({\raisebox{-.1cm}{\includegraphics[width=.4cm]{unknot}}};x, y;\lambda)=1-x \lambda^2 -y\lambda\left(1 -x\right)~.
\ee

 \subsection{Torus knots and $SL(2,\bZ)$ transformations}  
As explained in \S \ref{sec:CS}, there is a systematic procedure of determining quantum invariants with an arbitrary representation of the $(Q,P)$-torus knot. This was introduced by Rosso and Jones for $SU(2)$ Chern-Simons invariants \cite{Rosso:1993vn}, and has been further generalized
 through the construction of torus knot operators for $SU(N)$ \cite{Labastida:1990bt,lin2010hecke,Brini:2011wi} and $SO(N)/Sp(N)$ Chern-Simons invariants \cite{Stevan:2010jh}. Let us briefly review the procedure. 
 
 The Chern-Simons invariant of the unknot with winding number $Q$ carrying the representation $R$ can be written as $\left\langle\Tr_R~\mathcal{U}^Q_{{\raisebox{-.1cm}{\includegraphics[width=.3cm]{unknot}}}}\right\rangle$. Using the fact that the Hilbert space on the torus in Chern-Simons theory is  isomorphic to the space of conformal blocks, one can write that 
 \bea\label{Adams}
 \left\langle {\Tr}_{R} ~ \mathcal{U}^Q_{{\raisebox{-.1cm}{\includegraphics[width=.3cm]{unknot}}}}\right\rangle=\sum_V C_{R,Q}^V \dim_q V~,
 \eea
 which is called the Adams  operation. Since the $(Q,P)$ torus knot can be obtained by performing the modular $T$ transformation with the fractional power $P/Q$ to  the unknots with winding number $Q$, the quantum invariant of the  $(Q,P)$-torus knot can be written as
  \bea\label{rosso-jones}
 \left\langle {\Tr}_{R} ~ \mathcal{U}_{T_{Q,P}}\right\rangle=\sum_V C_{R,Q}^V q^{\frac{P}{Q} \kappa_V} \dim_q V~.
 \eea
 
It was shown in \cite{Brini:2011wi} that a closed form expression of the uncolored HOMFLY polynomial of the $(Q,P)$-torus knot can be obtained by this procedure since the Adams operation \eqref{Adams} involves only hook representations when the representation $R$ is a single box $\yng(1)$. Therefore, let us try to obtain a closed form expression of the uncolored Kauffman polynomial of the $(Q,P)$-torus knot in the same manner. It follows from \eqref{Kauffman-unknot} that the $\frakso(N)$ quantum invariant of the unknot is 
\bea
\overline J_{\yng(1)}^{\frakso(N)}({\raisebox{-.1cm}{\includegraphics[width=.4cm]{unknot}}};q)=1+\frac{q^{N-1}-q^{-N+1}}{q-q^{-1}}~.
\eea
Therefore, the holonomy matrix can be written  as
  \be
 \left\langle \mathcal{U}_{{\raisebox{-.1cm}{\includegraphics[width=.3cm]{unknot}}}}\right\rangle={\rm diag}(1,q^{N-2},q^{N-4},\cdots,q^{4-N},q^{2-N})~.
  \ee
 Subsequently, one can convince oneself that the expectation value of the unknot with winding number $m$ is given by
 \bea
\left\langle\rm{Tr}_{\yng(1)}~\mathcal{U}^m_{{\raisebox{-.1cm}{\includegraphics[width=.3cm]{unknot}}}}\right\rangle&=&\tfrac12(e^{\pi i m}+1)+\sum_R \chi_R(C_m){\overline F}_R({\raisebox{-.1cm}{\includegraphics[width=.4cm]{unknot}}})\cr
&=&\tfrac12(e^{\pi i m}+1)+\sum_{s=0}^{m-1} (-1)^s{\overline F}_{[m-s,1^s]}({\raisebox{-.1cm}{\includegraphics[width=.4cm]{unknot}}})~,
 \eea
 where the conjugacy class of length $m$ in the rank-$m$ symmetric group $S_m$ is expressed by $C_m$. 
From the first to the second line, we use the fact that the character $\chi_R(C_m)$  of the symmetric group $S_m$ with representation $R$ at the conjugacy class $C_m$ of length $m$ vanishes except for the hook representations $R=[m-s,1^s]$ of $m$ boxes $(s=0,\cdots,m-1)$:
\bea
\chi_R(C_m)=\left\{\begin{array}{lll} (-1)^s & & R=[m-s,1^s]\\ 0&& {\rm otherwise} \end{array}\right. ~.
\eea
For the framed unknot ${\raisebox{-.1cm}{\includegraphics[width=.4cm]{unknot}}}_f$ with framing $f$, we can just perform the $SL(2,\bZ)$ transformation $T^f$
\be
\left\langle\rm{Tr}~\mathcal{U}^m_{{\raisebox{-.1cm}{\includegraphics[width=.3cm]{unknot}}}_f}\right\rangle=\tfrac12(e^{\pi i m}+1)+\sum_{s=0}^{m-1}(-1)^sq^{-f\kappa_{[m-s,1^s]}}{\overline F}_{[m-s,1^s]}({\raisebox{-.1cm}{\includegraphics[width=.4cm]{unknot}}})~.
\ee
The quantum dimension and the quadratic Casimir of the representation $[m-s,1^s]$ of $\frakso(N)$ are given by  
\bea
{\overline F}_{[m-s,1^s]}({\raisebox{-.1cm}{\includegraphics[width=.4cm]{unknot}}})&=&{\rm dim}_q^{\frakso(N)}([m-s,1^s])\cr
&=&  \left(\tfrac{1}{q^{m} - q^{-m}}+ 
   \tfrac{1}{ \lambda q^{-1 + m - 2 s} -\lambda^{-1} q^{1 - m + 2 s}}\right)\tfrac{\prod_{i=s+1-m}^{s}( \lambda q^{-i} -\lambda^{-1} q^{i})}{\prod_{j=1}^{s}(q^{j} - q^{-j})\prod_{k=1}^{m-1-s}(q^{k} - q^{-k})}~,\\
   \kappa_{[m-s,1^s]}&=&m ((N-1)+m -  1 - 2s)~.
\eea
As in the case of the HOMFLY polynomial \cite{Brini:2011wi}, with appropriate normalization, the substitution of $f=\frac{P}{Q}$, $m=Q$ will give us the  Kauffman polynomial for the $(Q,P)$-torus knot $T_{Q,P}$:
\bea\label{torusknots}
&&F(T_{Q,P};\lambda,q)\\
&=&\tfrac{(q-q^{-1})\lambda^{QP}}{(q-q^{-1})+(\lambda-\lambda^{-1})}\Bigg[\tfrac12(e^{\pi iQ}+1)\cr
&&+\sum_{s=0}^{Q-1}(-1)^s \lambda^{-P}  q^{P(2s+1-Q) } \left(\tfrac{1}{q^{Q} - q^{-Q}}+ 
   \tfrac{1}{ \lambda q^{-1 + Q - 2 s} -\lambda^{-1} q^{1 - Q + 2 s}}\right)\tfrac{\prod_{i=s+1-Q}^{s}( \lambda q^{-i} -\lambda^{-1} q^{i})}{\prod_{j=1}^{s}(q^{j} - q^{-j})\prod_{k=1}^{Q-1-s}(q^{k} - q^{-k})}\Bigg]~.\nonumber
\eea
This expression proves  to be consistent with (4.4) of \cite{Labastida:1995kf}. Although the HOMFLY polynomials of the torus knots can be written in terms of the $q$-hypergeometric function ${}_2\phi_1$, the term $e^{\pi iQ}+1$ in \eqref{torusknots} as well as the symmetry $F(T_{Q,P};\lambda,q)=F(T_{P,Q};\lambda,q)$ under the interchange of $Q$ and $P$ prevent us from writing it in a similar manner.


\section{Interpretations in topological strings}\label{sec:string-theory}
\subsection{A-model description}\label{sec:string-theory-a}
It was shown in \cite{Witten:1992fb} that $SU(N)$ Chern-Simons theory on $S^3$ can be realized in the A-model topological string on the deformed conifold $T^*S^3$, where $N$ A-branes wraps the Lagrangian submanifold $S^3$. For $SO(N)/Sp(N)$ gauge groups, an orientifold has to be introduced \cite{Sinha:2000ap} in this setting. The deformed conifold $T^*S^3$ which can be expressed by
\be
\sum_{i=1}^4 x_i^2=\mu,
\ee
admits the anti-holomorphic involution $I: x_i \rightarrow \bar x_i$ where the set of the fixed points under the involution $I$ is $S^3$. Whether the gauge group is $SO(N)$ or $Sp(N)$ depends on the choice of
orientifold action on the gauge group.

Generalizing the Gopakumar-Vafa duality \cite{Gopakumar:1998ki},
 Chern-Simons theory with $SO(N)/Sp(N)$ gauge groups on $S^3$ at large $N$ is equivalent
to closed topological string theory on the resolved conifold $X=\cO (-1) \oplus \cO (-1) \to  \bC{\bf P}^1$ in the presence of an orientifold \cite{Sinha:2000ap}.  To specify the anti-holomorphic involution, let us briefly recall the geometry of the resolved conifold.
The resolved conifold $X$ can be described as a toric variety $X = \bC^4 / \bC^*$,
where $\bC^4$ is parametrized by $X_i$, $i=1,\ldots,4$,
with charges $(1,1,-1,-1)$ with respect to the $\bC^*$ action.
In these variables, the resolved conifold $X$ is
\bea
X = \{ |X_1|^2 + |X_2|^2 - |X_3|^2 - |X_4|^2 = r \}
/ U(1)~,
\eea
where the size of the $ \bC{\bf P}^1$ is set by the FI parameter $r$. This
is complexified by the theta angle of the gauged linear sigma model to give the complexified K\"ahler
parameter $t = r - i\theta$ on which the A-model amplitudes depend. 
The action of the anti-holomorphic involution $\tau$ on the space $X$ is defined as follows:
\bea
\tau~:~~ (X_1, X_2, X_3, X_4) \to (\bar X_2, - \bar
X_1, \bar X_4, - \bar X_3)~.
\eea
In particular, it acts freely on $X$, so that the quotient space
$X/\tau$ contains a 2-cycle $\bR{\bf P}^2$ instead of $ \bC{\bf P}^1$. 

With this setting, the large $N$ duality \cite{Sinha:2000ap} can be concretely depicted in the following way: the free energy of Chern-Simons theory  with $SO(N)/Sp(N)$ gauge groups on $S^3$ provides the closed topological string  partition function on the orientifolded resolved conifold
\be
\log Z_{CS}^{SO/Sp}(k) = {1 \over 2} {\cal F}^{\rm oriented}(t,g_s) + {\cal F}^{\rm  unoriented}(t,g_s)\label{partition}~.
\ee
Note that the variables in Chern-Simons theory are identified with the parameters in the closed topological string  as
\be
  e^{g_s/2} =q = \exp \left( \tfrac{\pi i }{ k +
h^{\vee}} \right) ~,\quad \quad e^{t/2} = \lambda =q^{N+ \tilde c} ~,
\ee
where $\tilde c$ expresses the charge on the orientifold plane.
Actually, the right hand side of \eqref{partition} illustrates the fact that the closed topological string  partition function receives contributions from an oriented sector and an unoriented 
sector where the factor $1/2$ in the first term takes care of modding by the anti-holomorphic involution $\bZ_2$. 
More specifically, the oriented sector is written in terms of $g_s^{2-2g}$, and the unoriented sector contains only odd powers of $g_s$ indicating Riemann surfaces with genus $g$  and one cross-cap. 

To incorporate a Wilson loop along a knot $K$ in Chern-Simons theory, another stack of $M$ Lagrangian branes have to be inserted on the conormal bundle $\cN_K$ to the knot $K$ in $S^3$ \cite{Ooguri:1999bv}. Furthermore, the large $N$ duality can be naturally extended to open topological string on the resolved conifold by wrapping $M$ branes on the Lagrangian submanifold $\cL_K$ associated to the knot $K$ which is the geometric transition of the submanifold $\cN_K$. Instead of the partition function, the insertion of the Wilson loop is captured by the Ooguri-Vafa operator
\bea
\cO_{CS}^{SO/Sp}(K;U,V)=\sum_R \Tr_R(U) \Tr_R(V)~,
\eea
where $U$ is the holonomy matrix of the $SO(N)/Sp(N)$ gauge field along the knot $K$ and  $V$ is the holonomy matrix of the $SU(M)$ gauge group associated to the probe branes on $\cN_K$.
Therefore, on the deformed conifold side, the free energy is given by the logarithm of the expectation value of   the Ooguri-Vafa operator
\bea
\log Z_{CS}^{SO/Sp}(K;\lambda,q,v)= \log \Big[ \sum_R \overline F_R(K;\lambda,q) s_R(v)\Big]~
\eea
 where the expectation value of $\Tr_R(U)$ provides the unreduced Kauffman polynomial $\overline F_R(K;\lambda,q)$ colored by a Young diagram $R$ while $\Tr_R(V)$ gives  the Schur polynomial $s_R(v)$ labeled by the Young diagram $R$ for $SU(M)$.  In a similar manner to \eqref{partition}, the free energy can be reformulated in terms of the open topological string partition function on the resolved conifold 
\begin{equation}
\log Z_{CS}^{SO/Sp}(K;\lambda,q,v)= {1 \over 2} {\cal  F}_K^{c=0}(t,g_s,v) + {\cal F}_K^{c=1}(t,g_s,v) +{\cal F}_K^{c=2}(t,g_s,v)~, \label {observ}
\end{equation}
where the superscript $c$ denotes the contribution from Riemann surfaces with $c$ cross-caps \cite{Bouchard:2004iu,Bouchard:2004ri}. Meanwhile, it had been hard to separate the $c=0$ and $c=2$ contributions, since their genus expansions  are similar,  whereas the $c=1$ contribution can be extracted using parity argument in the variable ${\lambda}$ \cite{Bouchard:2004iu,Bouchard:2004ri,Borhade:2005pw}.

In order to isolate the $c=0$ contribution, it was proposed in \cite{Marino:2009mw} that one has to take into account two sets of Lagrangian branes on $\cN_{K,\epsilon}$ and $\cN_{K,-\epsilon}$, which are related by the anti-holomorphic involution in the covering geometry. If you deform the conormal bundle $\cN_K$  to the fiber direction by  $\epsilon$, the anti-holomorphic involution creates the two stacks of probe branes $\cN_{K,\epsilon}$ and $\cN_{K,-\epsilon}=I(\cN_{K,\epsilon})$ (See Figure 9 in \cite{Marino:2009mw}). 
Since the $U(N)$ invariants account for the partition function of the covering geometry, the corresponding invariants are described by  HOMFLY polynomials 
$\bar P_{(R,S)}(K)$ carrying a composite representation $(R,S)$. Here, the composite representation $(R,S)$ can be considered as a representation with highest weight $\Lambda_R+ \Lambda_{\bar S}$ where $\bar S$ is the conjugate representation of $S$. 
Using this set-up,
the $c=0$ contribution  in \eqref{observ} is given by \cite{Marino:2009mw}
\be
{\cal  F}_K^{c=0}(t,g_s,v)=\log\Big[ \sum_{R,S}\overline P_{(R,S)}(K;\lambda,q) s_R (v) s_S(v)\Big]~.
\ee

These open-string topological amplitudes can be related to counting degeneracies of M2-M5 bound
states in M-theory on the resolved conifold $X$
\cite{Gopakumar:1998ii,Gopakumar:1998jq,Ooguri:1999bv,Labastida:2000yw,Marino:2009mw,Witten:2011zz}, where the configurations of $M$ M5-branes are as follows:
\bea
\text{space-time} &:& \quad S^1\times TN \times X~,\cr
M~ \text{M5-branes}&:&  \quad S^1\times D\times {\cL_K}~,
\eea
where $D\cong \bR^2$ is the cigar of  the Taub-NUT space $TN\cong \bR^4$. 
The M2-branes wrap a two-cycle $\beta\in H_2(X,\cL_K)$ of $X$ and end on the stack of the M5-branes. In the orientifold background, a two-cycle can be either an orientable ($c=0$) or a non-orientable ($c=1,2$) Riemann surface of genus $g$ with $h$ boundaries. The boundary condition is specified by the $h$-tuple winding number $w=(w_1,\cdots,w_h)$ where the total number $|R|$ of boxes in the Young diagram $R$ for $SU(M)$ is equal to $|R|=\sum_{i=1}^h w_i$. The K\'ahler parameter $\lambda$ becomes fugacity for the charge $\beta$  and  the variable $z=q-q^{-1}$ corresponds to the fugacity of the charge $g$ in the index which counts M2-M5 bound states. Therefore, one can define the reformulated invariants by the number $\wh N_{R;g,\beta}^c(K)$ of the M2-M5 bound states
\bea
\sum_{g\ge0} \sum_{\beta\in\bZ} \wh N_{R;g,\beta}^c(K) z^{2g-1+c}\lambda^\beta~.
\eea
Via the geometric transition, the reformulated invariants can be written in terms of the Chern-Simons invariants of the knot $K$, which is discussed in more detail in Appendix \ref{sec:integrality}. Most importantly, since $\wh N_{R;g,\beta}^c(K)$ is the number of  the M2-M5 bound states,  it is conjectured to be an integer \cite{Ooguri:1999bv,Labastida:2000yw,Marino:2009mw}. In Appendix \ref{sec:integrality}, we verify this conjecture for the figure-eight with $R=\yng(2)$ or $\yng(1,1)$.

It is conjectured in \cite{Gukov:2004hz} that the space of the M2-M5 bound states is isomorphic to the knot homologies
$\scH_{\rm BPS}\cong \scH_{\rm knot}$.
More precisely, we count the BPS states weighted by the charge $\beta$ as well as the charge $(s,r)$ of the rotation group $U(1)_q \times U(1)_t$  of the non-compact space $TN\cong \bR^4$ where the $q$- and $t$-gradings correspond to the equivariant action $U(1)_q \times U(1)_t$ on the tangent and normal bundle of $D\cong \bR^2$ in $TN\cong \bR^4$. With an appropriate change of basis, the space of the BPS states can be identified with the triply-graded homology, so-called Kauffman homology, which categorifies the Kauffman polynomial. The large $N$ duality predicts that the Kauffman homology is isomorphic to $\frakso(N)/\fraksp(N)$ homology at large $N$ \cite{Dunfield:2005si,Gukov:2005qp}. However, when 
 the K\"ahler parameter $N\sim \log(\lambda)$ varies, the BPS spectrum jumps. Therefore, as argued in \cite{Gukov:2011ry}, it is anticipated that there exist differentials in the knot homology which capture jumps of the BPS spectrum. In the uncolored case \cite{Gukov:2005qp}, the structure of the Kauffman homology has been studied. Moreover, it is natural to expect that the colored Kauffman homology  incorporates rich differential structure as in colored HOMFLY homology \cite{Gukov:2011ry,Gorsky:2013jxa}. Hence, the main goal of this paper is to investigate the differential structure of the colored Kauffman homology.


\subsection{B-model description}
Mirror symmetry relates the A-model on $M$  to B-model on the  mirror manifold $ M^\vee$.
For non-compact toric Calabi-Yau $M$, the mirror manifold $ M^\vee$ \cite{Hori:2000ck}
is given by 
\be
uv= H(x,y;a)\label{mirror}
\ee
where the spectral  (holomorphic) curve $H(x,y;a)=0$ with complex structure $a$ can be viewed as the moduli space of the canonical Lagrangian brane \cite{Aganagic:2000gs}. For instance, the spectral curve for the mirror  manifold of the resolved conifold \cite{Aganagic:2000gs} is expressed by
\bea\label{unknot-curve}
H_{\includegraphics[width=.3cm]{unknot}}(x,y;a)= 1-ax-y+a^{-1}xy~,
\eea
where  the canonical Lagrangian brane wraps the submanifold $\cL_{\includegraphics[width=.3cm]{unknot}}$ corresponding to the unknot. It was pointed out in  \cite{Aganagic:2001nx} that  there is an  ambiguity that preserves the geometry of the brane at infinity. It turns out that this corresponds to the mirror geometry for the configuration of the framed unknot ${\raisebox{-.05cm}{\includegraphics[width=.3cm]{unknot}}}_f$, where the spectral curve can be obtained \cite{Aganagic:2001nx,Brini:2011wi}  by the modular transformation $T^f$ of the curve \eqref{unknot-curve}
\bea\label{framed-unknot-curve}
H_{\includegraphics[width=.3cm]{unknot}_f}(x,y;a)= y^f(1-y)+a^{f-1}x(y-a^2)~.
\eea
Generalizing this, it was shown in \cite{Brini:2011wi} that the spectral curve corresponding to the configuration for the torus knot $T_{Q,P}$ can be derived by the $SL(2,\bZ)$ transformation of the curve \eqref{unknot-curve} as we obtain the Rosso-Jones formula \eqref{rosso-jones}:
\bea
H_{T_{Q,P}}(x,y;a)= y^P(1-y)^Q+a^{P-Q}x(y-a^2)^Q~.
\eea

The next step is to understand the mirror geometry of the  resolved conifold  with the Lagrangian submanifold $\cL_{T_{Q,P}}$ for the torus knot $T_{Q,P}$ in the presence of an orientifold. 
Mirror symmetry maps an anti-holomorphic involution of the A-model into a holomorphic involution of the B-model. The holomorphic involution on the manifold 
\be
{X}^\vee=\{(u,v,x,y) \in\bC\times\bC\times\bC^\times\times\bC^\times| uv=H_{\includegraphics[width=.3cm]{unknot}}(x,y;a)\}
\ee
mirror to the orientifold action on the resolved conifold $X$ was explicitly written in  \cite{Acharya:2002ag}:
\bea
(u,v,x,y) \to (-u,-ax^{-1}y^{-1}v,x^{-1},a^2 y^{-1}) ~.
\eea
This holomorphic involution can be extended to the geometry $uv=H_{T_{Q,P}}(x,y;a)$ mirror to the resolved conifold with M5-branes wrapping on the Lagrangian submanifold $\cL_{T_{Q,P}}$ associated to the torus knot $T_{Q,P}$ in such a way that
\bea\label{involution-torus}
(u,v,x,y) \to (-u,-a^{P+Q}x^{-1}y^{-(P+Q)}v,x^{-1},a^2 y^{-1}) ~.
\eea 
Hence, the geometry mirror to the configuration for the torus knot $T_{Q,P}$ in the presence of an orientifold is $uv=H_{T_{Q,P}}(x,y;a)$  with the involution \eqref{involution-torus}.

On the other hand, in \cite{Aganagic:2012jb}, the B-model description has been considered in the context of the SYZ formulation \cite{Strominger:1996it}. Given a Lagrangian brane whose topology is $S^1 \times \bR^2$, the moduli space receives the disc instanton corrections depending on the Lagrangian brane. Thus, even with the same resolved conifold background, the disc corrected moduli space of $\cL_K$ is dependent of a knot $K$. Furthermore, it is conjectured in \cite{Aganagic:2012jb} that the disc-corrected moduli space of $\cL_K$ is given by the $a$-deformed $A$-polynomial of $SU$-type for a knot $K$
\be
A^{SU}(K;x,y; a)=0~,
\ee
and the corresponding mirror manifold ${X}^\vee_K$ \eqref{mirror} is expressed as
\be
uv=A^{SU}(K;x,y; a)~.
\ee
The detailed explanation for the $a$-deformed $A$-polynomial of $SU$-type will be given in \S \ref{sec:super-a-polynomials}. Note that this conjecture encompasses any knots including non-torus knots. 

Although  the $a$-deformed $A$-polynomial of $SU$-type for  the framed unknot ${\raisebox{-.05cm}{\includegraphics[width=.3cm]{unknot}}}_f$ coincides with the spectral curve $H_{\includegraphics[width=.3cm]{unknot}_f}(x,y;a)$ given in \eqref{framed-unknot-curve} with suitable change of variables, they are no longer the same for general torus knots. For instance, the spectral curve $H_{T_{2,3}}(x,y;a)=0$ for the trefoil is of genus zero  while the zero locus of the $a$-deformed $A$-polynomial of $SU$-type for  the trefoil  $A^{SU}({\bf 3_1};x,y; a)=0$ determines a curve of genus one. Further study has to be undertaken in order to understand the relation between the two descriptions in the case of torus knots. In particular, it is important to study whether the application of  the topological recursions \cite{Eynard:2007kz} to the $a$-deformed $A$-polynomial of $SU$-type $A^{SU}(K;x,y; a)=0$ would provide the large color asymptotic expansion of the colored HOMFLY polynomial $P_{[r]}(K;a,q)$ as done in the case of colored Jones polynomials \cite{Dijkgraaf:2010ur,Borot:2012cw}.

Following the generalized SYZ formulation \cite{Aganagic:2012jb}, it would be easy to conjecture that the disc-corrected moduli space of the Lagrangian brane associated to a knot $K$ in the presence of an orientifold is given by  
the zero locus of the $\lambda$-deformed  $A$-polynomial $A^{SO}(K;x,y;\lambda)$ of $SO$-type. Nevertheless, the authors would like to emphasize that it is desirable to provide some support for the generalized SYZ conjecture \cite{Aganagic:2012jb}  involving a non-trivial knot first in the $SU$ context.


\section{Quadruply-graded Kauffman homology}\label{sec:quad-differential}
\subsection{Review of quadruply-graded HOMFLY homology}
In the case of the categorifications, the realization of knot homologies as the space of certain BPS states has given  rise to various predictions on the structure of the colored HOMFLY homology.
First, it was predicted in \cite{Dunfield:2005si} that there exists a triply-graded ($(a,q,t)$-graded) HOMFLY homology  $(\scH^\HOMFLY_{\yng(1)}(K))_{i,j,k}$, whose graded Euler characteristic is given by the HOMFLY polynomial $P_{\yng(1)}(K;a,q)$. It is endowed  with a set of anti-commuting differentials $\{d_N\}_{N\in\bZ}$ where the homology with respect to $d_{N>0}$ is isomorphic to the $\fraksl(N)$ homology \cite{Khovanov:2004}, which categorifies the $\fraksl(N)$ quantum invariant $P^{\fraksl(N)}_{\yng(1)}(K;q)=P_{\yng(1)}(K;a=q^N,q)$:
\bea\label{dn}
H_*(\scH^\HOMFLY_{\yng(1)}(K),d_{N>0}) \cong \scH^{\fraksl(N)}_{\yng(1)}(K)~.
\eea
In the sequel, the uncolored triply-graded HOMFLY homology \cite{Khovanov:2005} and the differentials $d_N$ \cite{Rasmussen:2006} were put on mathematically rigorous footing.

In \cite{Gukov:2011ry}, this approach has been extended to the colored case. Especially, the concrete study has been carried out for the HOMFLY homology carrying the symmetric and anti-symmetric representations.  It turns out that the $[r]$-colored ($[1^r]$-colored) HOMFLY homology $\scH^\HOMFLY_{[r]}(K)$  ($\scH^\HOMFLY_{[1^r]}(K)$)  is endowed not only with the $\fraksl(N)$ differentials $d_N$ but also with the colored differentials  $d^{\pm}_{[{r}]\to [{\ell}]}$  ($d^{\pm}_{[{1^r}]\to [{1^\ell}]}$). The colored differentials descend the original homology to those with lower-rank representations
\bea\label{colored-diff-sym}
H_*(\scH^\HOMFLY_{[r]}(K),d^\pm_{[{r}]\to [{\ell}]})&\cong& \scH^\HOMFLY_{[\ell]}(K)~,\cr
H_*(\scH^\HOMFLY_{[1^r]}(K),d^\pm_{[{1^r}]\to [{1^\ell}]})&\cong& \scH^\HOMFLY_{[1^\ell]}(K)~,
\eea
where $r>\ell\ge0$. In addition, it was realized in \cite{Gukov:2011ry,Morozov:2012am} that the $[r]$-colored HOMFLY homology for certain classes of knots, such as thin knots and torus knots, exhibits the \emph{exponential growth property}  
\bea
\sum_{i,j,k}a^i t^k \dim\;(\scH^\HOMFLY_{[r]}(K))_{i,j,k}=\Big[ \sum_{i,j,k } a^i t^k \dim\;(\scH^\HOMFLY_{\yng(1)}(K))_{i,j,k}\Big]^r~.
\eea
Furthermore, it was conjectured that  there exists an isomorphism between $\scH^\HOMFLY_{R}(K)$ and $\scH^\HOMFLY_{R^t}(K)$  for an arbitrary representation $R$
\bea
\phi:( \scH^\HOMFLY_R(K) )_{i,j,*} \xrightarrow{\cong}( \scH^\HOMFLY_{R^t}(K)  )_{i,-j,*}~,
\eea
Let us note that the representation $R^t$ is the transposition of the representation $R$. This involution $\phi$ is called the {\em mirror/transposition symmetry} in \cite{Gukov:2011ry}. Actually, the involution $\phi$ exchanges the positive and negative differentials
\be
\phi d_N = d_{-N} \phi~,
\ee 
when the representation $R$ is either a symmetric or an anti-symmetric representation.

In attempting to elucidate the mirror/transposition symmetry,  the two homological gradings denoted by $t_r$ and $t_c$ are introduced so that colored HOMFLY homology turns into quadruply-graded $(\scH^\HOMFLY_{R}(K))_{i,j,k,\ell}$: $(a,q,t_r,t_c)$-gradings \cite{Gorsky:2013jxa}. Particularly, to every generator $x$ of the $[r]$-colored quadruply-graded HOMFLY homology,
one can associate a $\delta$-grading by
\bea\label{delta-grading-HOMFLY}
\delta(x):= a(x)+\frac{q(x)}{2}-\frac{t_r(x)+t_c(x)}{2}~.
\eea
Although the four gradings are independent in general, a knot $K$ is called \emph{homologically-thin} if all generators of $\scH^\HOMFLY_{[r]}(K)$ have the same $\delta$-grading which is equal to $\frac{r}{2}S(K)$ where we denote the $S$-invariant of the knot $K$ by $S(K)$  \cite{Rasmussen:2004}.
Moreover, it became apparent that all of the structural properties and isomorphisms
become particularly elegant with the introduction of the $Q$-grading defined by
\be\label{Q}
Q(x) := \frac{q(x)+t_r(x)-t_c(x)}{\rho}~.
\ee
when the representation is specified by a rectangular  Young diagram $[r^\rho]$. While it is just a regrading of $(\scH^\HOMFLY_{[r^\rho]}(K))_{i,j,k,\ell}$,  it is named the {\it tilde-version} of colored HOMFLY homology $\wt\scH^\HOMFLY_{[r^\rho]}(K)$ due to its importance:
\be
(\wt\scH^\HOMFLY_{[r^\rho]}(K) )_{i,j,k,l} \; := \; (\scH^\HOMFLY_{[r^\rho]}(K))_{i,\rho j-k+l,k,l}~.
\ee
It is the uncolored case only when the two $t$-gradings coincide and therefore the resulting homology is triply-graded in agreement with \cite{Dunfield:2005si}. It simply follows from \eqref{Q} that the $q$- and $Q$-gradings of the uncolored homology are the same.

 The definite advantage of the quadruply-graded theory is that it makes
all of the structural features and isomorphisms completely explicit.  To see them, let us define the Poincar\'e polynomial of the quadruply-graded homology:
\bea
\scP_{[r^\rho]}(K;a,q, t_r,t_c)&:=& \sum_{i,j,k,\ell} a^i q^j t_r^k t_c^\ell ~\dim\;(\scH^\HOMFLY_{[r^\rho]}(K))_{i,j,k,\ell}~,\cr
\wt\scP_{[r^\rho]}(K;a,Q, t_r,t_c)&:=& \sum_{i,j,k,\ell} a^i Q^j t_r^k t_c^\ell ~\dim\;(\wt\scH^\HOMFLY_{[r^\rho]}(K))_{i,j,k,\ell}~,
\eea
where they are related by
\bea\label{Q-to-q-HOMFLY}
\wt\scP_{[r^\rho]}(K;a,q^\rho,t_rq^{-1},t_cq)=\scP_{[r^\rho]}(K;a,q,t_r,t_c)~.
\eea
Now, let us briefly describe the structural properties of the quadruply-graded colored HOMFLY homology. We refer the reader to \cite{Gorsky:2013jxa} for more detail.

\begin{itemize}
\item {\bf Self-symmetry} (Conjecture 3.1 \cite{Gorsky:2013jxa})\hfill \\
Once we use the tilde-version of the colored HOMFLY homology, a new  symmetry in $\wt\scH^{\HOMFLY}_{[r^\rho]}(K)$ becomes manifest:
\bea
(\wt\scH^{\HOMFLY}_{[r^\rho]}(K))_{i,j,k,\ell}&\cong&(\wt\scH^{\HOMFLY}_{[r^\rho]}(K))_{i,-j,k-\rho j,\ell-rj}~,
\eea
which can be stated at the level of the Poincar\'e polynomial 
\bea\label{self-symmetry} 
\wt\scP_{[r^\rho]}(K;a,Q,t_r,t_c)&=&\wt\scP_{[r^\rho]}(K;a,Q^{-1}t_r^{-\rho}t_c^{-r},t_r,t_c)~.
\eea

\item {\bf  Mirror/Transposition symmetry}  (Conjecture 3.3 and 3.4 \cite{Gorsky:2013jxa})\hfill \\  
The $[\rho^r]$-colored quadruply-graded HOMFLY homology enjoys the mirror/transposition symmetry
\bea
(\wt\scH^{\HOMFLY}_{[\rho^r]}(K))_{i,j,k,\ell}\cong(\wt\scH^{\HOMFLY}_{[r^\rho]}(K))_{i,j,\ell,k}\cong(\wt\scH^{\HOMFLY}_{[r^\rho]}(K))_{i,-j,\ell-\rho j,k-r j}~,~~~~~
\eea
which can be expressed in terms of the Poincar\'e polynomial 
\bea\label{mirror-HOMFLY}
\wt\scP_{[\rho^r]}(K;a,Q,t_r,t_c)&=&\wt\scP_{[r^\rho]}(K;a,Q,t_c,t_r)=\wt\scP_{[r^\rho]}(K;a,Q^{-1}t_c^{-\rho}t_r^{-r},t_c,t_r)~.~~~~
\eea
This lifts the following relation between the colored HOMFLY polynomials
\bea
P_{R^t}(K;a,q)=P_{R}(K;a,-q^{-1})=P_{R}(K;-a,q^{-1})~,
\label{PPmir}
\eea
for any representation $R$.
\item {\bf Refined exponential growth property}  (Conjecture 3.8 and 3.9 \cite{Gorsky:2013jxa})\hfill \\
Let $K$ be either a thin knot or a torus knot. The $[r^\rho]$-colored quadruply-graded HOMFLY homology of the knot $K$ obeys the refined exponential growth property 
\bea
\wt\scP_{[r^\rho]}(K;a,Q,t_r,t_c=1)&=&\left[\wt\scP_{[1^\rho]}(K;a,Q,t_r,t_c=1)\right]^r\label{exp-growth-HOMFLY-1}~,\\
\wt\scP_{[r^\rho]}(K;a,Q,t_r=1,t_c)&=&\left[\wt\scP_{[r]}(K;a,Q,t_r=1,t_c)\right]^\rho\label{exp-growth-HOMFLY-2}~.
\eea
It follows immediately that
\bea
\dim \scH^\HOMFLY_{[r^\rho]}(K)=\left[\dim \scH^\HOMFLY_{\yng(1)}(K)\right]^{r\rho}~.
\eea

The analogous statement at the polynomial level is as follows. For any knot $K$ and an arbitrary representation $R$, the following identity holds \cite{DuninBarkowski:2011yx}:
\bea
P_{R}(K;a,q=1)=\left[P_{\yng(1)}(K;a,q=1)\right]^{|R|}~,
\eea
where $|R|$ is the total number of the Young diagram corresponding to the representation $R$.

\item  {\bf $\fraksl(n|m)$ differentials}  \hfill \\ 
It was proposed in \cite{Gorsky:2013jxa} that the colored HOMFLY homology is actually gifted with a collection of the differentials  $\{d_{n|m}\}$ labeled by two non-negative integers $(n,m)$ associated to the Lie superalgebra $\fraksl(n|m)$. These are generalizations of the differentials $\{d_N\}$ \eqref{dn}. It appears that the representation theory of $\fraksl(n|m)$ explains the behavior of the colored differentials. In this paper, we will not go into the detail about the $\fraksl(n|m)$ differentials. 

\item  {\bf Colored differentials}  \hfill \\ 
For each rectangular Young diagram $[r^\rho]$, one can define colored differentials that remove any number of columns or rows from the original Young diagram $[r^\rho]$. For every $\ell$ with $r>\ell\ge 0$, there are two different column-removing  differentials  $d^\pm_{[r^\rho]\to[\ell^\rho]}$, and for  every $\sigma$ with $\rho>\sigma\ge0$, there are two different row-removing differentials  $d^\pm_{[r^\rho]\to[r^\sigma]}$ on $\wt\scH^\HOMFLY_{[r^\rho]}(K)$. These are generalizations of \eqref{colored-diff-sym}:
\bea\label{colored-diff-sym-gen}
H_*(\wt\scH^\HOMFLY_{[r^\rho]}(K),d^\pm_{[{r^\rho}]\to [{\ell^\rho}]})&\cong& \wt\scH^\HOMFLY_{[\ell^\rho]}(K)~,\cr
H_*(\wt\scH^\HOMFLY_{[r^\rho]}(K),d^\pm_{[{r^\rho}]\to [{r^\sigma}]})&\cong& \wt\scH^\HOMFLY_{[r^\sigma]}(K)~.
\eea
The isomorphisms above involve regrading. One of the striking features of the quadruply-graded homology is that it makes the regrading very explicit. 

\item  {\bf Universal colored differentials}  \hfill \\ 
If the representation is specified either by  $[r,r]$ or by $[2^r]$,  there exists yet another set of colored differentials $d^\uparrow$ or $d^\leftarrow$ so that 
\bea
H_*(\wt\scH^\HOMFLY_{[r,r]}(K),d^\uparrow)&\cong& \wt\scH^\HOMFLY_{[r]}(K)~,\cr
H_*(\wt\scH^\HOMFLY_{[2^r]}(K),d^\leftarrow)&\cong& \wt\scH^\HOMFLY_{[1^r]}(K)~.
\eea
They are called universal colored differentials because they are universal in the sense that their $a$-degree is equal to $0$. 
\end{itemize}

\subsection{Properties of quadruply-graded Kauffman homology}

Let us now discuss about the categorifications of Kauffman polynomials. The properties of the triply-graded homology that categorifies the uncolored Kauffman polynomials have been investigated in \cite{Gukov:2005qp,Khovanov:2007}. Like HOMFLY homology, it is gifted with a collection of the differentials $\{d_N\}_{N\in \bZ}$ so that the homology with respect to the differential $d_{N>1}$ is isomorphic to the $\frakso(N)$ homology, while the homology with respect to the differential $d_{N<0}$ is isomorphic to the $\fraksp(N)$ homology.  Furthermore, it was found in \cite{Gukov:2005qp} through the analysis of the Landau-Ginzburg theory that the most characteristic property of the Kauffman homology is that it contains the HOMFLY homology. More precisely, it is endowed with the so-called \emph{universal differential} $\duuu^\univ$ whose homology is isomorphic with the HOMFLY homology:
\bea\label{universal-uncolor}
H_*(\scH^\Kauffman_{\yng(1)}(K),\duuu^\univ)\cong \scH^\HOMFLY_{\yng(1)}(K)~.
\eea

From the perspective of topological string theory, it is natural to think that there exists the triply-graded homology theory categorifying Kauffman polynomials colored by arbitrary representations. Especially, it is expected that the structure becomes clear if we use quadruple-gradings, as in the case of colored HOMFLY homology, when the colors are specified by rectangular Young tableaux. Hence,
our goal in this section is to clarify all the structural features and isomorphisms in the colored quadruply-graded  Kauffman homology.

As we have seen in \S \ref{sec:CS}, for any representation $R$, there is the $R$-colored reduced Kauffman polynomial $F_R(K;\lambda,q)$ of a knot $K$. We conjecture the existence of the finite-dimensional homology $\scH^\Kauffman_R(K)$ of a knot $K$ categorifying the $R$-colored reduced Kauffman polynomial $F_R(K;\lambda,q)$ of the knot $K$. 
 In this paper, we focus on the case that the representation $R$ is specified by a rectangular Young tableau $[r^\rho]$. In this case, we further conjecture that the $[r^\rho]$-colored Kauffman homology  $(\scH^\Kauffman_{[r^\rho]}(K))_{i,j,k,\ell}$ of a knot $K$ is quadruply-graded so that its Poincar\'e polynomial,
\bea
\scF_{[r^\rho]}(K;\lambda,q, t_r,t_c)&:=& \sum_{i,j,k,\ell} \lambda^i q^j t_r^k t_c^\ell ~\dim\;(\scH^\Kauffman_{[r^\rho]}(K))_{i,j,k,\ell}~,
\eea
reduces to the $[r^\rho]$-colored Kauffman polynomial in the following way:
\bea
\scF_{[r^\rho]}(K;\lambda,q, t_r=-1,t_c=1)=\scF_{[r^\rho]}(K;\lambda,q, t_r=1,t_c=-1)=F_{[r^\rho]}(K;\lambda,q)~.
\eea
As in the case of the HOMFLY homology,
one can associate a $\delta$-grading to every generator $x$ of the $[r]$-colored quadruply-graded Kauffman homology by
\bea\label{delta-grading-Kauffman}
\delta(x):=\frac{3\lambda(x)}{2}+\frac{q(x)}{2}-\frac{t_r(x)+t_c(x)}{2}~.
\eea
For a homologically-thin knot $K_{\rm thin}$, the $\delta$-gradings of all the generators are equal to $r\,S(K_{\rm thin})$.
In addition, by introducing the $Q$-grading \eqref{Q}, we define the tilde-version of the $[r^\rho]$-colored quadruply-graded Kauffman homology
\bea\label{tilde-version-Kauffman}
(\wt\scH_{[r^\rho]}^{\Kauffman}(K))_{i,j,k,\ell}:=(\scH_{[r^\rho]}^{\Kauffman}(K))_{i,\rho j-k+\ell,k,\ell}~,
\eea
and its Poincar\'e polynomial
\bea
\wt\scF_R(K;\lambda,Q, t_r,t_c):= \sum_{i,j,k,\ell} \lambda^i Q^j t_r^k t_c^\ell ~\dim\;(\wt\scH^\Kauffman_{R}(K))_{i,j,k,\ell}~.
\eea
In terms of the Poincar\'e polynomials, the relation \eqref{tilde-version-Kauffman} can be rephrased by
\bea\label{Q-to-q-Kauffman}
\wt\scF_{[r^\rho]}(K;\lambda,q^\rho,t_rq^{-1},t_cq)=\scF_{[r^\rho]}(K;\lambda,q,t_r,t_c)~.
\eea
It is the uncolored case only when the two $t$-gradings coincide and therefore the resulting homology is triply-graded in agreement with \cite{Gukov:2005qp}. It clearly follows from \eqref{Q} that the $q$- and $Q$-gradings of the uncolored homology are the same. 

In what follows, we conjecture the structural properties of the $[r^\rho]$-colored Kauffman homology. Although they are very similar, we predict that there are two differences between the $[r^\rho]$-colored Kauffman homology and the $[r^\rho]$-colored HOMFLY homology. One of the difference is that the $[r^\rho]$-colored Kauffman homology does not enjoy the self-symmetry. The other is that there are differentials which relate the $[r]$-colored Kauffman homology to the $[r]$-colored HOMFLY homology.

\begin{itemize}
\item {\bf  Mirror/Transposition symmetry}\hfill\\ 
The $[r^\rho]$-colored Kauffman homology enjoys mirror/transposition symmetry
\bea\label{mirror-Kauffman-homology}
(\wt\scH^{\Kauffman}_{[\rho^r]}(K))_{i,j,k,\ell}&\cong&(\wt\scH^{\Kauffman}_{[r^\rho]}(K))_{i,-j,\ell-\rho j,k-r j}~,
\eea
which can be rephrased in terms of the Poincar\'e polynomial
\bea\label{mirror-Kauffman}
\wt\scF_{[\rho^r]}(K;\lambda,Q,t_r,t_c)&=&\wt\scF_{[r^\rho]}(K;\lambda,Q^{-1}t_c^{-\rho}t_r^{-r},t_c,t_r)~.
\eea
At the decategorified level, for any representation $R$, there is the comparable symmetry between the $R$-colored  and the $R^t$-colored Kauffman polynomial 
\bea
F_{R^t}(K;\lambda,q)=F_{R}(K;-\lambda,q^{-1})=F_{R}(K;\lambda,-q^{-1})~.
\eea
Only in the uncolored case can the mirror/transposition symmetry be regarded as the self-symmetry.
\item {\bf Refined exponential growth property}  \hfill \\
Let $K$ be a thin knot or a torus knot. Then, the Kauffman homology carrying a rectangular Young tableau possesses the refined exponential growth property
\bea\label{exp-growth}
\wt\scF_{[r^\rho]}(K;\lambda,Q,t_r,t_c=1)&=&\left[\wt\scF_{[1^\rho]}(K;\lambda,Q,t_r,t_c=1)\right]^r~,\cr
\wt\scF_{[r^\rho]}(K;\lambda,Q,t_r=1,t_c)&=&\left[\wt\scF_{[r]}(K;\lambda,Q,t_r=1,t_c)\right]^\rho~.
\eea
The analogous statement at the polynomial level is as follows. For any knot $K$ and an arbitrary representation $R$, the following identity holds:
\bea
F_{R}(K;a,q=1)=\left[F_{\yng(1)}(K;a,q=1)\right]^{|R|}~.
\eea

\item {\bf  $\frakso/\fraksp$ differentials} \hfill \\
The colored Kauffman homology is gifted with a set of the differentials $d_N$ so that the homology of $\scH^\Kauffman_{R}(K)$ with
respect to $d_N$ is isomorphic either to the $\frakso(N)$ homology carrying the representation $R$ for $N\ge2$ or to  the $\fraksp(N)$ homology carrying the representation $R^t$ for $N\le-2$ 
\bea\label{sosp-diff}
H_*(\scH^\Kauffman_{R}(K),d_N)\cong \left\{ \begin{array}{ll} \scH^{\frakso(N)}_R& (N\ge2) \\  \scH^{\fraksp(-N)}_{R^t} & (N\le-2) \end{array} \right. ~.
\eea

\item {\bf Universal differentials} \hfill \\
The $[r^\rho]$-colored  Kauffman homology is endowed with the universal differential $d^\univ$  so that the homology with respect to the universal differential is isomorphic to the  $[r^\rho]$-colored   HOMFLY homology
\bea
H_*(\wt\scH^\Kauffman_{[r^\rho]}(K),d^\univ)&\cong&\wt \scH^\HOMFLY_{[r^\rho]}(K)~.
\eea
They are  universal in the sense that their $\lambda$-degree is equal to $0$.

\item {\bf Diagonal differentials} \hfill \\
We conjecture  the existence of other differentials whose homologies in the $[r^\rho]$-colored Kauffman homology are isomorphic to the $[r^\rho]$-colored HOMFLY homology. We call them diagonal differentials, $d^\pm_{\rm diag}$, so that
\bea
H_*(\wt\scH^\Kauffman_{[r^\rho]}(K),d^\pm_{\rm diag})&\cong&\wt \scH^\HOMFLY_{[r^\rho]}(K)~.
\eea
They are not universal since their $\lambda$-degree is equal to $-1$. They are diagonal in the sense that every generator $x\in H_*(\wt\scH^\Kauffman_{[r^\rho]}(K),d^\pm_{\rm diag})$ obeys the grading relation
\bea
\lambda(x)\mp Q(x)=r \;S(K)~.
\eea

\item {\bf  Colored differentials} \hfill \\
There exists a collection of colored differentials that send the colored Kauffman homology to those with the lower-rank representations.
\bea
H_*(\wt\scH^\Kauffman_{[r^\rho]}(K),d^\pm_{[r^\rho]\to[\ell^\rho]})&\cong&\wt \scH^\Kauffman_{[\ell^\rho]}(K)~,\cr
H_*(\wt\scH^\Kauffman_{[r^\rho]}(K),d^\pm_{[r^\rho]\to[r^\sigma]})&\cong&\wt \scH^\Kauffman_{[r^\sigma]}(K)~.
\eea
It should be stressed that the existence of the colored differentials becomes manifest only when we use the tilde-version of the colored Kauffman homology.

\item {\bf Universal colored differentials} \hfill \\
If the representation is specified either by  $[r,r]$ or by $[2^r]$,  there exists yet another set of colored differentials $d^\uparrow$ or $d^\leftarrow$ so that 
\bea
H_*(\wt\scH^\Kauffman_{[r,r]}(K),d^\uparrow)&\cong& \wt\scH^\Kauffman_{[r]}(K)~,\cr
H_*(\wt\scH^\Kauffman_{[2^r]}(K),d^\leftarrow)&\cong& \wt\scH^\Kauffman_{[1^r]}(K)~.
\eea
They are called universal colored differentials because they are universal in the sense that their $\lambda$-degree is equal to $0$.

\end{itemize}

In the following subsections, we shall explicate all the differentials in detail. Since the size of the colored Kauffman homology is too large for concrete study of arbitrary rectangular Young tableaux, we restrict ourselves to the case that the representations are specified by the Young tableaux $[r]$ and their transpositions $[1^r]$. 

Before moving on to the next subsection, let us define the Poincar\'e polynomial of the homology with respect to a differential $d$ in  the HOMFLY homology 
\bea
\scP(\scH^\HOMFLY_{R}(K),d)(a,q,t_r,t_c)&:=& \sum_{i,j,k,\ell} a^i q^j t_r^k t_c^\ell ~\dim H_*(\scH^\HOMFLY_{R}(K),d)_{i,j,k,\ell}~,\cr
\wt\scP(\wt\scH^\HOMFLY_{R}(K),d)(a,Q,t_r,t_c)&:=& \sum_{i,j,k,\ell} a^i Q^j t_r^k t_c^\ell ~\dim H_*(\wt\scH^\HOMFLY_{R}(K),d)_{i,j,k,\ell}~,~~~~~~~~~
\eea
and the Poincar\'e polynomial of  the homology  with respect to a differential $d$ in the Kauffman homology  
\bea
\wt\scF(\wt\scH^\Kauffman_{R}(K),d)(\lambda,Q,t_r,t_c)&:=& \sum_{i,j,k,\ell} \lambda^i Q^j t_r^k t_c^\ell ~\dim H_*(\wt\scH^\Kauffman_{R}(K),d)_{i,j,k,\ell}~,\cr
\scF(\scH^\Kauffman_{R}(K),d)(\lambda,q,t_r,t_c)&:=& \sum_{i,j,k,\ell} \lambda^i q^j t_r^k t_c^\ell ~\dim H_*(\scH^\Kauffman_{R}(K),d)_{i,j,k,\ell}~.~~~~~~~~~~
\eea

\subsection{$\frakso/\fraksp$ differentials}
There is a set of  the differentials $\{d_N\}_{N\in\bZ}$ inherent in the colored Kauffman homology so that the homology with respect to $d_N$ is isomorphic to the 
colored $\frakso/\fraksp$ homology \eqref{sosp-diff}.\footnote{There is a certain issue in the $d_N$ differential on Kauffman homology of a thick knot, which we discuss in the beginning of  \S\ref{sec:discussions}. However, the examples we deal with in this paper are all thin knots so that it is not relevant in this paper.} To see the isomorphism \eqref{sosp-diff} explicitly, it is convenient  to use the Poincar\'e polynomials in the $(\lambda,q,t_r,t_c)$-gradings.  Specifically, in the case of the $[r]$-colored Kauffman homology, we have the following identities at the level of the Poincar\'e polynomials
\bea
\scF(\scH^\Kauffman_{[r]}(K),d_N)(\lambda=q^{N-1},q,t,1)=\left\{ \begin{array}{ll} \scP^{{\frakso}(N)}_{[r]}(K;q,t) & (N\ge2) \\ t^{rS(K)} \scP^{{\fraksp}(-N)}_{[1^r]}(K;q^{-1},t) & (N\le -2r) \end{array} \right.,~~~~
\eea
where the $(\lambda,q,t_r,t_c)$-degrees of the differential $d_N$ acting on $\scH^\Kauffman_{[r]}(K)$ are
\bea\label{dN-degree}
\deg d_N\Big|_{\scH^\Kauffman_{[r]}(K)}=\left\{ \begin{array}{ll} (-1,N-1,-1,*) & (N\ge2) \\  (-1,N-1,N-1,*) & (N\le-2r) \end{array} \right.~.
\eea
The differential $d_{-2}$, which specializes the uncolored Kauffman homology to the $\fraksp(2)$ homology, acts nontrivially even on the Kauffman homology of a thin knot \cite{Gukov:2005qp,Gukov:2011ry}.

On the other hand, the $(\lambda,q,t_r,t_c)$-degrees of the differential $d_N$ acting on $\scH^\Kauffman_{[1^r]}(K)$ are
\bea
\deg d_N\Big|_{\scH^\Kauffman_{[1^r]}(K)}=\left\{ \begin{array}{ll} (-1,N-1,*,-1) & (N\ge r+2) \\  (-1,N-1,*,N-1) & (N\le-2) \end{array} \right.~,
\eea
so that  we can see the isomorphism \eqref{sosp-diff} in terms of the Poincar\'e polynomials in the following way:
\bea
\scF(\scH^\Kauffman_{[1^r]}(K),d_N)(\lambda=q^{N-1},q,t,1)=\left\{ \begin{array}{ll} \scP^{{\frakso}(N)}_{[1^r]}(K;q,t) & (N\ge r+2) \\ t^{rS(K)} \scP^{{\fraksp}(-N)}_{[r]}(K;q^{-1},t) & (N\le-2) \end{array} \right. .~~~~~~
\eea

\subsection{Relations from representation theory}
The $\frakso/\fraksp$ specializations by using the differentials $d_N$ are useful to determine the colored Kauffman homology. In fact, there are several isomorphisms of  representations which lead to nontrivial identities among the homological invariants.
\begin{itemize}
\item It is well-known that the vector representation $\yng(1)$ of $\frakso(3)$ is isomorphic to the spin-1 representation $\yng(2)$ of $\fraksl(2)$. Moreover, since this relation can be extended to the symmetric product
\be
(\frakso(3), [r]) \cong (\fraksl(2), [2r])~,
\ee
we have the isomorphism between the $[r]$-colored $\frakso(3)$ homology and the $[2r]$-colored $\fraksl(2)$ homology in the $t_r$-grading
\bea
&&\scF( \scH^{\Kauffman}_{[r]}(K),d_3)(\lambda=q^2,q,t_r,t_c=1) \cr
&&\hspace{2cm}=\scP( \scH^{\HOMFLY}_{[2r]}(K),d_2)(a=q,q^{1/2},t_r,t_c=1)~.
\eea
Particularly, since the differentials act trivially for a thin knot $K_{\rm thin}$, the naive substitutions lead to the identity 
\bea\label{so3-sl2-thin}
\scF_{[r]}(K_{\rm thin};q^2,q,t_r,1)=\scP_{[2r]}(K_{\rm thin};q,q^{1/2},t_r,1)~.
\eea
\item In addition, since $\frakso(4)$ is isomorphic to $\fraksl(2)\oplus \fraksl(2)$ as Lie algebras, we have
\bea
&&\scF( \scH^{\Kauffman}_{[r]}(K),d_4)(\lambda=q^3,q,t_r,t_c=1) \cr
&&\hspace{2cm}= \left[\scP( \scH^{\HOMFLY}_{[r]}(K),d_2)(a=q^2,q,t_r,t_c=1)\right]^2~.
\eea
In particular, for a thin knot $K_{\rm thin}$, the identity holds even with the $t_c$-gradings
\bea\label{so4-sl2sl2-thin}
\scF_{[r]}(K_{\rm thin};q^3,q,t_r ,t_c )=\left[\scP_{[r]}(K_{\rm thin};q^2,q,t_r,t_c )\right]^2~.
\eea

\item  The isomorphism between $\fraksp(2)$ and $\fraksl(2)$ leads to
\bea
&&\scF( \scH^{\Kauffman}_{[1^r]}(K),d_{-2})(\lambda=q^{3},q^{-1},1,t)  \cr
&&\hspace{3cm}=t^{r S(K)} \scP( \scH^{\HOMFLY}_{[r]}(K),d_2)(a=q^{4},q^{2},t,1)~.
\eea
Moreover,  the differential $d_{-2}$ can be evident in the $(\lambda,Q,t_r,t_c)$-grading whose degree is $(-1,-3,-3r,-3)$ on $\wt\scH_{[1^r]}^\Kauffman(K_{\rm thin})$. We predict that the Poincar\'e polynomial of the homology $H_*(\wt\scH^\Kauffman_{[1^r]}(K),d_{-2})$ with respect to the differential $d_{-2}$ can be expressed in terms of the $[r]$-colored HOMFLY homology
\bea\label{d-2}
&&\wt\scF( \wt\scH^{\Kauffman}_{[1^r]}(K_{\rm thin}),d_{-2})(\lambda,Q,t_r,t_c)\\
&& =\lambda^{\frac{rS(K)}{2}} Q^{\frac{3rS(K)}{2}} t_r^{\frac{r(3r-1)S(K)}{2}}  t_c^{rS(K)} \wt \scP_{[r]}(K_{\rm thin};\lambda Q^{-1}t_r^{1-r},(\lambda Q^{-1} t_r^{1-r})^{\frac12},t_r^{-1}t_c, t_r^{2} )~.\nonumber
\eea

\item Furthermore, the isomorphism of the representations, 
\be
(\frakso(6), [r]) \cong (\fraksl(4), [r,r])~,
\ee
provides us with the isomorphism between the $[r]$-colored $\frakso(6)$ homology and the $[r,r]$-colored $\fraksl(4)$ homology in the $t_c$-grading
\bea \label{sl4-so6}
&&\scF( \scH^{\Kauffman}_{[r]}(K),d_6)(\lambda=q^5,q,t_r=1,t_c) \cr
&&\hspace{2.5cm}= \scP( \scH^{\HOMFLY}_{[r,r]}(K),d_4)(a=q^4,q,t_r=1,t_c )~.
\eea
Specifically for a thin knot $K_{\rm thin}$, we have the following relation:
\bea\label{thin-so6-sl4}
\scF_{[r]}(K_{\rm thin};q^5,q,1,t_c)=\scP_{[r,r]}(K_{\rm thin};q^4,q^2,1,t_c)~.
\eea
Interestingly, the relation \eqref{sl4-so6} from representation theory sheds  new light on the structure of the $[r]$-colored Kauffman homology. In fact, the relation \eqref{sl4-so6} implies that the differential structure of $\scH^{\fraksl(4)}_{[r,r]}(K)$ can be mapped to that of $\scH^{\frakso(6)}_{[r]}(K)$. Furthermore, the $[r]$-colored Kauffman homology of a thin knot is expected to have a similar differential structure to the $[r,r]$-colored HOMFLY homology because there is a one-to-one correspondence between the generators  of both  the homologies through \eqref{thin-so6-sl4}. Actually, in the $[r]$-colored Kauffman homology, one can find the counterparts of all the colored differentials inherent in the $[r,r]$-colored HOMFLY homology. This can be seen in Table \ref{comparison}, where the differentials are related by
\bea
(5\lambda+Q-t_r+t_c)[d^\Kauffman]=\pm(4a+2Q-t_r+t_c)[d^\HOMFLY]~,
\eea
with the same $t_c$-degrees $t_c[d^\Kauffman]=t_c[d^\HOMFLY]$ since \eqref{thin-so6-sl4} can be written in terms of the tilde-version
\bea
&&\wt\scF_{[r]}(K_{\rm thin};\lambda=q^5,Q=q,t_r=q^{-1},t_c= qt )\cr
&&\hspace{3cm}=\wt\scP_{[r,r]}(K_{\rm thin};a=q^4,Q=q^2,t_r=q^{-1},t_c=  qt )~.
\eea

\end{itemize}

\begin{table}[h]\centering
\begin{tabular}{c|c||c|c}
\multicolumn{2}{c||}{$\wt\scH^{\rm Kauffman}_{[r]}$(K)} & \multicolumn{2}{c}{$\wt\scH^{\rm HOMFLY}_{[r,r]}$(K)} \\
\hline
differentials & $(\lambda,Q,t_r,t_c)$-degrees& differentials&  $(a,Q,t_r,t_c)$-degrees\\
\hline
$ d^{\rm univ}_{\to}$& $(0,2,1,1)$ & $d^+_{[r,r]\to[r]}$ & $(-2,2,-3,-1)$ \\ 
\hline
$ d^-_{\rm diag}$& $(-1,-3,-3,-2r-1)$ & $d^-_{[r,r]\to[r]}$ & $(-2,-2,-7,-2r-1)$ \\ 
\hline
$d_{[r]\to[\ell]}^+$ &$(-1,1,-1,-2\ell-1)$ & $d^+_{[r,r] \to [\ell,\ell]}$ & $(-2,2,-1,-2\ell-1)$ \\
\hline
$d_{[r]\to[\ell]}^-$ &$(-2,0,-3,-2\ell-2r-1)$ & $d^-_{[r,r] \to [\ell,\ell]}$ & $(-2,-2,-5,-2\ell-2r-1)$ \\
\hline
$d^\leftarrow \ (r=2)$ &$(0,0,0,2)$ & $d^{\leftarrow} \ (r=2)$ & $(0,0,0,2)$ \\
\hline
\end{tabular}
\caption{Comparison of the differential structure of $[r]$-colored Kauffman homology with that of $[r,r]$-colored HOMFLY homology. In the last four rows, we have the grading relation  $5\lambda+Q-t_r+t_c=4a+2Q-t_r+t_c$, and the $t_c$-degrees are the same. If we reverse the direction of the universal differential $ d^{\rm univ}_{\to}$, theses relations still hold in the first row.} \label{comparison}
\end{table}


\subsection{Universal differentials}

In this subsection we discuss the universal differential acting on colored Kauffman homology. The most typical feature in the differential structure of uncolored Kauffman homology is the existence of the universal differentials that relate the Kauffman homology to the HOMFLY homology \cite{Gukov:2005qp}. It is interesting to ask if there are extensions of the universal differentials to the higher rank representations.
In \cite{Gukov:2011ry}, by using the relation \eqref{sl4-so6}, the differential $d^+_{[1,1]\to[1]}$ on the $[1,1]$-colored HOMFLY homology was constructed from the universal differential $d_{\to}^{\rm univ}$ on the uncolored Kauffman homology. Reversing the direction, the differential $d^+_{[r,r]\to[r]}$ now accounts for the existence of the universal differential $d^{\rm univ}_{\rightarrow}$ on $\wt\scH^{\Kauffman}_{[r]}(K)$. Additionally, the mirror/transposition symmetry ensures that there exists  the differential $d^{\rm univ}_{\leftarrow}$ on $\wt\scH^{\Kauffman}_{[1^r]}(K)$. It is the uncolored case only when the Kauffman homology is endowed with both of the universal differentials $\duuu^\univ$ \eqref{universal-uncolor}.

It turns out that the $(\lambda, Q, t_r, t_c)$-degrees of the universal differentials $\duuu^{\rm univ}$  are
\bea
\wt \deg ~d^{\rm univ}_{\rightarrow}&=&(0,2,1,1)~,\cr
\wt \deg ~ d^{\rm univ}_{\leftarrow}&=&(0,-2,-2r+1,-1)~.
\eea
Under the action of the universal differentials, it becomes manifest that the $[r]$-colored ($[1^r]$-colored) Kauffman homology contains the $[r]$-colored ($[1^r]$-colored) HOMFLY homology in such a way that 
\bea\label{univ-diff-regrading}
\wt \scF(\wt\scH^{\Kauffman}_{[r]}(K),d^{\rm univ}_{\to})(\lambda, Q, t_r, t_c)&=&Q^{rS(K)}\wt\scP_{[r]}(K;\lambda Q^{-1}, Q, t_r, t_c)~,\\
\wt \scF(\wt\scH^{\Kauffman}_{[1^r]}(K),d^{\rm univ}_{\leftarrow})(\lambda, Q, t_r, t_c)&=&Q^{-rS(K)}t_r^{-r^2S(K)}t_c^{-rS(K)} \wt\scP_{[1^r]}(K;\lambda Q t_r^rt_c, Q, t_r, t_c)~.\nonumber
\eea

\subsection{Diagonal differentials}
The relation \eqref{sl4-so6} predicts the existence of  the differential in $\wt\scH^{\Kauffman}_{[r]}(K)$ which corresponds to the differential $d^-_{[r,r]\to[r]}$ in $\wt\scH^{\HOMFLY}_{[r,r]}(K)$. In fact, it is easy to find such a differential $d^-_{\rm diag}$ as well as its cousin $d^+_{\rm diag}$  whose  $(\lambda,Q,t_r,t_c)$-degrees on $\wt\scH^{\Kauffman}_{[r]}(K)$ are
\bea
\wt \deg ~d^+_{\rm diag}\Big|_{\wt\scH^{\Kauffman}_{[r]}(K)}&=&(-1,3,0,0)~,\cr
\wt \deg ~ d^-_{\rm diag}\Big|_{\wt\scH^{\Kauffman}_{[r]}(K)}&=&(-1,-3,-3,-2r-1)~.
\eea
We call them the diagonal differentials since every generator $x\in H_*(\wt\scH^\Kauffman_{[r]}(K),d^\pm_{\rm diag})$ is subject to the grading relation
\bea
\lambda(x)\mp Q(x)=r \;S(K)~.
\eea
Similar to the universal differentials, the homology $ H_*(\wt\scH^\Kauffman_{[r]}(K),d^\pm_{\rm diag})$ with respect to the diagonal differentials is isomorphic to the $[r]$-colored HOMFLY homology, where the precise grading changes are given by
\bea\label{diag-diff-regrading}
&&\wt \scF(\wt\scH^{\Kauffman}_{[r]}(K),d^+_{\rm diag})(\lambda, Q, t_r, t_c)\cr
&&\hspace{1cm}=\lambda^{\frac{rS(K)}{2}}Q^{-\frac{3rS(K)}{2}}t_r^{-\frac{rS(K)}{2}}t_c^{-\frac{rS(K)}{2}}\wt\scP_{[r]}(K;\lambda Q t_rt_c, (\lambda Q t_rt_c)^{\frac12}, t_rt_c^{-1}, t_c^2)~,\cr
&&\wt \scF(\wt\scH^{\Kauffman}_{[r]}(K),d^-_{\rm diag})(\lambda, Q, t_r, t_c)\cr
&&\hspace{1cm}=\lambda^{\frac{rS(K)}{2}} Q^{\frac{3rS(K)}{2}}   t_r^{rS(K)} t_c^{r^2S(K)}\wt\scP_{[r]}(K;\lambda Q^{-1} , (\lambda Q^{-1} )^{\frac12}, t_r, t_c)~.
\eea
It straightforwardly follows from the mirror/transposition symmetry that the  $(\lambda,Q,t_r,t_c)$-degrees of the diagonal differentials $d^\pm_{\rm diag}$ on $\wt\scH^{\Kauffman}_{[1^r]}(K)$ are
\bea
\wt \deg ~d^+_{\rm diag}\Big|_{\wt\scH^{\Kauffman}_{[1^r]}(K)}&=&(-1,3,r-1,0)~,\cr
\wt \deg ~ d^-_{\rm diag}\Big|_{\wt\scH^{\Kauffman}_{[1^r]}(K)}&=&(-1,-3,-3r,-3)~.
\eea
The homology $ H_*(\wt\scH^\Kauffman_{[1^r]}(K),d^\pm_{\rm diag})$ is isomorphic to the $[1^r]$-colored HOMFLY homology
 \bea
 &&\wt \scF(\wt\scH^{\Kauffman}_{[1^r]}(K),d^+_{\rm diag})(\lambda, Q, t_r, t_c)\cr
&&\hspace{1cm}=\lambda^{\frac{rS(K)}{2}} Q^{-\frac{3rS(K)}{2}}   t_r^{-\frac{r^2S(K)}{2}} t_c^{-\frac{rS(K)}{2}}\wt\scP_{[1^r]}(K;\lambda Q t_r^r t_c , (\lambda Qt_r^r t_c)^{\frac12}, t_r, t_c)~,\cr
&&\wt\scF( \wt\scH^{\Kauffman}_{[1^r]}(K),d^-_{\rm diag})(\lambda,Q,t_r,t_c)\cr
&& =\lambda^{\frac{rS(K)}{2}} Q^{\frac{3rS(K)}{2}} t_r^{\frac{r(3r-1)S(K)}{2}}  t_c^{rS(K)} \wt \scP_{[1^r]}(K;\lambda Q^{-1}t_r^{1-r},(\lambda Q^{-1} t_r^{1-r})^{\frac12}, t_r^{2},t_r^{-1}t_c )~.~~~~~
\eea
It turns out that the diagonal differential $d^-_{\rm diag}$ on $\wt\scH^\Kauffman_{[1^r]}(K)$ coincides with the  differential $d_{-2}$. (See \eqref{d-2}.)

\subsection{Colored differentials}\label{sec:color-diff}
Analogous to the colored differentials in HOMFLY homology, the colored Kauffman homology is also equipped with colored differentials which send to the lower rank colored Kauffman homology:
\bea\label{Kauffman-color-diff}
&&H_*(\wt\scH^\Kauffman_{[r]}(K),d^\pm_{[r]\to [\ell]})\cong\wt\scH^\Kauffman_{[\ell]}(K) ~,\cr
&& H_*(\wt\scH^\Kauffman_{[1^r]}(K),d^\pm_{[1^r]\to [1^\ell]})\cong\wt\scH^\Kauffman_{[1^\ell]}(K)~.
\eea
It turns out that, in the context of \eqref{sl4-so6}, the differentials $d^\pm_{[r]\to[\ell]}$ inherent in the $[r]$-colored Kauffman homology  correspond to the differentials $d^\pm_{[r,r] \to [\ell,\ell]}$ innate in the $[r,r]$-colored HOMFLY homology. The $(\lambda, Q, t_r, t_c)$-degrees of the colored differentials $d^\pm_{[r]\to[\ell]}$ 
\bea
\wt{\deg}~ d^+_{[r]\to[\ell]}&=& (-1,1,-1,-2\ell-1) ~,\cr
\wt{\deg} ~d^-_{[r]\to[\ell]}&=& (-2,0,-3,-2\ell-2r-1)~.
\eea
The mirror/transposition symmetry  \eqref{mirror-Kauffman-homology} tells us the $(\lambda, Q, t_r, t_c)$-degrees of the colored differentials $d^\pm_{[1^r]\to[1^\ell]}$ 
\bea
\wt{\deg}~ d^-_{[1^r]\to[1^\ell]}&=& (-1,-1,-r-2\ell-1,-2) ~,\cr
\wt{\deg} ~d^+_{[1^r]\to[1^\ell]}&=& (-2,0,-2\ell-2r-1,-3)~.
\eea

When $\ell=0$, these differentials become canceling differentials. The differentials $d^+_{[r]\to[0]}$ and $d^-_{[1^r]\to[0]}$ are generalizations of the differentials $d_2$ and $d_0$, respectively in (6.16) of \cite{Gukov:2005qp}, and the differentials $d^-_{[r]\to[0]}$ and $d^+_{[1^r]\to[0]}$ correspond to generalizations of the differential $d_1$ to higher rank representations. 
The  homology with respect to the canceling differential is one-dimensional, and its grading can be written in terms of the $S$-invariant $S(K)$  \cite{Rasmussen:2004}
\bea
\wt\scF(\wt\scH^\Kauffman_{[r]},d^+_{[r]\to[0]})(K;\lambda, Q, t_r, t_c)&=& \lambda^{r S(K)}Q^{-r S(K)} ~,\cr
\wt\scF(\wt\scH^\Kauffman_{[r]},d^-_{[r]\to[0]})(K;\lambda, Q, t_r, t_c)&=& \lambda^{2r S(K)} t_r^{2r S(K)} t_c^{2r^2 S(K)}~,\cr
\wt\scF(\wt\scH^\Kauffman_{[1^r]},d^-_{[1^r]\to[0]})(K;\lambda, Q, t_r, t_c)&=& \lambda^{r S(K)}Q^{r S(K)} t_r^{r^2 S(K)} t_c^{r S(K)} ~,\cr
\wt\scF(\wt\scH^\Kauffman_{[1^r]},d^+_{[1^r]\to[0]})(K;\lambda, Q, t_r, t_c)&=& \lambda^{2r S(K)} t_r^{2r^2 S(K)} t_c^{2r S(K)}~.
\eea

At general values of $\ell\in\bZ$, the isomorphisms \eqref{Kauffman-color-diff}
involve the grading changes. For the  colored differentials $d^-_{[r]\to[\ell]}$ and $d^+_{[1^r]\to[1^\ell]}$, the grading changes can be explicitly spelled out by
\bea\label{colored-diff-regrading-1}
&&\wt\scF(\wt\scH^\Kauffman_{[r]}(K),d^-_{[r]\to[\ell]})(\lambda, Q, t_r, t_c)= (\lambda t_r)^{2(r-\ell) S(K)} t_c^{2(r^2-\ell^2) S(K)} \wt\scF_{[\ell]}(K;\lambda, Q, t_r, t_c)~,\cr
&&\wt\scF(\wt\scH^\Kauffman_{[1^r]}(K),d^+_{[1^r]\to[1^\ell]})(\lambda, Q, t_r, t_c)\cr
&&\hspace{4cm}=  (\lambda t_c)^{2(r-\ell) S(K)}t_r^{2(r^2-\ell^2) S(K)} \wt\scF_{[1^\ell]}(K;\lambda, Q t_r^{r-\ell}, t_r, t_c)~.
\eea
Here, we stress that these grading changes become clear only when we use the tilde-version of the colored Kauffman homology.
On the other hand, the grading changes of the other colored differentials are not as straightforward. They are only evident  at  $t_c=1$ (for symmetric representations) or $t_r=1$ (for anti-symmetric representations), where
\bea\label{colored-diff-regrading-2}
\wt\scF(\wt\scH^\Kauffman_{[r]}(K),d^+_{[r]\to[\ell]})(\lambda, Q, t_r, t_c=1)&=& (\lambda Q^{-1})^{(r-\ell) S(K)} \wt\scF_{[\ell]}(K;\lambda, Q, t_r, t_c=1)~,\cr
\wt\scF(\wt\scH^\Kauffman_{[1^r]}(K),d^-_{[1^r]\to[1^\ell]})(\lambda, Q, t_r=1, t_c)&=&(\lambda Q t_c)^{(r-\ell) S(K)} \wt\scF_{[1^\ell]}(K;\lambda, Q, t_r=1, t_c)~.\cr
&&
\eea
The way in which the $t_c$-gradings in the homology $H_*(\wt\scH^\Kauffman_{[r]}(K),d^+_{[r]\to[\ell]})$ are changed from those in $\wt\scH^\Kauffman_{[\ell]}(K)$ is rather intricate. To see that, one can use the differentials. The homology $H_*(\wt\scH^\Kauffman_{[r]}(K),d^+_{[r]\to[\ell]})$ with respect to the differentials $d^+_{[r]\to[\ell]}$  inherits all the differentials in  the $[\ell]$-colored Kauffman homology $\wt\scH^\Kauffman_{[r]}(K)$. However, the  $t_c$-gradings of the differentials acting on the homology $H_*(\wt\scH^\Kauffman_{[r]}(K),d^+_{[r]\to[\ell]})$ are different from the original ones. The re-gradings of  the  $t_c$-degrees of the differentials are summarized in Table \ref{regrading-diff}. The same statement holds for the $t_r$-gradings in the homology $H_*(\wt\scH^\Kauffman_{[1^r]}(K),d^+_{[1^r]\to[1^\ell]})$.

\begin{table}[h]
\begin{tabular}{c|c||c|c}
\multicolumn{2}{c||}{$H_*(\wt\scH^\Kauffman_{[r]}(K),d^+_{[r]\to[\ell]})$} & \multicolumn{2}{c}{$H_*(\wt\scH^\Kauffman_{[1^r]}(K),d^-_{[1^r]\to[1^\ell]})$} \\
\hline
differentials & $(\lambda,Q,t_r,t_c)$-degrees& differentials&  $(\lambda,Q,t_r,t_c)$-degrees\\
\hline
$ d^{\rm univ}_{\to}$& $(0,2,1,1-2(r-\ell))$ & $d^{\rm univ}_{\leftarrow}$ & $(0,-2,1-2(2r-\ell),-1)$ \\ 
\hline
$ d^+_{\rm diag}$& $(-1,3,0,0)$ & $d^+_{\rm diag}$ & $(-1,3,r-1,0)$ \\ 
\hline
$ d^-_{\rm diag}$& $(-1,-3,-3,-2r-1)$ & $d^-_{\rm diag}$ & $(-1,-3,-3r,-3)$ \\ 
\hline
$d_{[\ell]\to[k]}^+$ &$(-1,1,-1,2(r-\ell-k)-1)$ & $d^+_{[1^\ell]\to[1^k]}$ & $(-2,0,-2(r+k)-1,-3)$ \\
\hline
$d_{[\ell]\to[k]}^-$ &$(-2,0,-3,-2(r+k)-1)$ & $d^-_{[1^\ell]\to[1^k]}$ & $(-1,-1,r-2(\ell+k)-1,-2)$ \\
\hline
\end{tabular}
\caption{The degree changes of the differentials in $H_*(\wt\scH^\Kauffman_{[r]}(K),d^+_{[r]\to[\ell]})$ and $H_*(\wt\scH^\Kauffman_{[1^r]}(K),d^-_{[1^r]\to[1^\ell]})$. } \label{regrading-diff}
\end{table}

Another way to check the $t_c$-gradings in the homology 
$H_*(\wt\scH^\Kauffman_{[r]}(K_{\thin}),d^+_{[r]\to[\ell]})$ for a thin knot $K_{\thin}$ is by using the $[\ell]$-colored HOMFLY homology:
\bea\label{check-color-diff+}
&&\wt\scF(\wt\scH^\Kauffman_{[r]}(K_\thin),d^+_{[r]\to[\ell]})(q^3, q,q^{-1} t_r, q t_c)\cr
&&\hspace{4cm}=t_c^{2(r-\ell)\ell S(K)}\left[\wt\scP_{[\ell]}(K_{\thin};q^2,q t_c^{r-\ell},q^{-1}t_r,qt_c) \right]^2~.
\eea

\subsection{Universal colored differentials}

There exists yet another set of  colored differentials, called universal colored differentials, when the color involves the Young tableaux with double boxes. The analogous differentials in HOMFLY homology are $d^\leftarrow$ and $d^\uparrow$. By using the same symbols in the Kauffman homology, the homology with respect to the universal colored differentials is isomorphic to the uncolored Kauffman homology
\bea\label{universal-color-diff}
H_*(\wt\scH^\Kauffman_{\yng(2)}(K),d^\leftarrow)\cong \wt\scH^\Kauffman_{\yng(1)} ~,\quad  H_*(\wt\scH^\Kauffman_{\yng(1,1)}(K),d^\uparrow)\cong \wt\scH^\Kauffman_{\yng(1)} ~.
\eea
The $(\lambda, Q, t_r, t_c)$-degrees  of the differentials $d^\leftarrow$ and $d^\uparrow$ are given by
\bea
\wt\deg ~d^\leftarrow=(0,0,0,2)~, \quad
\wt\deg ~d^\uparrow=(0,0,-2,0)~,
\eea
where the re-gradings in the isomorphisms \eqref{universal-color-diff} are provided by 
\bea
\wt \scF(\wt\scH^{\Kauffman}_{\yng(2)}(K),d^\leftarrow)(\lambda, Q, t_r, t_c)&=&\wt\scF_{\yng(1)}(K;\lambda^2, Q^2, t_r^2, t_c^4)~,\cr
\wt \scF(\wt\scH^{\Kauffman}_{\yng(1,1)}(K),d^\uparrow)(\lambda, Q, t_r, t_c)&=&\wt\scF_{\yng(1)}(K;\lambda^2, Q^2, t_r^4, t_c^2)~.
\eea


\section{Homological invariants: \\ the power of refined exponential growth property}\label{sec:homologi}
In this section, we will investigate the $[r]$-colored Kauffman homology and the $[r,r]$-colored HOMFLY homology of  both the trefoil and the figure-eight in order to make the properties summarized in \S\ref{sec:quad-differential} explicit. To obtain the Poincar\'e polynomials, the refined exponential growth property is very helpful.

\subsection{Trefoil}
Before considering the colored Kauffman homology, let us recall the $[r]$-colored HOMFLY homology of the trefoil.
The Poincar\'e polynomial of the quadruply-graded $[r]$-colored HOMFLY homology of the trefoil knot can be written as 
\bea
\wt\scP_{[r]}({\bf 3_1};a,Q,t_r,t_c)&=&a^{2 r}Q^{-2r}  \sum_{k=0}^{r}  Q^{4k}t_r^{2 k}  t_c^{2 rk} \left[\begin{array}{c}r \\ k \end{array}\right]_{t_c^2} (-a ^2Q^{-2}  t_rt_c;t_c^2)_{k} ~,\label{quad-HOMFLY-trefoil-1row-1} \\
&=&a^{2 r} Q^{2 r} t_r^{2r}  t_c^{2 r^2} \sum_{k=0}^{r}  Q^{-4 k} t_c^{-2 k r} t_r^{-4 k} \left[\begin{array}{c}r \\ k \end{array}\right]_{t_c^2}(- a^2 Q^2t_r^3t_c^{1 + 2 r}  ;t_c^2)_k~.~~~ \label{quad-HOMFLY-trefoil-1row-2}
\eea
The identity of the first line \eqref{quad-HOMFLY-trefoil-1row-1} with the second line \eqref{quad-HOMFLY-trefoil-1row-2} represents the self-symmetry  \eqref{self-symmetry}. Translating into the triply-graded homology $\wt\scP_{[r]}({\bf 3_1};a,q,tq^{-1},q)$ in the $t_r$-grading, the expression in the first line \eqref{quad-HOMFLY-trefoil-1row-1}  is equal to  (3.2) in \cite{Fuji:2012pi}.

\begin{figure}[h]
 \centering
    \includegraphics[width=6cm]{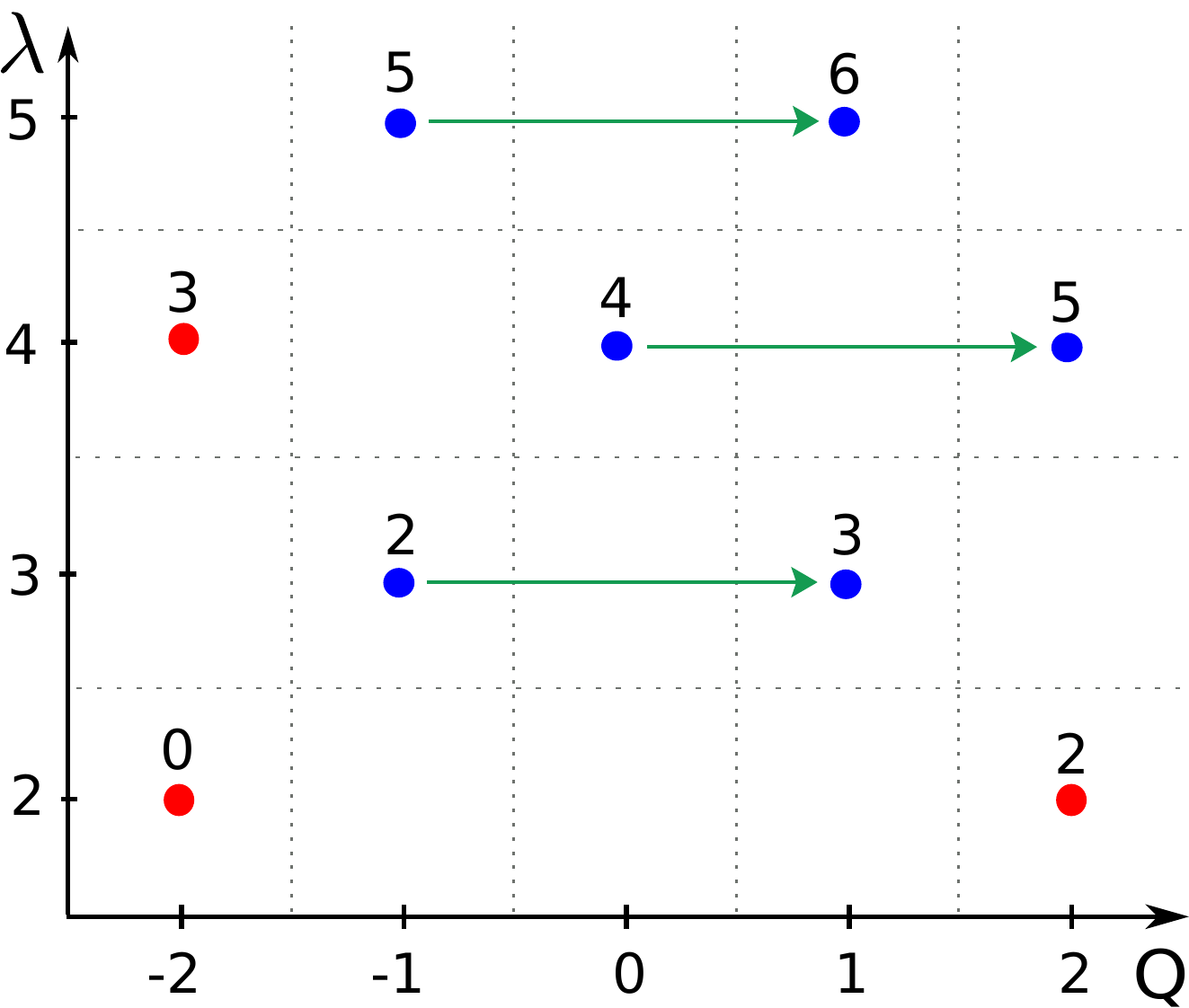}
    \caption{The uncolored Kauffman homology of the trefoil. The label with each dot stands for the homological $t_r$-degree and $t_c$-degree. The green arrows represent the universal differential $d^{\rm univ}_{\to}$ where the red dots represent the elements in $H_*(\wt\scH^\Kauffman_{[1]}({\bf 3_1}),d^{\rm univ}_{\to})$. The red and blue dots correspond to the terms colored  with red and blue in \eqref{Kauffman-uncolor-trefoil} respectively.}
    \label{fig:Kauffman-uncolor-trefoil}
\end{figure}

We now move on to the Kauffman homology. The uncolored Kauffman homology has been indeed obtained in \cite{Gukov:2005qp}. Here, we write the Poincar\'e polynomial  in the $(\lambda,Q,t_r,t_c)$-degrees
\bea\label{Kauffman-uncolor-trefoil}
\wt\scF_{\yng(1)}({\bf 3_1};\lambda,Q,t_r,t_c)&=&{\color{red} \lambda ^2 \left(Q^{-2}+Q^2 t_r^2t_c^2\right)+\lambda^4Q^{-2}t_r^3t_c^3}\\
&&+{\color{blue}\lambda ^3\left(Q^{-1}t_r^2t_c^2+Qt_r^3t_c^3\right) + \lambda ^4\left(t_r^4t_c^4+Q^2 t_r^5t_c^5\right)+ \lambda ^5\left(Q^{-1}t_r^5t_c^5+Q t_r^6t_c^6\right)}~.\nonumber
\eea
The respective homology diagram is drawn in Figure \ref{fig:Kauffman-uncolor-trefoil}. In the uncolored case, the $q$-degree is equal to the $Q$-degree for each element since the $t_r$-degree is the same as the $t_c$-degree. In order to see the universal differential explicitly, we color the homology $H_*(\wt\scH^\Kauffman_{[1]}({\bf 3_1}),d^{\rm univ}_{\to})$ with respect to the differential $d^{\rm univ}_{\to}$ with red, and the element exact under the differential $d^{\rm univ}_{\to}$  with blue in \eqref{Kauffman-uncolor-trefoil} and Figure \ref{fig:Kauffman-uncolor-trefoil}.

Next, let us investigate the $[2]$-colored Kauffman homology using the properties in \S\ref{sec:quad-differential}. In fact, the $[2]$-colored Kauffman  polynomial can be simply computed by the Rosso-Jones formula. In addition, it is easy to obtain the $[2]$-colored $\frakso(3)$ homology and $\frakso(4)$ homology  of the trefoil from \eqref{quad-HOMFLY-trefoil-1row-2}
\bea
\scF_{\yng(2)}({\bf 3_1};q^2,q,t_r=t,t_c=1)&=&\scP^{\frakso(3)}_{\yng(2)}({\bf 3_1};q,t)=\wt\scP_{\yng(4)}({\bf 3_1};q,q^{1/2},t q^{-1/2},q^{1/2})~,\cr
\scF_{\yng(2)}({\bf 3_1};q^3,q,t_r=t,t_c=1)&=&\scP^{\frakso(4)}_{\yng(2)} ({\bf 3_1};q,t)=\left[\wt\scP_{\yng(2)}({\bf 3_1};q^2,q,t q^{-1},q)\right]^2~.
\eea 
With the canceling differential $d_{[2]\to[0]}$, these data uniquely determine the Poincar\'e polynomial \eqref{triple-Kauffman-trefoil-2} of the $[2]$-colored triply-graded Kauffman homology in the $t_r$-grading. Note that Figure \ref{fig:Kauffman-trefoil-2to1} shows how the universal colored differential $d^\leftarrow$ acts on the triply-graded homology. Since the trefoil is a thin knot, it is easy to find the $t_c$-grading in \eqref{triple-Kauffman-trefoil-2} by using $\delta$-grading \eqref{delta-grading-Kauffman}, where the $S$-invariant $S({\bf 3_1})$ of the trefoil is $2$. Then, through \eqref{Q}, one can write the tilde-version of  the $[2]$-colored quadruply-graded Kauffman homology with the $(\lambda,Q,t_r,t_c)$-gradings \eqref{quad-Kauffman-trefoil-2}.

Proceeding further, we will try to obtain the $[r]$-colored Kauffman homology, making use of the refined exponential growth property \eqref{exp-growth}. Actually, the refined exponential growth property is so powerful that it specifies the form of the $[r]$-colored  quadruply-graded Kauffman homology at the $t_c=1$ specialization 
\bea\label{quad-Kauffman-trefoil-special}
&&\wt\scF_{[r]}({\bf 3_1};\lambda,Q,t_r,t_c=1)=\left[\wt\scF_{[1]}({\bf 3_1};\lambda,Q,t_r,t_c=1)\right]^r\cr
&=&\sum_{k=0}^{r} \left(\begin{array}{c}r\\ k \end{array}\right) {\color{red} \left[\lambda^2 Q^2 t_r^2 (1+Q^{-4}t_r^{-2}(1+\lambda^2t_r^3))\right]^k}\cr
&&\hspace{5cm}\times {\color{blue}\left[\lambda ^3Q^{-1}t_r^2(1+Q^2t_r)(1+\lambda Q t_r^2(1+\lambda Q^{-1}t_r))\right]^{r-k}}~,\cr
&=&\sum_{k=0}^{r}\sum_{j=0}^{k}\sum_{i=0}^{r-k} \left(\begin{array}{c}r\\ k \end{array}\right) {\color{red}\lambda^{2 k} Q^{2 ( k-2 j )} t_r^{2 (k-j)} \left(\begin{array}{c}k \\ j \end{array}\right)(1+\lambda^2 t_r^3)^j }\cr
&&\hspace{2cm} \times {\color{blue} \lambda^{ 3 (r-k)+i}Q^{i + k - r} t_r^{
 2 ( r-k+i)}  \left(\begin{array}{c}r-k\\ i \end{array}\right) (1+Q^2 t_r )^{r-k} (1+\lambda Q^{-1} t_r )^i}~.
\eea
To obtain the full expression, $t_c$-gradings remain to be determined. First of all, the binomials  in \eqref{quad-Kauffman-trefoil-special} are replaced by the $t_c^2$-binomials: {\it e.g.} $\left(\begin{array}{c}r\\ k \end{array}\right)$  in \eqref{quad-Kauffman-trefoil-special} is restored to $\left[\begin{array}{c}r\\ k \end{array}\right]_{t_c^2}$. In addition, the factors with red color at $k=r$ accord to the homology $ H_*(\wt\scH^\Kauffman_{[r]}({\bf 3_1}),d^{\univ}_{\to})$ with respect to the universal differential $d^{\rm univ}_{\to}$ and therefore is identical to the form
\bea
\wt\scF(\wt\scH_{[r]}^\Kauffman({\bf 3_1}),d^{\rm univ}_{\to})(\lambda,Q,t_r,t_c)=Q^{2r}\wt\scP_{[r]}({\bf 3_1};\lambda Q^{-1}, Q, t_r, t_c)~.
\eea
 Moreover, the form for the red factor is roughly of the form $Q^{2k}\wt\scP_{[k]}({\bf 3_1};\lambda Q^{-1}, Q, t_r, t_c)$ while the $t_c$-grading has to be modified in general. On the other hand, the factors colored in blue are killed by the universal differential $d_{\to}^{\rm univ}$. (See Figure \ref{fig:quad-Kauffman-trefoil-2-universal}.) This can be realized by uplifting the term $(1+Q^2 t_r )^{r-k} $ in \eqref{quad-Kauffman-trefoil-special}  to the $t_c^2$-Pochhammer symbol $(-Q^2 t_rt_c;t_c^2 )_{r-k}$, which is very natural, judging from the homological elements in the top $\lambda$-degree in Figure \ref{fig:quad-Kauffman-trefoil-2-universal}. In a similar fashion,  the $t_c^2$-Pochhammer symbol $(-\lambda Q^{-1} t_r t_c^*;t_c^2)_i$ is substituted for the term $(1+\lambda Q^{-1} t_r )^i$ in \eqref{quad-Kauffman-trefoil-special}, although the $t_c$-degrees in the argument has to be fixed. To incorporate $t_c$-gradings appropriately, the explicit expression \eqref{quad-Kauffman-trefoil-2-color} of  the $[2]$-colored quadruply-graded  Kauffman homology is inevitable.\footnote{What is written in this paragraph was explained to S.N. by Marko Sto$\check{\text{s}}$i$\acute{\text{c}}$. S.N. would like to thank him.}

By fixing $t_c$-gradings in such a way that all the properties in  \S\ref{sec:quad-differential} are satisfied, we find 
the Poincar\'e polynomial of the $[r]$-colored quadruply-graded Kauffman homology of the trefoil
\bea\label{quad-Kauffman-trefoil}
&&\wt\scF_{[r]}({\bf 3_1};\lambda,Q,t_r,t_c)\cr
&=&\sum_{k=0}^{r}\sum_{j=0}^{k}\sum_{i=0}^{r-k} \left[\begin{array}{c}r\\ k \end{array}\right]_{t_c^2}{\color{red}\lambda^{2 k} Q^{2 ( k-2 j )} t_r^{2 (k-j)}  t_c^{2 r(k - j) } \left[\begin{array}{c}k \\ j \end{array}\right]_{t_c^2}(- \lambda^2 t_r^3t_c^{1 + 2 r}  ;t_c^2)_j }\cr
&& \times {\color{blue} \lambda^{ 3 (r-k)+i}Q^{i + k - r} t_r^{
 2 ( r-k+i)}   t_c^{2 r (r- k + i) + 2 i j} \left[\begin{array}{c}r-k\\ i \end{array}\right]_{t_c^2} (-Q^2 t_r t_c;t_c^2)_{r-k} (-\lambda Q^{-1} t_r t_c;t_c^2)_i}\cr
&=&\sum_{k=0}^{r}\sum_{j=0}^{k}\sum_{i=0}^{r-k}\lambda^{i - k + 3 r}Q^{i - 4 j + 3 k - r} t_r^{ 2 (i - j + r)} t_c^{2 r ( r-j) + 2 i (j + r)}  \left[\begin{array}{c}r\\ k \end{array}\right]_{t_c^2}\left[\begin{array}{c}k \\ j \end{array}\right]_{t_c^2}\left[\begin{array}{c}r-k\\ i \end{array}\right]_{t_c^2}\cr
 &&\hspace{3cm}\times(- \lambda^2 t_r^3t_c^{1 + 2 r}  ;t_c^2)_j  (-Q^2 t_r t_c;t_c^2)_{r-k} (-\lambda Q^{-1} t_r t_c;t_c^2)_i~.
\eea
Apart from the refined exponential growth property and the universal differential, one can check that the formula has the following properties.
\begin{itemize}
\item $\scH^{\frakso(3)}_{[r]}\cong \scH^{\fraksl(2)}_{[2r]}$  [\eqref{so3-sl2-thin}]
\begin{eqnarray*}
\wt\scF_{[r]}({\bf 3_1};q^2,q,t_r q^{-1},q)=\wt\scP_{[2r]}({\bf 3_1};q,q^{1/2},t_r q^{-1/2},q^{1/2} )
\end{eqnarray*}
\item  $\scH^{\frakso(4)}_{[r]}\cong \scH^{\fraksl(2)\oplus \fraksl(2)}_{[r]}$  [\eqref{so4-sl2sl2-thin}]
\begin{eqnarray*}
\wt\scF_{[r]}({\bf 3_1};q^3,q,t_r q^{-1},t_c q)=\left[\wt\scP_{[r]}({\bf 3_1};q^2,q,t_r q^{-1},t_cq )\right]^2
\end{eqnarray*}
\item {\bf Diagonal differentials $d^\pm_{\rm diag}$ } [\eqref{univ-diff-regrading}, Figure \ref{fig:quad-Kauffman-trefoil-diagonal+} and Figure \ref{fig:quad-Kauffman-trefoil-diagonal-}] \hfill \\
\begin{eqnarray*}
\wt \scF(\wt\scH^{\Kauffman}_{[r]}({\bf 3_1}),d^+_{\rm diag})(\lambda, Q, t_r, t_c)&=&\tfrac{\lambda^r }{Q^{3r} t_r^rt_c^r}\wt\scP_{[r]}({\bf 3_1};\lambda Q t_rt_c, (\lambda Q t_rt_c)^{\frac12}, t_rt_c^{-1}, t_c^2)\cr
\wt \scF(\wt\scH^{\Kauffman}_{[r]}({\bf 3_1}),d^-_{\rm diag})(\lambda, Q, t_r, t_c)&=&\lambda^{r} Q^{3r}   t_r^{2r} t_c^{2r^2}\wt\scP_{[r]}({\bf 3_1};\lambda Q^{-1} , (\lambda Q^{-1} )^{\frac12}, t_r, t_c)
\end{eqnarray*}
\item {\bf Colored differential $d^-_{[r]\to[\ell]}$ } [\eqref{colored-diff-regrading-1} and Figure \ref{fig:quad-Kauffman-trefoil-2to1-}] \hfill \\
\begin{eqnarray*}
\wt\scF(\wt\scH^\Kauffman_{[r]}({\bf 3_1}),d^-_{[r]\to[\ell]})(\lambda, Q, t_r, t_c)= (\lambda t_r)^{4(r-\ell) } t_c^{4(r^2-\ell^2) } \wt\scF_{[\ell]}({\bf 3_1};\lambda, Q, t_r, t_c)
\end{eqnarray*}

\item {\bf Colored differential $d^+_{[r]\to[\ell]}$ } [Figure \ref{fig:quad-Kauffman-trefoil-2to1+}] \hfill \\
The Poincar\'e polynomial of the homology with respect to the colored differential $d^+_{[r]\to[\ell]}$ is given by
\begin{eqnarray*}
&&\wt\scF(\wt\scH^\Kauffman_{[r]}({\bf 3_1}),d_{[r]\to{[\ell]} }^+ )(\lambda,Q,t_r,t_c)\cr
&=&\lambda^{2(r-\ell)}Q^{-2(r-\ell)}\cr
&&\times\sum_{k=0}^{\ell}\sum_{j=0}^{k}\sum_{i=0}^{\ell-k} \left[\begin{array}{c}\ell\\ k \end{array}\right]_{t_c^2}{\color{red}\lambda^{2 k} Q^{2 ( k-2 j )} t_r^{2 (k-j)}  t_c^{2 (2r-\ell)(k - j) } \left[\begin{array}{c}k \\ j \end{array}\right]_{t_c^2}(- \lambda^2 t_r^3t_c^{1 + 2 r}  ;t_c^2)_j }\cr
&& \times {\color{blue} \lambda^{ 3 (\ell-k)+i}Q^{i + k - \ell} t_r^{
 2 ( \ell-k+i)}   t_c^{2 (2r-\ell)(\ell- k + i) + 2 i j} \left[\begin{array}{c}\ell-k\\ i \end{array}\right]_{t_c^2}}\cr
 && \times {\color{blue} (-Q^2 t_r t_c^{1-2(r-\ell)};t_c^2)_{\ell-k} (-\lambda Q^{-1} t_r t_c^{1-2(r-\ell)};t_c^2)_i}~.
\end{eqnarray*}
Then, one can verify \eqref{check-color-diff+}:
\begin{eqnarray*}
\wt\scF(\wt\scH^\Kauffman_{[r]}({\bf 3_1}),d^+_{[r]\to[\ell]})(q^3, q,q^{-1} t_r, q t_c)=t_c^{4(r-\ell)\ell}\left[\wt\scP_{[\ell]}({\bf 3_1};q^2,q t_c^{r-\ell},q^{-1}t_r,qt_c) \right]^2~.
\end{eqnarray*}
\end{itemize}
It is straightforward from \eqref{Q-to-q-Kauffman} to obtain the Poincar\'e polynomial of the $[r]$-colored triply-graded Kauffman homology of the trefoil in the $t_r$-grading 
\bea\label{triple-Kauffman-trefoil}
&&\scF_{[r]}({\bf 3_1};\lambda,q,t_r=t,t_c=1)=\wt\scF_{[r]}({\bf 3_1};\lambda,q,q^{-1}t,q)\cr
&=&\sum_{k=0}^{r}\sum_{j=0}^{k}\sum_{i=0}^{r-k}\lambda^{i - k + 3 r}q^{3 k - 2 j (1 + r) + r (2 r-3) + i (2 j + 2 r-1)} t^{ 2 (i - j + r)}   \left[\begin{array}{c}r\\ k \end{array}\right]_{q^2}\left[\begin{array}{c}k \\ j \end{array}\right]_{q^2}\left[\begin{array}{c}r-k\\ i \end{array}\right]_{q^2}\cr
 &&\hspace{3cm}\times(- \lambda^2 t^3 q^{ 2 r-2}  ;q^2)_j  (-q^2 t;q^2)_{r-k} (-\lambda q^{-1} t;q^2)_i~.
\eea
We verify that the expression reduces to colored Kauffman polynomial computed by the Rosso-Jones formula \cite{Stevan:2010jh} at $t=-1$ up to 4 boxes. 

In addition, the mirror/transposition symmetry \eqref{mirror-Kauffman} tells us  the Poincar\'e polynomial of the $[1^r]$-colored triply-graded Kauffman homology of the trefoil in the $t_c$-grading
\bea
&&\scF_{[1^r]}({\bf 3_1};\lambda,q,t_r,=1,t_c=t)=\wt\scF_{[1^r]}({\bf 3_1};\lambda,q^{r},q^{-1},qt)=\wt\scF_{[r]}({\bf 3_1};\lambda,q^{-1}t^{-1},qt,q^{-1})\cr
&=&\sum_{k=0}^{r}\sum_{j=0}^{k}\sum_{i=0}^{r-k}\lambda^{i - k + 3 r}q^{-3 k + 2 j (1 + r) - r (2 r-3) - i (2 j + 2 r-1)} t^{ i + 2 j - 3 k + 3 r}   \left[\begin{array}{c}r\\ k \end{array}\right]_{q^{-2}}\left[\begin{array}{c}k \\ j \end{array}\right]_{q^{-2}}\left[\begin{array}{c}r-k\\ i \end{array}\right]_{q^{-2}}\cr
 &&\hspace{3cm}\times(- \lambda^2 t^3 q^{ 2-2 r}  ;q^{-2})_j  (-q^{-2} t^{-1};q^{-2})_{r-k} (-\lambda q t^2;q^{-2})_i~.
\eea
For instance, the homology diagrams of the $[1^2]$-colored Kauffman homology of the trefoil are depicted in Figure \ref{fig:Kauffman-trefoil-11} for the triple-gradings and  Figure \ref{fig:quad-Kauffman-trefoil-11-cancel} for the quadruple-gradings. Especially, in Figure \ref{fig:Kauffman-trefoil-11}, one can see the action of the differential $d_{-2}$, providing the $[2]$-colored $\fraksp(2)$ homology of the trefoil
\bea
\scP_{\yng(2)}^{\fraksp(2)}({\bf 3_1};q,t)&=&t^{-4} \scF(\scH^\Kauffman_{\yng(1,1)}({\bf 3_1}),d_{-2})(\lambda=q^3,q^{-1},1,t)=  \wt \scP_{\yng(2)}({\bf 3_1};q^4,q^2,tq^{-2},q^2)\cr
&=&q^8 + q^{20} t^2 + q^{24} t^2 + q^{24} t^3 + q^{28} t^3+ q^{32} t^4+ 
 q^{36} t^5 + q^{40} t^5 + q^{44} t^{6}~.~~~~~
\eea 
(See also \eqref{triple-Kauffman-trefoil-sp2}.)

\begin{figure}[h]
 \centering
    \includegraphics[width=9cm]{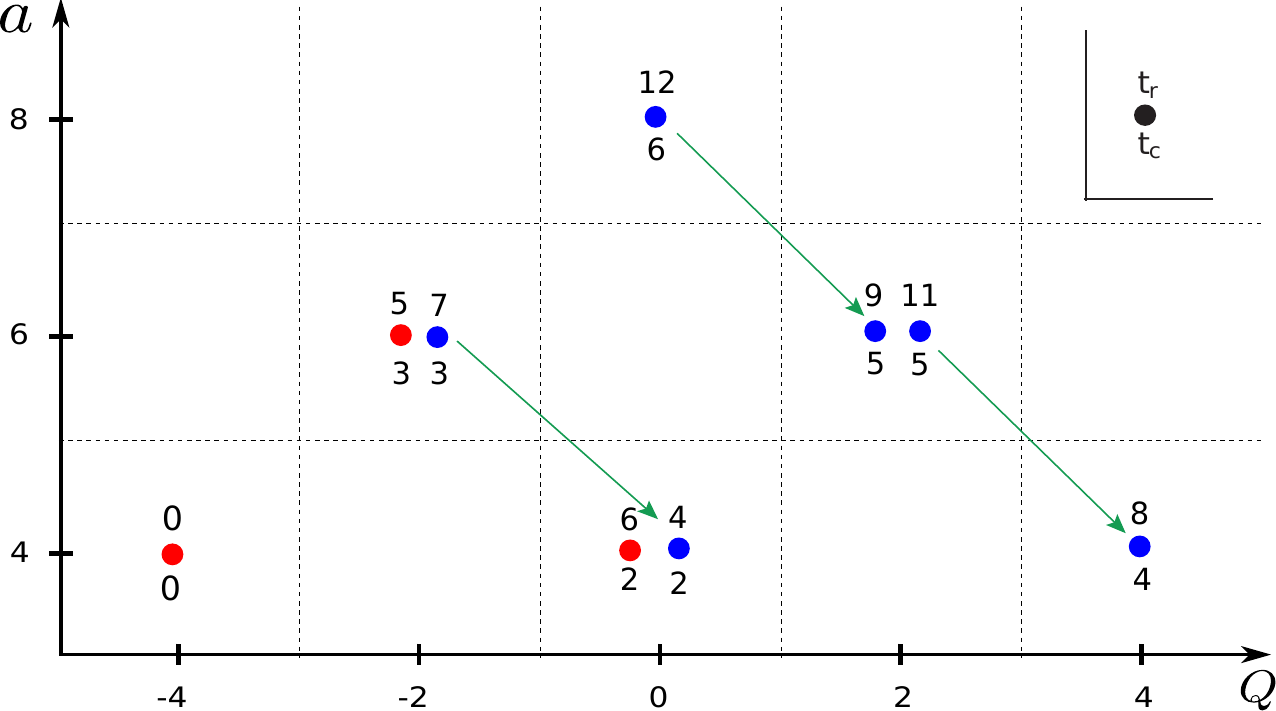}
    \caption{The $[1,1]$-colored HOMFLY homology of the trefoil. The green arrows represent the colored differential $d^+_{[1,1]\to[1]}$ where the red dots are the elements in $H_*(\wt\scH^\HOMFLY_{[1,1]}({\bf 3_1}),d^+_{[1,1]\to[1]})$. The red and blue dots correspond to the factors colored  with red and blue in \eqref{HOMFLY-11-trefoil} respectively.}\label{fig:HOMFLY-trefoil-11}
\end{figure}

Having studied the $[r]$-colored Kauffman homology of the trefoil, the next goal is to obtain the $[r,r]$-colored HOMFLY homology of the trefoil using the relation to the $[r]$-colored Kauffman homology predicted in \S\ref{sec:quad-differential}. First, it is useful to review the case of $r=1$ \cite{Gukov:2011ry}. The expression for the $[1,1]$-colored HOMFLY homology follows from \eqref{quad-HOMFLY-trefoil-1row-1} via the mirror/transposition symmetry
\bea\label{HOMFLY-11-trefoil}
\wt\scP_{\yng(1,1)}({\bf 3_1};a,Q,t_r,t_c)&=&{\color{red} a ^4 \left(Q^{-4}+ t_r^6t_c^2\right)+a^6Q^{-2}t_r^5t_c^3}\\
&&+{\color{blue}a^4\left(t_r^4t_c^2+Q^4t_r^8t_c^4\right) + a^6\left(Q^{-2}t_r^7t_c^3+Q^2 (t_r^9+t_r^{11})t_c^5\right)+ a^8 t_r^{12}t_c^6}~.\nonumber
\eea 
A simple calculation involving \eqref{Kauffman-uncolor-trefoil} and \eqref{HOMFLY-11-trefoil} confirms that 
\bea
\wt\scF_{\yng(1)}({\bf 3_1};\lambda=q^5,q,q^{-1},qt)=\wt\scP_{\yng(1,1)}({\bf 3_1};a=q^4,q,q^{-1},qt)~.
\eea
In fact, when comparing Figure \ref{fig:HOMFLY-trefoil-11} with Figure \ref{fig:Kauffman-uncolor-trefoil} in the $t_c$-grading, it is easy to see the one-to-one correspondence between the generators of Kauffman homology and those of HOMFLY homology. In addition, the differential $d^+_{[1,1]\to[1]}$ in HOMFLY homology clearly corresponds to $d^{\univ}_{\to}$ in Kauffman homology. Thus, to make an analogy to Kauffman homology, it is convenient to separate the homology $H_*(\scH^{\HOMFLY}_{[1,1]},d^+_{[1,1]\to[1]})$ with respect to the differential $d^+_{[1,1]\to[1]}$ (red color) from the exact elements under the differential $d^+_{[1,1]\to[1]}$ (blue color).

Like the colored Kauffman homology \eqref{quad-Kauffman-trefoil-special},  with the great help of the refined exponential growth property \eqref{exp-growth-HOMFLY-1}, one can evaluate  the $t_c=1$ specialization of the $[r,r]$-colored quadruply-graded HOMFLY homology 
\bea\label{quad-HOMFLY-trefoil-special}
&&\wt\scP_{[r,r]}({\bf 3_1};a,Q,t_r,t_c=1)=\left[\wt\scP_{[1,1]}({\bf 3_1};\lambda,Q,t_r,t_c=1)\right]^r\cr
&=&\sum_{k=0}^{r} \left(\begin{array}{c}r\\ k \end{array}\right) {\color{red} \left[a^4t_r^6 (1+Q^{-4}t_r^{-6}(1+a^2Q^2t_r^5))\right]^k} \cr
&&\hspace{5cm}{\color{blue}\left[a^4t_r^4(1+a^2 Q^{-2} t_r^3)(1+ Q^{4} t_r^4(1+a^2Q^{-2} t_r))\right]^{r-k}}~,\cr
&=&\sum_{k=0}^{r}\sum_{j=0}^{k}\sum_{i=0}^{r-k}  \left(\begin{array}{c}r\\ k \end{array}\right) {\color{red}a^{4 k} Q^{-4 j} t_r^{6(k- j)} \left(\begin{array}{c}k \\ j \end{array}\right)(1+a^2Q^2t_r^5)^j }\cr
&&\hspace{2cm} \times {\color{blue} a^{4 (r-k)} Q^{4 i}t_r^{4 (r- k + i)}  \left(\begin{array}{c}r-k\\ i \end{array}\right) (1+a^2Q^{-2} t_r^3)^{r-k} (1+a^2 Q^{-2} t_r )^i}~.
\eea
On the other hand, the isomorphism  $ (\fraksl(4),[r,r])\cong (\frakso(6),[r])$ can be seen in the identity with the $t_c$-grading at the naive specialization $\lambda=q^5$ and $a=q^4$
\bea
\wt\scF_{[r]}({\bf 3_1};\lambda=q^5,q,q^{-1},qt_c)=\wt\scP_{[r,r]}({\bf 3_1};a=q^4,q^2,q^{-1},qt_c)~,
\eea
since the trefoil is homologically thin. Thus, this relation helps us determine the $t_c$-degrees in \eqref{quad-HOMFLY-trefoil-special} by using the formula \eqref{quad-Kauffman-trefoil}. Consequently, 
the Poincar\'e polynomial of the $[r,r]$-colored quadruply-graded HOMFLY homology of the trefoil can be written as
\bea\label{quad-HOMFLY-2row-trefoil}
&&\wt\scP_{[r,r]}({\bf 3_1};a,Q,t_r,t_c)\cr
&=&\sum_{k=0}^{r}\sum_{j=0}^{k}\sum_{i=0}^{r-k} \left[\begin{array}{c}r\\ k \end{array}\right]_{t_c^2}{\color{red}a^{4 k} Q^{-4 j} t_r^{6(k- j)}t_c^{2 r ( k-j)}  \left[\begin{array}{c}k \\ j \end{array}\right]_{t_c^2}(- a^2 Q^2t_r^5t_c^{1 + 2 r}  ;t_c^2)_j }\cr
&& \times {\color{blue} a^{4 (r-k)} Q^{4 i}t_r^{4 (r- k + i)} t_c^{2 i (j + r) + 2 r ( r-k)} \left[\begin{array}{c}r-k\\ i \end{array}\right]_{t_c^2} (-a^2Q^{-2} t_r^3 t_c;t_c^2)_{r-k} (-a^2 Q^{-2} t_r t_c;t_c^2)_i}\cr
&=&\sum_{k=0}^{r}\sum_{j=0}^{k}\sum_{i=0}^{r-k}a^{4 r} Q^{4 (i - j)}  t_r^{2 (2 i - 3 j + k + 2 r)} t_c^{2 r ( r-j) + 2 i (j + r)}\left[\begin{array}{c}r\\ k \end{array}\right]_{t_c^2}\left[\begin{array}{c}k \\ j \end{array}\right]_{t_c^2}\left[\begin{array}{c}r-k\\ i \end{array}\right]_{t_c^2}\cr
 &&\hspace{3cm}\times(- a^2Q^2 t_r^5t_c^{1 + 2 r}  ;t_c^2)_j  (-a^2Q^{-2 }t_r^3 t_c;t_c^2)_{r-k} (-a^2 Q^{-2} t_r t_c;t_c^2)_i~.
\eea
It is easy to confirm that the formula reproduces the $[2,2]$-colored HOMFLY homology of the trefoil obtained in \S4.4 of \cite{Gorsky:2013jxa}. By using \eqref{quad-HOMFLY-trefoil-1row-1}, one can check that the formula \eqref{quad-HOMFLY-2row-trefoil} satisfies the other refined exponential growth property \eqref{exp-growth-HOMFLY-2}
\bea
\wt\scP_{[r,r]}({\bf 3_1};a,Q,1,t_c)=\left[\wt\scP_{[r]}({\bf 3_1};a,Q,1,t_c)\right]^2~,
\eea
with small values of $r$. In addition, the formula shows the behaviors of the colored differentials predicted in \cite{Gorsky:2013jxa}:
\bea
\wt\scP(\scH^\HOMFLY_{[r,r]}({\bf 3_1}),d^+_{[r,r]\to[\ell,\ell]})(a,Q,t_r,t_c)&=&[aQ^{-1}t_c^{\ell}]^{4(r-\ell)}\wt\scP_{[\ell,\ell]}({\bf 3_1};a,Qt_c^{r-\ell},t_r,t_c)~,\cr
\wt\scP(\scH^\HOMFLY_{[r,r]}({\bf 3_1}),d^-_{[r,r]\to[\ell,\ell]})(a,Q,t_r,t_c)&=&[aQt_r^{2}t_c^{(r+\ell)}]^{4(r-\ell)}\wt\scP_{[\ell,\ell]}({\bf 3_1};a,Q,t_r,t_c)~,\cr
\wt\scP(\scH^\HOMFLY_{[r,r]}({\bf 3_1}),d^+_{[r,r]\to[r]})(a,Q,t_r,t_c)&=&a^{2r}Q^{-2r}t_r^{2r}\wt\scP_{[r]}({\bf 3_1};a,Q t_r,t_r,t_c)~,\cr
\wt\scP(\scH^\HOMFLY_{[r,r]}({\bf 3_1}),d^-_{[r,r]\to[r]})(a,Q,t_r,t_c)&=&a^{2r}Q^{2r}t_r^{6r}t_c^{2r^2}\wt\scP_{[r]}({\bf 3_1};a,Q,t_r,t_c)~.
\eea
This is actually expected since the colored differentials in the $[r,r]$-colored HOMFLY homology have their own counterparts in the $[r]$-colored  Kauffman homology as we see in \S \ref{sec:quad-differential}, and we have seen that the formula \eqref{quad-Kauffman-trefoil} analogous to \eqref{quad-HOMFLY-2row-trefoil} is endowed with the correct differential structure. Furthermore, we have verified that the Poincar\'e polynomials with the $t_c$-grading $\wt\scP_{[r,r]}({\bf 3_1};a,q,q^{-1},qt)$ agree with the corresponding refined Chern-Simons invariants computed in \cite{Shakirov:2013moa} up to $r=3$.

As in the Kauffman homology, the Poincar\'e polynomial of the $[r,r]$-colored triply-graded HOMFLY homology of the trefoil in the $t_r$-grading immediately follows:
\bea
\label{triple-HOMFLY-[r,r]}
&&\scP_{[r,r]}({\bf 3_1};a,q,t_r=t,t_c=1)=\wt\scP_{[r,r]}({\bf 3_1};a,q^2,q^{-1}t,q)\cr
&=&\sum_{k=0}^{r}\sum_{j=0}^{k}\sum_{i=0}^{r-k} a^{4 r} q^{-2 (j + k + (2 + j) r - r^2 - i (2 + j + r))} t^{
 2 (2 i - 3 j + k + 2 r)}\left[\begin{array}{c}r\\ k \end{array}\right]_{q^2}\left[\begin{array}{c}k \\ j \end{array}\right]_{q^2}\left[\begin{array}{c}r-k\\ i \end{array}\right]_{q^2}\cr
 &&\hspace{3cm}\times(- a^2q^{2 r}  t^5 ;q^2)_j  (-a^2q^{-6}t^3 ;q^2)_{r-k} (-a^2 q^{-4} t ;q^2)_i~.
\eea

The mirror/transposition symmetry yields the Poincar\'e polynomial of the $[2^r]$-colored triply-graded HOMFLY homology of the trefoil in the $t_c$-grading
\bea
&&\scP_{[2^r]}({\bf 3_1};a,q,t_r=1,t_c=t)=\wt\scP_{[2^r]}({\bf 3_1};a,q^r,q^{-1},qt)=\wt\scP_{[r,r]}({\bf 3_1};a,q^r,qt,q^{-1})\cr
&=&\sum_{k=0}^{r}\sum_{j=0}^{k}\sum_{i=0}^{r-k}a^{4 r} q^{-2 (-k + i (-2 + j - r) - 2 r + r^2 + j (3 + r))} t^{
 2 (2 i - 3 j + k + 2 r)}\left[\begin{array}{c}r\\ k \end{array}\right]_{q^{-2}}\left[\begin{array}{c}k \\ j \end{array}\right]_{q^{-2}}\left[\begin{array}{c}r-k\\ i \end{array}\right]_{q^{-2}}\cr
 &&\hspace{3cm}\times(- a^2q^{4}  t^5 ;q^{-2})_j  (-a^2q^{2-2r}t^3 ;q^{-2})_{r-k} (-a^2 q^{-2r} t ;q^{-2})_i~.
\eea
Setting $r=3$, the formula reproduces the  $[2,2,2]$-colored HOMFLY homology of the trefoil in \S4.5 of \cite{Gorsky:2013jxa}.

\subsection{Figure-eight}
In this subsection, we obtain  the $[r]$-colored Kauffman homology and the $[r,r]$-colored HOMFLY homology of the figure-eight.
The strategy is the same as the case of the trefoil although the size of the homology is bigger and therefore the computations are more tedious. Hence, we will not repeat the detailed explanations for the method.

\begin{figure} [h]
 \centering
    \includegraphics[width=10cm]{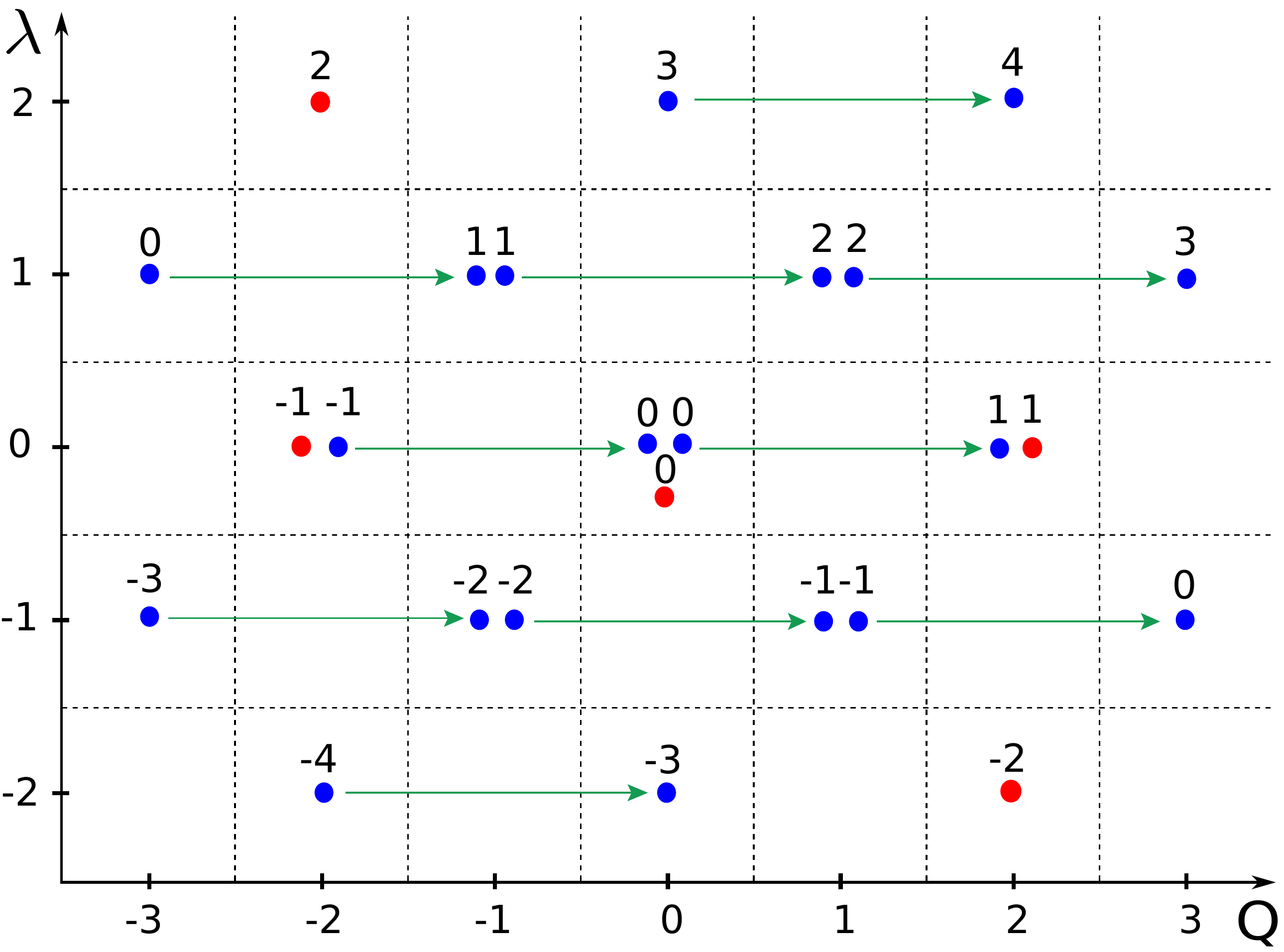}
    \caption{The uncolored Kauffman homology of the figure-eight. The green arrows represent the universal differential $d^{\rm univ}_{\to}$ and  the red dots are the elements in $H_*(\wt\scH^\Kauffman_{[1]}({\bf 4_1}),d^{\rm univ}_{\to})$. The red and blue dots correspond to the factors colored  by red and blue in \eqref{Kauffman-uncolor-fig8} respectively.}
    \label{fig:fig8-uncolor}
\end{figure}

As in the case of the trefoil, let us start with writing the Poincar\'e polynomial of the $[r]$-colored quadruply-graded HOMFLY homology of the figure-eight 
\begin{small}
\bea\label{quad-HOMFLY-fig8-1row}
\wt\scP_{[r]}({\bf 4_1};a,Q,t_r,t_c)&=& \sum_{k=0}^{r} a^{-2 k}  t_r^{-2 k}  t_c^{-2 rk} \left[\begin{array}{c}r \\ k \end{array}\right]_{t_c^2} (-a ^2Q^{-2}  t_rt_c;t_c^2)_{k} (- a ^2Q^2t_r^3 t_c^{1+2r};t_c^2)_{k }~.\quad
\eea
\end{small}
The expression  $\wt\scP_{[r]}({\bf 4_1};a,q,tq^{-1},q)$  for the triply-graded homology  in the $t_r$-grading is equal to (3.3) in \cite{Fuji:2012pi}.
The uncolored HOMFLY homology of the figure-eight is given in \cite{Gukov:2005qp}
\bea\label{Kauffman-uncolor-fig8}
&&\wt\scF_{\yng(1)}({\bf 4_1};\lambda,Q,t_r,t_c)\\
&=&{\color{red}Q^2t_r^{-2}t_c^{-2}{\lambda ^{-2}}+Q^{-2} t_r^{-1}t_c^{-1}+1+Q^2t_rt_c+Q^{-2}t_r^2t_c^2\lambda^2}\cr
&&+{\color{blue} \left(Q^{-2} t_r^{-4}t_c^{-4}+t_r^{-3}t_c^{-3}\right){\lambda ^{-2}}+\left(Q^{-3} t_r^{-3}t_c^{-3}+2Q^{-1} t_r^{-2}t_c^{-2}+2 Qt_r^{-1}t_c^{-1}+Q^3\right){\lambda^{-1} }}\cr
&&{\color{blue}+2+Q^{-2} t_r^{-1}t_c^{-1}+Q^2 t_rt_c+\left(Q^{-3}+2Q^{-1} t_rt_c+2 Q t_r^2t_c^2+Q^3 t_r^3t_c^3\right) \lambda +\left(t_r^3t_c^3+Q^2 t_r^4t_c^4\right) \lambda ^2}~,\nonumber
\eea
whose homology diagram is presented in Figure \ref{fig:fig8-uncolor}.

Unlike the case of the trefoil, the colored Kauffman polynomials of the figure-eight are not available to date. However, using the refined exponential growth property, the representation theoretic relation and the differential property, one can uniquely determine the $[2]$-colored Kauffman homology of the figure-eight
\begin{footnotesize}
\bea\label{Kauffman-triple-fig8-2}
&&\scF_{\yng(2)}({\bf 4_1};\lambda,q,t_r=t,t_c=1)\cr
&=&\lambda ^{-4}\left(\tfrac{1}{q^{12} t^8}+\tfrac{1}{q^{10} t^7}+\tfrac{1}{q^8 t^7}+\tfrac{2}{q^6 t^6}+\tfrac{1}{q^4 t^6}+\tfrac{1}{q^4 t^5}+\tfrac{1}{q^2 t^5}+\tfrac{1}{t^4}\right)\cr
&&+\lambda ^{-3}\Big(\tfrac{1}{q^{13} t^7}+\tfrac{1}{q^{11} t^7}+\tfrac{1}{q^{11} t^6}+\tfrac{3}{q^9 t^6}+\tfrac{2}{q^7 t^6}+\tfrac{4}{q^7 t^5}+\tfrac{5}{q^5 t^5}+\tfrac{1}{q^3 t^5}+\tfrac{2}{q^5 t^4}+\tfrac{5}{q^3 t^4}+\tfrac{3}{q t^4}+\tfrac{3}{q t^3}\cr
&& \hspace{2cm}+\tfrac{3 q}{t^3}+\tfrac{q}{t^2}+\tfrac{q^3}{t^2}\Big)\cr
&&+\lambda ^{-2}\Big(q^6+\tfrac{1}{q^{12} t^6}+\tfrac{3}{q^{10} t^5}+\tfrac{4}{q^8 t^5}+\tfrac{1}{q^6 t^5}+\tfrac{3}{q^8 t^4}+\tfrac{9}{q^6 t^4}+\tfrac{5}{q^4 t^4}+\tfrac{1}{q^2 t^4}+\tfrac{3}{t^3}+\tfrac{1}{q^6 t^3}+\tfrac{9}{q^4 t^3}\cr
&&\hspace{2cm}+\tfrac{11}{q^2 t^3}+\tfrac{10}{t^2}+\tfrac{4}{q^2 t^2}+\tfrac{4 q^2}{t^2}+\tfrac{1}{t}+\tfrac{4 q^2}{t}+\tfrac{3 q^4}{t}\Big)\cr
&&+\lambda^{-1 }\Big(6 q+10 q^3+4 q^5+\tfrac{1}{q^{11} t^4}+\tfrac{3}{q^9 t^4}+\tfrac{2}{q^7 t^4}+\tfrac{1}{q^9 t^3}+\tfrac{7}{q^7 t^3}+\tfrac{9}{q^5 t^3}+\tfrac{3}{q^3 t^3}+\tfrac{7}{q^5 t^2}\cr
&&\hspace{2cm}+\tfrac{16}{q^3 t^2}+\tfrac{10}{q t^2}+\tfrac{q}{t^2}+\tfrac{3}{q^3 t}+\tfrac{14}{q t}+\tfrac{14 q}{t}+\tfrac{3 q^3}{t}+q^3 t+3 q^5 t+2 q^7 t\Big)\cr
&&+21+\tfrac{1}{q^4}+\tfrac{10}{q^2}+10 q^2+q^4+\tfrac{1}{q^{10} t^3}+\tfrac{1}{q^8 t^3}+\tfrac{2}{q^8 t^2}+\tfrac{7}{q^6 t^2}+\tfrac{4}{q^4 t^2}+\tfrac{1}{q^2 t^2}+\tfrac{4}{t}+\tfrac{2}{q^6 t}+\tfrac{12}{q^4 t}+\tfrac{14}{q^2 t}\cr
&&\hspace{2cm}+4 t+14 q^2 t+12 q^4 t+2 q^6 t+q^2 t^2+4 q^4 t^2+7 q^6 t^2+2 q^8 t^2+q^8 t^3+q^{10} t^3\cr
&&+\lambda\Big(\tfrac{4}{q^5}+\tfrac{10}{q^3}+\tfrac{6}{q}+\tfrac{2}{q^7 t}+\tfrac{3}{q^5 t}+\tfrac{1}{q^3 t}+\tfrac{3 t}{q^3}+\tfrac{14 t}{q}+14 q t+3 q^3 t+\tfrac{t^2}{q}+10 q t^2\cr
&&\hspace{2cm}+16 q^3 t^2+7 q^5 t^2+3 q^3 t^3+9 q^5 t^3+7 q^7 t^3+q^9 t^3+2 q^7 t^4+3 q^9 t^4+q^{11} t^4\Big)  \cr
&&+\lambda^2 \Big(\tfrac{1}{q^6}+t+\tfrac{3 t}{q^4}+\tfrac{4 t}{q^2}+10 t^2+\tfrac{4 t^2}{q^2}+4 q^2 t^2+3 t^3+11 q^2 t^3+9 q^4 t^3+q^6 t^3\cr
&&\hspace{2cm}+q^2 t^4+5 q^4 t^4+9 q^6 t^4+3 q^8 t^4+q^6 t^5+4 q^8 t^5+3 q^{10} t^5+q^{12} t^6\Big) \cr
&&+\lambda ^3\Big(\tfrac{t^2}{q^3}+\tfrac{t^2}{q}+\tfrac{3 t^3}{q}+3 q t^3+3 q t^4+5 q^3 t^4+2 q^5 t^4+q^3 t^5+5 q^5 t^5+4 q^7 t^5+2 q^7 t^6+3 q^9 t^6\cr
&&\hspace{2cm} +q^{11} t^6+q^{11} t^7+q^{13} t^7 \Big) \cr
&&+\lambda ^4\left(t^4+q^2 t^5+q^4 t^5+q^4 t^6+2 q^6 t^6+q^8 t^7+q^{10} t^7+q^{12} t^8\right) 
\eea
\end{footnotesize}
Here, the expression is written in the triple-grading with the $t_r$-grading, and it has 625 generators. Using the $\delta$-grading \eqref{delta-grading-Kauffman} where the $S$-invariant $S({\bf 4_1})$ of the figure-eight is $0$, one can assign the $t_c$-gradings in \eqref{Kauffman-triple-fig8-2}. Note that the colored Kauffman homology obeys the following identity since the figure-eight is the same as its mirror image:
\bea
\scF_{[r]}({\bf 4_1};\lambda,q,t_r,t_c)=\scF_{[r]}({\bf 4_1};\lambda^{-1},q^{-1},t_r^{-1},t_c^{-1})~.
\eea

The method to obtain the $[r]$-colored Kauffman homology of the figure-eight is the same as in the case of the trefoil  although it is  more tedious due to its size.
The refined exponential growth property determines the $t_c=1$ specialization of the Poincar\'e polynomial of the $[r]$-colored quadruply-graded  Kauffman homology of the figure-eight. Then, the $t_c$-gradings are fixed by the differential structure and the $[2]$-colored Kauffman homology, yielding the full expression
\bea\label{quad-Kauffman-fig8}
&&\wt\scF_{[r]}({\bf 4_1};\lambda,Q,t_r,t_c)\cr
&=&\sum_{k=0}^{r}\sum_{j=0}^{k}\sum_{i=0}^{r-k} \left[\begin{array}{c}r\\ k \end{array}\right]_{t_c^2}{\color{red} \lambda^{-2 j} Q^{
 2 j} t_r^{-2 j}  t_c^{-2 rj} \left[\begin{array}{c}k \\ j \end{array}\right]_{t_c^2} (-\lambda ^2Q^{-4}  t_rt_c^{2\left(r-k-i\right)+1} ;t_c^2)_{j} (- \lambda ^2t_r^3 t_c^{1+4 r-2k};t_c^2)_{j}}\cr
&&\times{\color{blue} \lambda^{i + 2 k - 2 r}  Q^{i + 2 k - 2 r} t_r^{ 2 (i + 2 k - 2 r)} t_c^{ 2 i (j - k + r) + 2 (k - r) (j- k + 2 r)} \left[\begin{array}{c}r-k \\ i \end{array}\right]_{t_c^2}}\cr
&&\times {\color{blue}  (-Q^2 t_r t_c;t_c^2)_{r-k}(-\lambda^2t_r^3 t_c^{1+2 r}  ;t_c^2)_{r-k} (- \lambda Q^{-1}t_r t_c;t_c^2)_{r-k - i }(-\lambda Q^3 t_r^3  t_c^{1+2 (r-k+j)}  ;t_c^2)_{r-k-i}}\cr
&=&\sum_{k=0}^{r}\sum_{j=0}^{k}\sum_{i=0}^{r-k} \lambda^{2 (k-r-j)+i}Q^{
 2 (k +  j -  r)+ i } t_r^{ 2 (2 k + i - j - 2 r)}t_c^{-2 (k + i - 2 r) (k - j - r)}  \left[\begin{array}{c}r\\ k \end{array}\right]_{t_c^2}\left[\begin{array}{c}r-k \\ i \end{array}\right]_{t_c^2} \left[\begin{array}{c}k \\ j \end{array}\right]_{t_c^2}\cr
&&\times  (-\lambda ^2Q^{-4}  t_rt_c^{2\left(r-k-i\right)+1} ;t_c^2)_{j}(- \lambda ^2t_r^3 t_c^{1+4 r-2k};t_c^2)_{j} (-Q^2 t_r t_c;t_c^2)_{r-k}(-\lambda^2t_r^3 t_c^{1+2 r}  ;t_c^2)_{r-k}\cr
 &&\times (- \lambda Q^{-1}t_r t_c;t_c^2)_{r-k - i }(-\lambda Q^3 t_r^3  t_c^{1+2 (r-k+j)}  ;t_c^2)_{r-k-i}~.
\eea 
By construction, the red factors in \eqref{quad-Kauffman-trefoil} are very close to $Q^{2k}\wt\scP_{[k]}({\bf 4_1};\lambda Q^{-1}, Q, t_r, t_c)$ and the blue factors are killed by the universal differential $d^\univ_{\to}$ due to the presence of the $t_c^2$-Pochhammer $(-Q^2 t_r t_c;t_c^2)_{r-k}$.
One can check that the formula satisfies all the structural properties innate in the $[r]$-colored Kauffman homology. Subsequently, the Poincar\'e polynomial of the $[r]$-colored triply-graded Kauffman homology of the figure-eight in the $t_r$-grading can be expressed by
\bea\label{triple-Kauffman-fig8}
&&\scF_{[r]}({\bf 4_1};\lambda,q,t_r=t,t_c=1)=\wt\scF_{[r]}({\bf 4_1};\lambda,q,q^{-1}t,q)\cr
&=&\sum_{k=0}^{r}\sum_{j=0}^{k} \sum_{i=0}^{r-k} \lambda ^{2 (k-r-j)+i}q^{4 j - 2 k^2 - 2 k (1 + i - j - 3 r)  + 
 (i-2r) (2 r+ 2 j -1)} t^{2 (2 k-2 r+i-j)}\cr
&&\times\left[\begin{array}{c}r\\ k \end{array}\right]_{q^2}\left[\begin{array}{c}r-k \\ i \end{array}\right]_{q^2}\left[\begin{array}{c}k \\ j \end{array}\right]_{q^2}(-\lambda ^2q^{2 (r-k-i)-4} t ;q^2)_{j}(-\lambda ^2q^{-2+4 r-2k} t^3 ;q^2)_{j} (-q^2 t;q^2)_{r-k}\cr
&&\times (-\lambda ^2q^{-2+2 r} t^3 ;q^2)_{r-k}(- \lambda q^{-1}t;q^2)_{r-k-i}(-\lambda q^{1+2 (r-k+j)} t^3  ;q^2)_{r-k-i}~.
\eea
The  mirror/transposition symmetry provides the Poincar\'e polynomial of the $[1^r]$-colored triply-graded Kauffman homology of the trefoil in the $t_c$-grading
\bea
&&\scF_{[1^r]}({\bf 4_1};\lambda,q,t_r=1,t_c=t)=\wt\scF_{[1^r]}({\bf 4_1};\lambda,q^{r},q^{-1},qt)=\wt\scF_{[r]}({\bf 4_1};\lambda,q^{-1}t^{-1},qt,q^{-1})\cr
&=&\sum_{k=0}^{r}\sum_{j=0}^{k} \sum_{i=0}^{r-k} \lambda ^{2 (k-r-j)+i}q^{-4 j + 2 k^2 + 2 k (1 + i - j - 3 r)  -
 (i-2r) (2 r+ 2 j -1)} t^{i + 2 k - 2 (2 j + r)}\\
&&\times\left[\begin{array}{c}r\\ k \end{array}\right]_{q^{-2}}\left[\begin{array}{c}r-k \\ i \end{array}\right]_{q^{-2}}\left[\begin{array}{c}k \\ j \end{array}\right]_{q^{-2}}(-\lambda ^2q^{-2 (r-k-i)+4} t^5 ;q^{-2})_{j}(-\lambda ^2q^{2-4 r+2k} t^3 ;q^{-2})_{j} \cr
&&\times (-q^{-2} t^{-1};q^{-2})_{r-k}(-\lambda ^2q^{2-2 r} t^3 ;q^{-2})_{r-k}(- \lambda qt^2;q^{-2})_{r-k-i}(-\lambda q^{-1-2 (r-k+j)}  ;q^{-2})_{r-k-i}~.\nonumber
\eea

\begin{figure} [h]
 \centering
    \includegraphics[width=9cm]{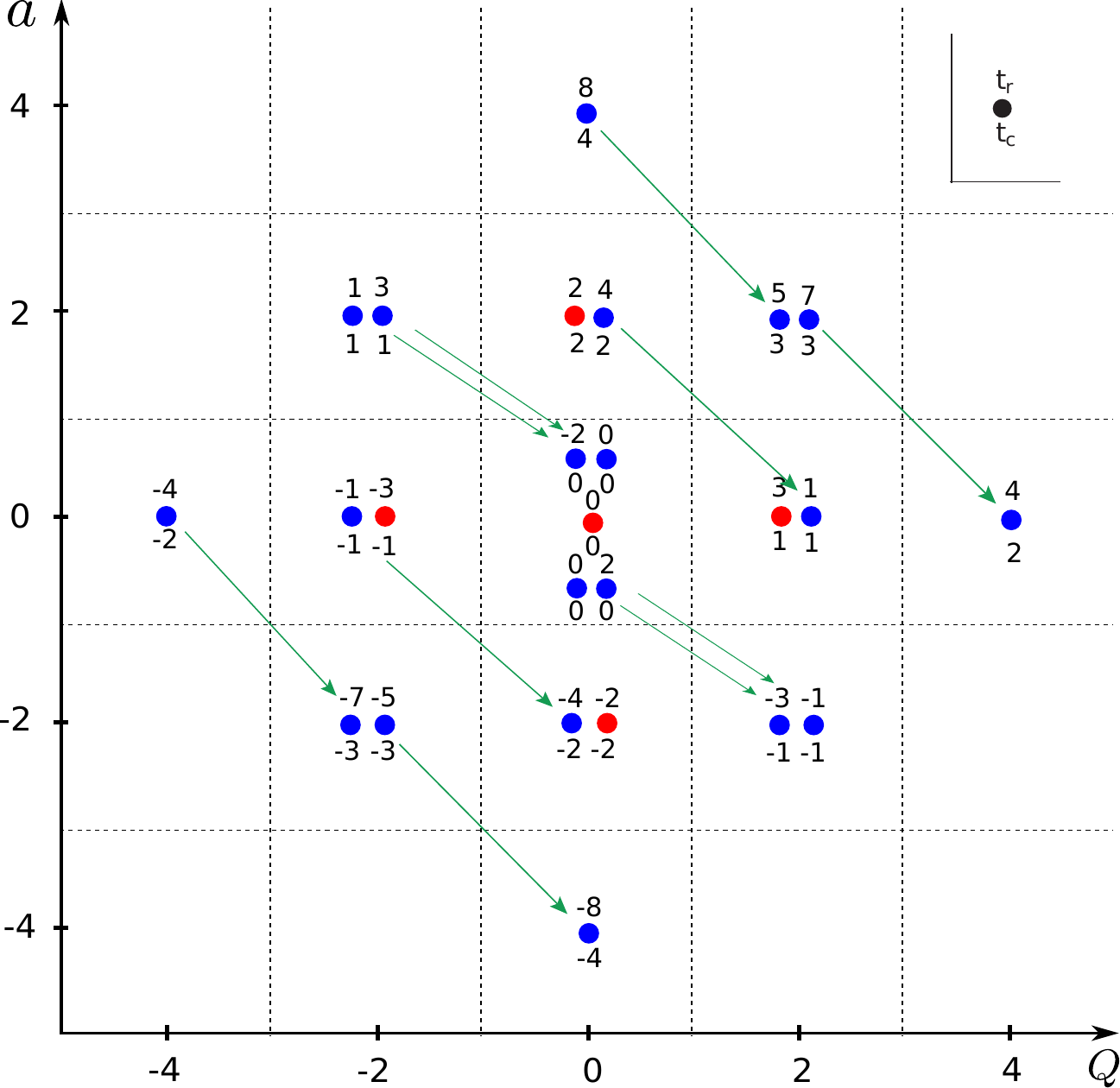}
    \caption{The $[1,1]$-colored HOMFLY homology of the figure-eight. The green arrows represent the colored differential $d^+_{[1,1]\to[1]}$ where the red dots are the elements in $H_*(\wt\scH^\HOMFLY_{[1,1]}({\bf 4_1}),d^+_{[1,1]\to[1]})$. The red and blue dots correspond to the factors colored  in red and blue in \eqref{quad-HOMFLY-2row-fig8} respectively.}
    \label{fig:fig8-[1,1]}
\end{figure}

Let us now try to obtain the $[r,r]$-colored HOMFLY homology of the figure-eight. The Poincar\'e polynomial of the $[1,1]$-colored HOMFLY homology (Figure \ref{fig:fig8-[1,1]}) which can be obtained from \eqref{quad-HOMFLY-fig8-1row} by the mirror/transposition symmetry 
\bea
\wt\scP_{\yng(1,1)}({\bf 4_1};a,Q,t_r,t_c)&=&{\color{red} \tfrac{1}{a^2  t_c^2 t_r^2 }+ \tfrac{1}{Q^2 t_c t_r^3} +1+ Q^2 t_c t_r^3 +a^2  t_c^2 t_r^2 }\cr
	&&{\color{blue} \tfrac{1}{ a^4 t_c^4 t_r^8} +\tfrac{1}{a^{2}} \left(\tfrac{1}{Q^2 t_c^3 t_r^7} + \tfrac{1}{Q^2 t_c^3 t_r^5} + \tfrac{1}{t_c^2 t_r^4} +  \tfrac{Q^2}{
  t_c t_r^3}+  \tfrac{1}{t_c^2 t_r^2} + \tfrac{Q^2}{t_c t_r}\right)  }\cr
  &&{\color{blue}+ \tfrac{1}{Q^4 t_c^2 t_r^4} +\tfrac{1}{
 Q^2 t_c t_r} +2+ \tfrac{1}{t_r^2 }+ t_r^2+ Q^2 t_c t_r  + Q^4 t_c^2 t_r^4 } \cr
 &&{\color{blue}
 + 
 a^2 \left(\tfrac{t_c t_r}{Q^2} + \tfrac{t_c t_r^3}{Q^2} + t_c^2 t_r^4 + 
    Q^2 t_c^3 t_r^5 + Q^2 t_c^3 t_r^7\right)+a^4 t_c^4 t_r^8 }~.
\eea
The $t_c=1$ specialization of the Poincar\'e polynomial of the $[r,r]$-colored quadruply-graded HOMFLY homology of the figure-eight is determined by the refined exponential growth property and the $t_c$-gradings can be eventually given by using \eqref{quad-Kauffman-fig8}. As a result, we can write a closed form expression
\bea\label{quad-HOMFLY-2row-fig8}
&&\wt\scP_{[r,r]}({\bf 4_1};a,Q,t_r,t_c)\cr
&=&\sum_{k=0}^{r}\sum_{j=0}^{k}\sum_{i=0}^{r-k} \left[\begin{array}{c}r\\ k \end{array}\right]_{t_c^2}{\color{red} a^{-2 j}  t_r^{-2 j}  t_c^{-2 rj} \left[\begin{array}{c}k \\ j \end{array}\right]_{t_c^2} (-a^2Q^{-2}  t_r^{-1}t_c^{2\left(r-k-i\right)+1} ;t_c^2)_{j} (- a^2Q^2t_r^5 t_c^{1+4 r-2k};t_c^2)_{j}}\cr
&&\times{\color{blue}a^{2 (i + 2 k - 2 r)} t_r^{4 (i + 2 k - 2 r)} t_c^{
 2 i (j - k + r) + 2 (k - r) (j - k + 2 r)}\left[\begin{array}{c}r-k \\ i \end{array}\right]_{t_c^2} (-a^2Q^{-2} t_r^3 t_c;t_c^2)_{r-k}}\cr
&&\times {\color{blue} (-a^2Q^2t_r^5 t_c^{1+2 r}  ;t_c^2)_{r-k} (- a^2 Q^{-2}t_r t_c;t_c^2)_{r-k - i }(-a^2 Q^2 t_r^7  t_c^{1+2 (r-k+j)}  ;t_c^2)_{r-k-i}}\cr
&=&\sum_{k=0}^{r}\sum_{j=0}^{k}\sum_{i=0}^{r-k} a^{2 (i - j + 2 k - 2 r)} t_c^{2 (i + k - 2 r) (j - k + r)} t_r^{
 4 i - 2 j + 8 k - 8 r}  \left[\begin{array}{c}r\\ k \end{array}\right]_{t_c^2}\left[\begin{array}{c}r-k \\ i \end{array}\right]_{t_c^2} \left[\begin{array}{c}k \\ j \end{array}\right]_{t_c^2}\cr
&&\times(-a^2Q^{-2}  t_r^{-1}t_c^{2\left(r-k-i\right)+1} ;t_c^2)_{j} (- a^2Q^2t_r^5 t_c^{1+4 r-2k};t_c^2)_{j}  (-a^2Q^{-2} t_r^3 t_c;t_c^2)_{r-k}\cr
 &&\times (-a^2Q^2t_r^5 t_c^{1+2 r}  ;t_c^2)_{r-k} (- a^2 Q^{-2}t_r t_c;t_c^2)_{r-k - i }(-a^2 Q^2 t_r^7  t_c^{1+2 (r-k+j)}  ;t_c^2)_{r-k-i} ~.
\eea
The Poincar\'e polynomial of the $[r,r]$-colored triply-graded HOMFLY homology of the figure-eight in the $t_r$-grading
\bea
&&\scP_{[r,r]}({\bf 4_1};a,q,t_r=t,t_c=1)=\wt\scP_{[r,r]}({\bf 4_1};a,q^2,q^{-1}t,q)\cr
&=&\sum_{k=0}^{r}\sum_{j=0}^{k}\sum_{i=0}^{r-k}  a^{2 (i - j + 2 k - 2 r)} q^{
 2 j (1 + k - 2 r) - 2 (4 + k - 2 r) (k - r) + 
  2 i (-2 + j - k + r)} t^{4 i - 2 j + 8 k - 8 r}\cr
  &&\times \left[\begin{array}{c}r\\ k \end{array}\right]_{q^2}\left[\begin{array}{c}r-k \\ i \end{array}\right]_{q^2} \left[\begin{array}{c}k \\ j \end{array}\right]_{q^2}(-a^2 q^{2\left(r-k-i-1\right)} t^{-1};q^2)_{j} (- a^2 q^{4 r-2k} t^5;q^2)_{j}  (-a^2q^{-6} t^3 ;q^2)_{r-k}\cr
 &&\times (-a^2 q^{2 r} t^5 ;q^2)_{r-k} (- a^2 q^{-4}t ;q^2)_{r-k - i }(-a^2  q^{2 (r-k+j-1)} t^7 ;q^2)_{r-k-i} ~.
\eea
Indeed, setting $r=2$, this formula decategorifies at $t=-1$ to the HOMFLY polynomial colored by $[2,2]$-representation written in (E.1) of \cite{Anokhina:2013ica}. 
The Poincar\'e polynomial of the $[2^r]$-colored triply-graded HOMFLY homology of the figure-eight in the $t_c$-grading can be written as
\bea
&&\scP_{[2^r]}({\bf 4_1};a,q,t_r=1,t_c=t)=\wt\scP_{[2^r]}({\bf 4_1};a,q^r,q^{-1},qt)=\wt\scP_{[r,r]}({\bf 4_1};a,q^r,qt,q^{-1})\cr
&=&\sum_{k=0}^{r}\sum_{j=0}^{k}\sum_{i=0}^{r-k}  a^{2 (i - j + 2 k - 2 r)} q^{
- 2 j (1 + k - 2 r) + 2 (4 + k - 2 r) (k - r) - 
  2 i (-2 + j - k + r)} t^{4 i - 2 j + 8 k - 8 r}\\
  &&\times \left[\begin{array}{c}r\\ k \end{array}\right]_{q^{-2}}\left[\begin{array}{c}r-k \\ i \end{array}\right]_{q^{-2}} \left[\begin{array}{c}k \\ j \end{array}\right]_{q^{-2}}(-a^2 q^{2 (-1 + i + k - 2 r)} t^{-1};q^{-2})_{j} (- a^2 q^{4 + 2 k - 2 r }t^5;q^{-2})_{j}  \cr
 &&\times  (-a^2q^{2-2r} t^3 ;q^{-2})_{r-k}(-a^2 q^{4} t^5 ;q^{-2})_{r-k} (- a^2 q^{-2r}t ;q^{-2})_{r-k - i }(-a^2  q^{6 - 2 j + 2 k} t^7 ;q^{-2})_{r-k-i} ~.\nonumber
\eea


\section{Super-$A$-polynomials} \label{sec:super-a-polynomials}
In the last fifteen years, remarkable results have been obtained by looking at the large color behaviors of colored Jones polynomials, {\it i.e.} the volume conjectures. (See a comprehensive review \cite{Murakami:2008} and references therein.) It is apparent that the volume conjecture \cite{Kashaev:1996kc,Murakami:1999} is a key to understanding the relationship between quantum invariants of a knot $K$ and classical geometry of the knot complement $S^3\backslash K$. Surprisingly, the large color behavior of colored Jones polynomials is dominated by $SL(2,\bC)$ flat connections rather than $SU(2)$ \cite{Gukov:2003na}. Hence, it is more directly related to analytically continued $SL(2,\bC)$ Chern-Simons theory \cite{Gukov:2003na,Gukov:2006ze,Witten:2010cx}. Let us briefly review the conjecture below.

Let $\frakg$ be  either $\fraksl(2)$ or $\frakso(3)$. Certainly, the corresponding gauge group $G$ in Chern-Simons theory is either $SU(2)$ or $SO(3)$ respectively. Since representations of $\frakg$ are specified by Young tableaux with a single row,  the colored $\frakg$ quantum invariants of a knot can be expressed by $J^\frakg_{[r]}(K,q)$. If one takes the double scaling limit $r\to \infty$ and $q=e^\hbar\to 1$ with $q^{2r}=x$ fixed, the invariant $J^{\frakg}_{[r]}(K;q)$ is conjectured to take the form
\be
J^{\frakg}_{[r]}(K;q)\;\overset{{r \to \infty \atop \hbar \to 0}}{\sim}\;
\exp\left( \frac{1}{2\hbar} \int \log y \frac{dx}{x} \,+\, \ldots \right)~,
\label{VC}
\ee
where the integral is carried out on the zero locus $A^{G_\bC}(K;x,y)=0$ of the $A$-polynomial. It is known that the $A$-polynomial $A^{G_\bC}(K;x,y)$ of a knot $K$ is the character variety of $G_\bC$-representation $\rho:\pi_1(S^3\backslash K) \to G_\bC$ of the fundamental group of the knot complement \cite{Cooper:1994}. Note that the complexification $G_{\bC}$ of the gauge group $G$ is either $SL(2,\bC)$ or $PSL(2,\bC)$, respectively. 

In fact, the moduli space $\scM_{\rm flat}(SL(2,{\bC}),T^2)$ of $SL(2,{\bC})$ flat connections on the boundary torus is a hyper-K\"ahler manifold $\bC^\times \times\bC^\times /\bZ_2$, where $\bC^\times  \times\bC^\times$ is spanned by the holonomy eigenvalues of the $SL(2,{\bC})$ gauge connection along the meridian $x$ and the longitude $y$, and $\bZ_2$ is the Weyl group symmetry of the gauge group $SL(2,{\bC})$. The moduli space of $SL(2,{\bC})$ flat connections on the knot complement is a Lagrangian submanifold of $\scM_{\rm flat}(SL(2,{\bC}),T^2)$, with respect to the symplectic form $\omega=\frac{i}{\hbar}d\log{x}\wedge d\log y$,  defined by the zero locus $A^{SL(2,{\bC})}(K;x,y)=0$ of the $A$-polynomial:
\bea
\scM_{\rm flat}(SL(2,{\bC}),S^3\backslash K)=\{ (x,y)\in\scM_{\rm flat}(SL(2,{\bC}),T^2)| A^{SL(2,{\bC})}(K;x,y)=0\}~.
\eea
It turns out that the moduli space of $PSL(2,{\bC})$ flat connections on the  torus is
\bea
\scM_{\rm flat}(PSL(2,{\bC}),T^2)=\scM_{\rm flat}(SL(2,{\bC}),T^2)/\Xi
\eea
where $\Xi=\bZ_2\times \bZ_2$ is generated by $(x,y)\mapsto (-x,y)$ and $(x,y)\mapsto (x,-y)$ \cite{Dimofte:2011jd}. In addition, it is shown in \cite{Dimofte:2011jd} that the ${PSL(2,\bC)}$ character variety can be written in terms of the ${SL(2,\bC)}$ character variety
\bea\label{PSL-SL}
A^{PSL(2,\bC)}(K;x,y)=A^{SL(2,\bC)}(K;x^{1/2},y^{1/2})A^{SL(2,\bC)}(K;x^{1/2},-y^{1/2})~.
\eea

Consecutively, the volume conjecture has been extended to the quantum version, called the quantum volume conjecture or the AJ conjecture \cite{Gukov:2003na,Garoufalidis:2003a,Garoufalidis:2003b}. Namely, the quantization of the $A$-polynomial becomes the $q$-holonomic function of the knot invariants:
\be\label{AJ}
\widehat{A}^{G_\bC}(K;\hat x,\hat y;q) J_{[r]}^{\frakg}(K;q) =0~,
\ee
where the operators $\hat x$ and  $\hat y$ act on the set of the colored quantum invariants as 
\be\label{xyactionJ}
\hat x J_{[r]}^{\frakg}(K,q)=q^{2r} J_{[r]}^{\frakg}(K,q) \ , \quad \hat y J_{[r]}^{\frakg}(K;q)=J_{[r+1]}^{\frakg}(K;q) \ .
\ee
Therefore,  the $q$-difference equation of the colored quantum invariants of minimal order
\be
b_k(\hat x,q)J_{[r+k]}^{\frakg}(K;q)+\cdots+b_0(\hat x,q)J_{[r]}^{\frakg}(K;q)=0
\ee
amounts to the quantum $A$-polynomial ${\widehat A}^{G_{\bC}}(K;\hat x, \hat y; q)=\sum_{j=0}^k b_j(\hat x,q)\hat y^j$ where taking $q=e^{\hbar}=1$ gives the classical $A$-polynomial $A^{G_{\bC}}(K;x,y)$.

Recently, generalizations of these conjectures have been proposed by incorporating $[r]$-colored HOMFLY polynomials and their categorifications \cite{Aganagic:2012jb,Fuji:2012pm,Fuji:2012nx,Garoufalidis:2012rt}. Specifically, the $q$-difference equation and the large color behavior of the Poincar\'e polynomial of a $[r]$-colored HOMFLY homology are called the quantum and classical super-$A$-polynomial \cite{Fuji:2012nx}. In this paper, we call it the super-$A$-polynomial $\scA^{SU}(K;x,y;a,t)$ of $SU$-type. The explicit computations have been performed for the $(2,2p+1)$-torus knots and the twist knots \cite{Fuji:2012nx,Nawata:2012pg,Fuji:2012pi}.

 Let us extend the notion of super-$A$-polynomials by including Poincar\'e polynomials of $[r]$-colored Kauffman homology.
In the limit
\bea
q = e^{\hbar} \to 1 \,, \qquad  \lambda = \text{fixed} \,, \qquad t = \text{fixed} \,, \qquad x = q^{2r} = \text{fixed} \ ,
\label{reflimit}
\eea
the Poincar\'e polynomial  $\scF_{[r]} (K;\lambda,q,t,1)$ of the  $[r]$-colored Kauffman homology asymptotes to the form 
\bea\label{VCsuper}
\scF_{[r]} (K;\lambda,q,t,1) \;\overset{{r \to \infty \atop \hbar \to 0}}{\sim}\;
\exp\left( \frac{1}{2\hbar} \int \log y \frac{dx}{x} \,+\, \ldots \right)~,
\eea
where the integral is carried out on the zero locus of the classical super-$A$-polynomial of $SO$-type 
\bea\label{supercurve}
\scA^{SO}(K;x,y;\lambda,t) =0~.
\eea

We conjecture that the $q$-difference equation of minimal order for the Poincar\'e polynomials of Kauffman homology,
\be
{\tt{b}}_k(\hat x;\lambda,q,t)\scF_{[r+k]}(K;\lambda,q,t,1)+\cdots+{\tt{b}}_0(\hat x;\lambda,q,t)\scF_{[r]}(K;\lambda,q,t,1)=0~,
\ee
provides the quantum super-$A$-polynomial of $SO$-type
\bea
\wh\scA^{SO}(K;\hat x,\hat y;\lambda,q,t)=\sum_k {\tt{b}}_k(\hat x;\lambda,q,t) \hat y^k~,
\eea
where the operators $\hat x$ and $\hat y$ act on $\scF_{[r]}(K;\lambda,q,t,1)$ as in \eqref{xyactionJ}, so that its classical limit $q\to1$ is equal to $\scA^{SO}(K;\hat x,\hat y;\lambda,t)$ up to factors.

The same procedure for the $[r]$-colored Kauffman polynomial $F_{[r]}(K;\lambda,q)$  leads to the quantum $\wh A^{SO}(K;\hat x,\hat y;\lambda,q) $ and classical $\lambda$-deformed $A$-polynomial $A^{SO}(K;x,y;\lambda) $ of $SO$-type. We emphasize that the $t=-1$ specialization of the super-$A$-polynomial is not necessarily equal to the $\lambda$-deformed $A$-polynomial though it always contains the $\lambda$-deformed $A$-polynomial. Furthermore, the $\lambda=1,t=-1$ specialization of the super-$A$-polynomial embraces the $PSL(2,\bC)$ character variety. The statement holds true for $SU$-type as well.
 
As conjectured in \cite{Fuji:2012pm,Gorsky:2013jxa}, we also predict that there is the relation between the super-$A$-polynomial of $SO$-type and the Poincar\'e polynomial of the uncolored Kauffman homology for any knot $K$:
\bea\label{x=1-uncolor}
\scA^{SO}(K;x=1,y;\lambda,t)\propto y-\scF_{\yng(1)}(K;\lambda,q=1,t,1)~.
\eea
 
\subsection{Trefoil}
Let us demonstrate the explicit calculation of the super-$A$-polynomial of $SO$-type for the trefoil, using the expression \eqref{triple-Kauffman-trefoil}. To implement it, we introduce three variables $w=q^{2i}$, $v=q^{2j}$ and $z=q^{2k}$. Then, in the limit \eqref{reflimit}, the sum over $i$, $j$ and $k$ in \eqref{triple-Kauffman-trefoil} for $\scF_{[r]} (3_1;\lambda,q,t)$ can be approximated by the integral over $w$, $v$ and $z$:
\be\label{asymptotics-trefoil}
\scF_{[r]}({\bf 3_1};\lambda,q,t,1)\sim \int e^{\frac{1}{2\hbar}\widetilde{\cW}^{SO}({\bf 3_1};x,w,v,z)}dwdvdz
\ee
where the twisted superpotential $\widetilde{\cW}$ is given by
\bea
\widetilde{\cW}^{SO}({\bf 3_1};x,w,v,z)&=&\log\left(\tfrac{wx^3}{z}\right)\log\lambda-\log v\log x+(\log x)^2+\log w\log (vx)\cr
		&&+2\log\left(\tfrac{wx}{v}\right)\log t-\tfrac{3\pi^{2}}{6}-\Li_{2}(x)+\Li_{2}(v)+\Li_{2}(zv^{-1})\cr
		&&+\Li_{2}(w)+\Li_{2}(xz^{-1}w^{-1})+\Li_{2}(-\lambda^{2}t^{3}x)-\Li_{2}(-\lambda^{2}t^{3}xv)\cr
		&&+\Li_{2}(-t)-\Li_{2}(-txz^{-1})+\Li_{2}(-\lambda t)-\Li_{2}(-\lambda tw)~.
		\label{W1trefoil}
\eea
Note that this can be obtained by using the asymptotics of the $q$-Pochhammer symbol \eqref{q-Poch-asymptotic}.
The leading asymptotic behavior \eqref{VCsuper} with respect to $\hbar$ comes from the saddle point
\be
\label{saddlepoint}
\left.\tfrac{\partial{\widetilde{\cW}^{SO}({\bf 3_1};x,w,v,z)}}{\partial{w}}\right|_{(w_0,v_0,z_0)}=\left.\tfrac{\partial{\widetilde{\cW}^{SO}({\bf 3_1};x,w,v,z)}}{\partial{v}}\right|_{(w_0,v_0,z_0)}=\left.\tfrac{\partial{\widetilde{\cW}^{SO}({\bf 3_1};x,w,v,z)}}{\partial{z}}\right|_{(w_0,v_0,z_0)}=0, 
\ee
and the zero locus \eqref{supercurve}  of the classical $A$-polynomial is determined by
\be
\label{zero locus}
y=\exp\left(x\tfrac{\partial{\widetilde{\cW}^{SO}({\bf 3_1};x,w_0,v_0,z_0)}}{\partial{x}}\right)~. 
\ee
Plugging \eqref{W1trefoil} into the above two equations, we have
\bea
1&=&\frac{t^2 x \lambda  v_0 \left(1+t \lambda  w_0\right) \left(x-w_0 z_0\right)}{\left(-1+w_0\right) w_0 z_0}~, \cr
1&=& -\frac{\left(1+t^3 x \lambda ^2 v_0\right) w_0 \left(v_0-z_0\right)}{t^2 x \left(-1+v_0\right) v_0}~,\cr
1&=&\frac{v_0 \left(x-w_0 z_0\right)}{\lambda  w_0 \left(-v_0+z_0\right)\left(t x+z_0\right)}~,\cr
y&=&\frac{t^2 (-1+x) x^2 \lambda ^3 \left(1+t^3 x \lambda ^2 v_0\right) w_0^2 \left(t x+z_0\right)}{\left(1+t^3 x \lambda ^2\right) v_0 \left(x-w_0 z_0\right)}~. 
\eea
Eliminating $w_0$, $v_0$ and $z_0$ from the set of the equations, we get the classical $A$-polynomial of $SO$-type
\bea
&&\scA^{SO}({\bf 3_1};x,y;\lambda,t)\cr
&=&(1+\lambda t^3 x)(1+\lambda^2 t^3 x)^2y^3\cr
&&-\lambda^2(1+\lambda^2 t^3 x)(1-t^2 x-\lambda t^2 x+2 t^2 x^2+2 \lambda t^2 x^2+2 \lambda t^3 x^2+2 \lambda^2 t^3 x^2\cr
&&\quad+\lambda t^4 x^2-\lambda t^4 x^3-2 \lambda^2 t^4 x^3+\lambda t^5 x^3-\lambda^2 t^5 x^3-2 \lambda^3 t^5 x^3+3 \lambda^2 t^4 x^4\cr
&&\quad+4 \lambda^2 t^5 x^4+4 \lambda^3 t^5 x^4+\lambda^2 t^6 x^4+4 \lambda^3 t^6 x^4+\lambda^4 t^6 x^4-\lambda^3 t^7 x^4+2 \lambda^3 t^7 x^{5}\cr
&&\quad+\lambda^4 t^7 x^{5}+2 \lambda^3 t^8 x^{5}+\lambda^4 t^8 x^{5}-\lambda^5 t^8 x^{5}+2 \lambda^4 t^8 x^{6}+2 \lambda^5 t^8 x^{6}\cr
&&\quad+2 \lambda^4 t^9 x^{6}+2 \lambda^5 t^9 x^{6}+\lambda^5 t^{10} x^{6}+\lambda^5 t^{11} x^{7}+\lambda^6 t^{11} x^{7}+\lambda^6 t^{12} x^{8})y^2\cr
&&-\lambda^4 t^4 (1-x) x^3(1-\lambda t^2 x-\lambda^2 t^2 x+2 \lambda^2 t^2 x^2+2 \lambda^3 t^2 x^2+2 \lambda^2 t^3 x^2\cr
&&\quad+2 \lambda^3 t^3 x^2+\lambda^3 t^4 x^2-2 \lambda^3 t^4 x^3-\lambda^4 t^4 x^3-2 \lambda^3 t^5 x^3-\lambda^4 t^5 x^3+\lambda^5 t^5 x^3\cr
&&\quad+3 \lambda^4 t^4 x^4+4 \lambda^4 t^5 x^4+4 \lambda^5 t^5 x^4+\lambda^4 t^6 x^4+4 \lambda^5 t^6 x^4+\lambda^6 t^6 x^4-\lambda^5 t^7 x^4\cr
&&\quad+\lambda^5 t^7 x^{5}+2 \lambda^6 t^7 x^{5}-\lambda^5 t^8 x^{5}+\lambda^6 t^8 x^{5}+2 \lambda^7 t^8 x^{5}+2 \lambda^6 t^8 x^{6}+2 \lambda^7 t^8 x^{6}\cr
&&\quad+2 \lambda^7 t^9 x^{6}+2 \lambda^8 t^9 x^{6}+\lambda^7 t^{10} x^{6}+\lambda^8 t^{11} x^{7}+\lambda^9 t^{11} x^{7}+\lambda^{10} t^{12} x^{8})y\cr
&&-\lambda^{11} t^{12} (1-x)^2 x^{9}(-1+\lambda x) ~.
\eea
Then, one can convince oneself that the relation \eqref{x=1-uncolor} holds in the case of the trefoil.

Since the expression \eqref{triple-Kauffman-trefoil} involves triple summations, it is very difficult to find the $q$-difference equation for it. Nevertheless, one can check whether the classical super-$A$-polynomial satisfies the condition for the quantizability \cite{Gukov:2011qp}.
It is argued in \cite{Gukov:2003na,Gukov:2011qp} that the integral of the one-form  $ \log y \frac{dx}{x}$  along a one-cycle $\gamma$ on the algebraic curve $\scA^{SO}(K;x,y;\lambda,t)=0$ must be subject to the Bohr-Sommerfeld condition in order for a classical super-$A$-polynomial $\scA^{SO}(K;x,y;\lambda,t)$ to be quantizable. If one writes the Newton polygon of $\scA^{SO}(K;x,y;\lambda,t)$, its faces correspond to punctures of the algebraic curve $\scA^{SO}(K;x,y;\lambda,t)=0$. Then, the Bohr-Sommerfeld condition around a puncture amounts to all roots of the corresponding face polynomial are roots of unity. Thus, the necessary condition for the quantizability is that the classical super-$A$-polynomial $\scA^{SO}(K;x,y;\lambda,t)$ is tempered \cite{Gukov:2011qp}.

  \begin{figure}[ht]
  \begin{minipage}[b]{8cm}\centering
\includegraphics[scale=.55]{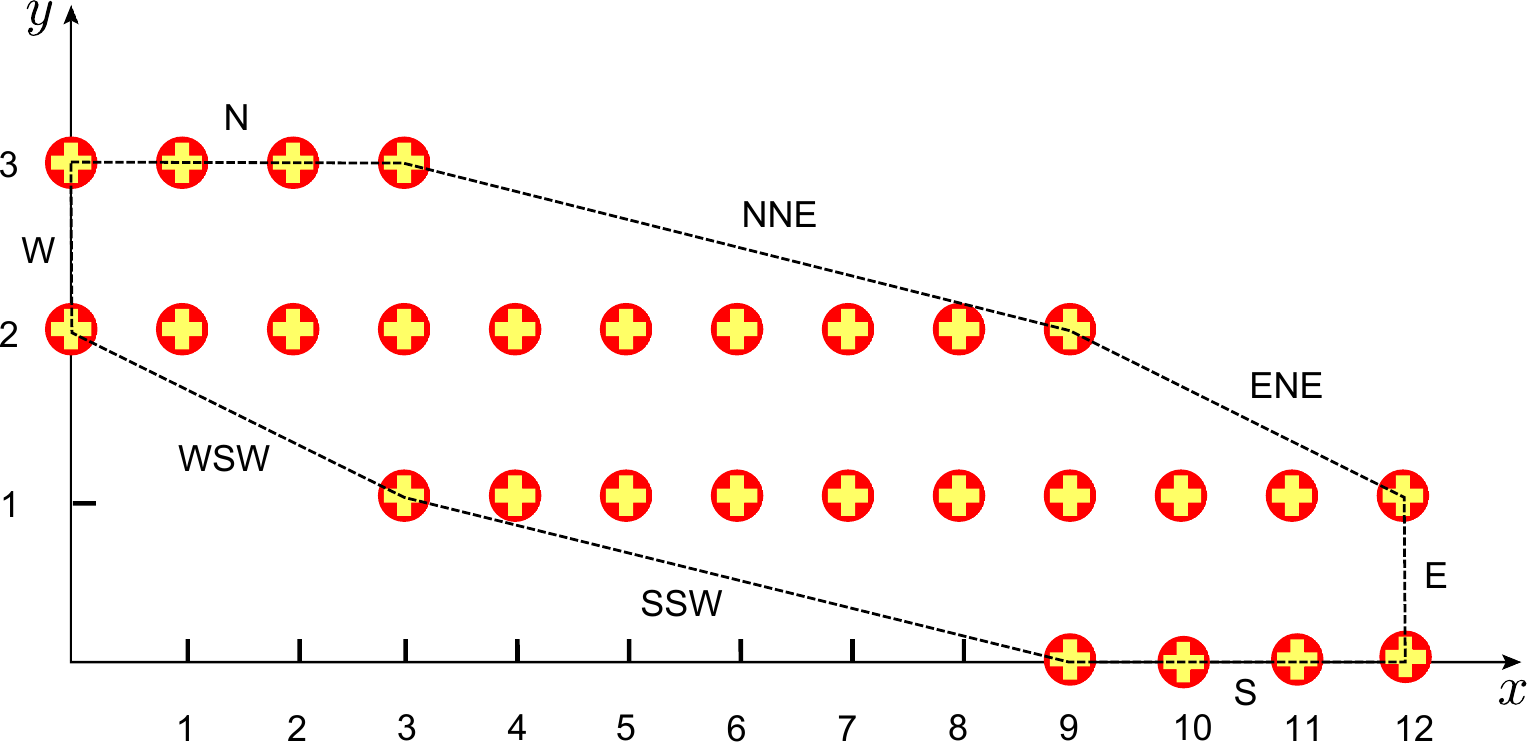}
\caption{Newton polygon of $\scA^{SO}({\bf 3_1};x,y;\lambda,t)$.}\label{fig:Newton-trefoil}
\end{minipage}
\hspace{.5cm}
  \begin{minipage}[b]{8cm}\centering
\begin{tabular}{|c|p{4cm}|}
\hline 
 $\text{\bf Face}$ &  $\text{\bf Face polynomials}$\tabularnewline
\hline 
\hline
 N & $\left(1+\lambda t^3 z\right) \left(1+\lambda^2 t^3 z\right)^2$
 \tabularnewline
\hline
NNE & $\lambda^5 t^9 \left(1-\lambda^5 t^6 z\right)$
\tabularnewline
\hline 
ENE & $\lambda^{10} t^{15} \left(-1+\lambda^4 t z\right)$
 \tabularnewline
\hline 
E & $\lambda^{12} t^{12} \left(\lambda^2 t^4-z\right)$
\tabularnewline
\hline 
S & $-\lambda^{11} t^{12} (\lambda-z) (-1+z)^2$
 \tabularnewline
\hline 
SSW & $\lambda^4 t^4 \left(\lambda^7 t^8-z\right)$
\tabularnewline
\hline 
WSW & $-\lambda^2 \left(\lambda^2 t^4+z\right)$
\tabularnewline
\hline 
W &$-\lambda^2+z$
\tabularnewline
\hline
 \end{tabular}\caption{Face polynomials of $\scA^{SO}({\bf 3_1};x,y;\lambda,t)$.}\label{fig:Face-trefoil}
\end{minipage}
\end{figure}

The Newton polygon of the super-$A$-polynomial $\scA^{SO}({\bf 3_1};x,y;\lambda,t)$ and its face polynomials are shown in Figure \ref{fig:Newton-trefoil} and Figure \ref{fig:Face-trefoil}. Writing  $\scA^{SO}({\bf 3_1};x,y;\lambda,t)= \sum_{i,j} c_{i,j} (\lambda,t) x^i y^j $, the Newton polygon is designed by plotting red circles for monomials $c_{ij}\neq0$ and yellow crosses for monomials at the special limit $c_{ij}(\lambda=1,t=-1)\neq0$. The faces of the Newton polygons are denoted by the dotted line in Figure \ref{fig:Newton-trefoil}. For a given face, we rename the monomial coefficients on the face as $c_k$. Then, the face polynomial is defined to be  $f(z)=\sum_k c_k z^k$. Assuming that the variables $\lambda,t$ are roots of unity, the quantizability condition requires that all roots of $f(z)$ constructed for all faces of the Newton polygon must be roots of unity. Therefore, it is easy to see from Figure \ref{fig:Face-trefoil} that the classical super-$A$-polynomials $\scA^{SO}({\bf 3_1};x,y;\lambda,t)$ satisfy the necessary condition of quantizability.

At $t=-1$, the super-$A$-polynomial becomes
\bea
&&\scA^{SO}({\bf 3_1};x,y;\lambda,t=-1)\\
&=&(-1 + x\lambda)(y - x y\lambda^2 - x^6\lambda^7 + 
   x^7\lambda^7) \cr
&&\times(-y^2 + y\lambda^2 - x y\lambda^2 + 
   2 x^2 y\lambda^2 + x y^2\lambda^2 + x^3\lambda^4 - 
   x^4\lambda^4 - 2 x^2 y\lambda^4 - x^3 y\lambda^4 + 
   x^4 y\lambda^6)~, \nonumber
\eea
while the analysis for the large color behavior of the Kauffman polynomials $F_{[r]}(K;\lambda,q)$ leads to the $\lambda$-deformed $A$-polynomial of $SO$-type
\bea
A^{SO}({\bf 3_1};x,y;\lambda)=y - x y\lambda^2 - x^6\lambda^7 + 
   x^7\lambda^7~.
\eea
Therefore, we can clearly see that $\scA^{SO}({\bf 3_1};x,y;\lambda,t=-1)\neq A^{SO}({\bf 3_1};x,y;\lambda)$ although $\scA^{SO}({\bf 3_1};x,y;\lambda,t=-1)$ contains $A^{SO}({\bf 3_1};x,y;\lambda)$. Furthermore, the $\lambda = 1, t=-1$ specialization of the super-$A$-polynomial can be written as  
\bea
\scA^{SO}({\bf 3_1};x,y;\lambda=1,t=-1)&=&( x-1)^3 (1- y)(y-x^{6} )  (x^3 + y) ~,
\eea
whereas the $PSL(2,\bC)$ character variety of the trefoil is expressed by
\bea\label{CV-trefoil}
A^{PSL(2,\bC)}({\bf 3_1};x,y)=(1-y)(y-x^{6} )~.
\eea
In fact, the large color limit of the $[r]$-colored $\frakso(3)$ quantum invariant of the trefoil, {\it i.e.} $\scF_{[r]}({\bf 3_1};\lambda=q^2,q,t=-1,1)$, provides only the non-abelian branch $y-x^{6}=0$. Hence, ignoring the trivial factor $1-x=0$, the specialization of the super-$A$-polynomial contains not only the abelian branch $1-y=0$ but also the extra non-abelian branch $x^3+y=0$. We postpone further study of understanding its meaning to future work. 

In the case of the colored $\frakso(3)$ quantum invariants, we can find the $q$-difference equation, providing the quantum $PSL(2,\bC)$ character variety of the trefoil
\bea
&&\wh{A}^{PSL(2,\bC)}({\bf 3_1};\hat{x},\hat{y};q)\cr
&=&(-1+q\hat{x}+\hat{x}^2q^{-3}+\hat{x}^3q^{-5}-\hat{x}^3q^{-2}-\hat{x}^4q^{-4}-\hat{x}^5q^{-10}+\hat{x}^6q^{-9})\hat{y}^2\cr
&&-(q^2-q^3 \hat{x}-q^7 \hat{x}^2+q^8 \hat{x}^3-q^9 \hat{x}^3+q^{10} \hat{x}^4+q^{12} \hat{x}^5-q^{14} \hat{x}^7-q^{14} \hat{x}^8\cr
&&-q^{14} \hat{x}^9+q^{15} \hat{x}^9+q^{15} \hat{x}^{10}+q^{13} \hat{x}^{11}-q^{14} \hat{x}^{12})\hat{y}\cr
&&-(q^{15} \hat{x}^6-q^{16} \hat{x}^7-q^{24} \hat{x}^8+q^{25} \hat{x}^9-q^{28} \hat{x}^9+q^{29} \hat{x}^{10}+q^{35} \hat{x}^{11}-q^{36} \hat{x}^{12})~.
\eea
A simple computation shows that it reduces to \eqref{CV-trefoil} up to trivial factors at $q=1$.

\subsection{Figure-eight}
\begin{figure}[ht]
  \begin{minipage}[b]{8cm}\centering
 \includegraphics[scale=.55]{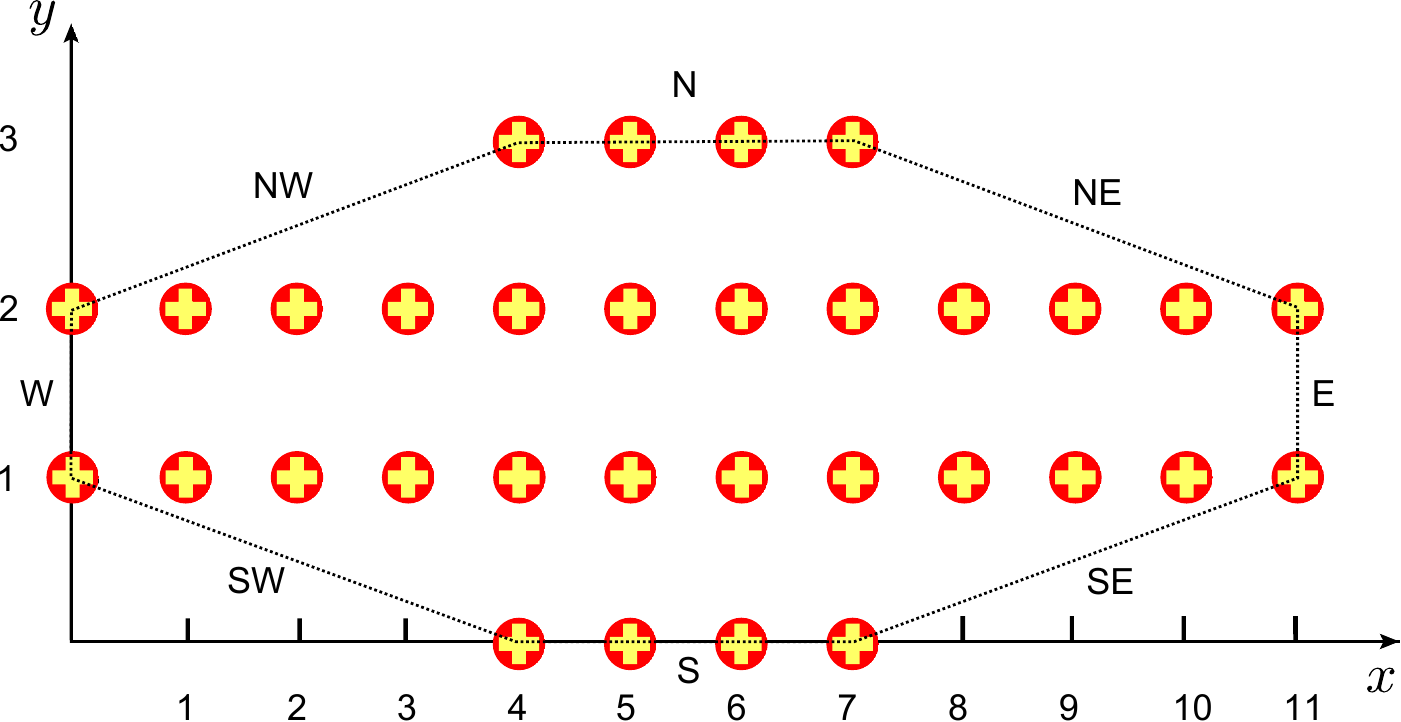}
\caption{Newton polygon of  $A^{SO}({\bf 4_1};x,y;\lambda)$.}\label{fig:Newton-fig8}
\end{minipage}
\hspace{.5cm}
  \begin{minipage}[b]{8cm}\centering
\begin{tabular}{|c|p{3.5cm}|}
\hline 
 $\text{\bf Face}$ &  $\text{\bf Face polynomials}$\tabularnewline
\hline 
\hline
 N & $\lambda ^3 \left(-1+z \lambda ^2\right)^3$
 \tabularnewline
\hline
NE & $-\lambda ^9 \left(-1+z \lambda ^2\right)$
\tabularnewline
\hline 
E & $\lambda ^{11} (-1+z \lambda )$
 \tabularnewline
\hline 
SE & $-\lambda ^6 \left(z-\lambda ^6\right)$
\tabularnewline
\hline 
S & $(-1+z)^3 \lambda ^6$
 \tabularnewline
\hline 
SW & $-z+\lambda ^6$
\tabularnewline
\hline 
W & $-1+z \lambda$
\tabularnewline
\hline 
NW &$-\lambda  \left(-1+z \lambda ^2\right)$
\tabularnewline
\hline
 \end{tabular}\caption{Face polynomials of $A^{SO}({\bf 4_1};x,y;\lambda)$.}\label{fig:Face-fig8}
\end{minipage}
\end{figure}

Now, let us consider the figure-eight.
In the limit \eqref{reflimit}, the expression \eqref{triple-Kauffman-fig8}  behaves as
\be \label{asymptotics-fig8}
\scF_{[r]}({\bf 4_1};\lambda,q,t,1)\sim \int e^{\frac{1}{2\hbar}\widetilde{\cW}^{SO}({\bf 4_1};x,w,v,z)}dwdvdz~,
\ee
with the twisted superpotential
\bea
\widetilde{\cW}^{SO}({\bf 4_1};x,w,v,z)&=&\log\left(\tfrac{z^2w}{v^2x^2}\right)\log{\lambda}-(\log{z})^2-\log{z}\log\left(\tfrac{w}{vx^3}\right)+\log{\left(\tfrac{w}{x^2}\right)}\log{(xv)}\cr
&&+\log\left(\tfrac{z^2w}{vx^2}\right)\log{t^2}-\tfrac{3\pi^2}{6}-\Li_2(x)+\Li_2(w)+\Li_2(xz^{-1}w^{-1})\cr
&&+\Li_2(v)+\Li_2(zv^{-1})+\Li_2(-\lambda^2xz^{-1}w^{-1}t)-\Li_2(-\lambda^2xz^{-1}w^{-1}vt )\cr
&&-\Li_2(-x^2z^{-1}t^3\lambda^2 v)+\Li_2(-t)-\Li_2(-xz^{-1}t)+\Li_2(-\lambda^2xt^3)\cr
&&+\Li_2(-\lambda t)-\Li_2(-\lambda xz^{-1}w^{-1}t)+\Li_2(-\lambda xz^{-1}vt^3)\cr
&&-\Li_2(-\lambda x^2z^{-2}vw^{-1}t^3 )~.
\label{Wfig8}
\eea
In this case, the saddle points are given by the following system of equations:
\bea
1&=&-\tfrac{t^2 v_0 w_0 x z_0 (-x+w_0 z_0) \lambda  \left(w_0 z_0+t x \lambda ^2\right)}{(-1+w_0) (w_0 z_0+t x \lambda ) \left(w_0 z_0^2+t^3 v_0 x^2 \lambda \right) \left(w_0 z_0+t v_0 x \lambda ^2\right)}~,\cr
1&=&-\tfrac{(v_0-z_0) \left(w_0 z_0^2+t^3 v_0 x^2 \lambda \right) \left(w_0 z_0+t v_0 x \lambda ^2\right) \left(z_0+t^3 v_0 x^2 \lambda ^2\right)}{t^2 (-1+v_0) v_0 w_0 x^2 z_0^2 \lambda ^2 \left(z_0+t^3 v_0 x \lambda \right)}~,\cr
1&=&\tfrac{t^4 v_0^2 w_0 x^3 z_0^3 (-x+w_0 z_0) \lambda ^2 \left(z_0+t^3 v_0 x \lambda \right) \left(w_0 z_0+t x \lambda ^2\right)}{(v_0-z_0) (t x+z_0) (w_0 z_0+t x \lambda ) \left(w_0 z_0^2+t^3 v_0 x^2 \lambda \right)^2 \left(w_0 z_0+t v_0 x \lambda ^2\right) \left(z_0+t^3 v_0 x^2 \lambda ^2\right)}~,\cr
y&=&-\tfrac{(-1+x) (t x+z_0) (w_0 z_0+t x \lambda ) \left(w_0 z_0^2+t^3 v_0 x^2 \lambda \right)^2 \left(w_0 z_0+t v_0 x \lambda ^2\right) \left(z_0+t^3 v_0 x^2 \lambda ^2\right)^2}{t^4 v_0^2 w_0 x^4 z_0^3 (-x+w_0 z_0) \lambda ^2 \left(z_0+t^3 v_0 x \lambda \right) \left(w_0 z_0+t x \lambda ^2\right) \left(1+t^3 x \lambda ^2\right)}~.
 \eea
With a current desktop computer, it is difficult to solve the above set of the equations for general value of $t$.\footnote{The authors would appreciate it if the reader could solve the equations.} Thus, we solve it only for the special case $t=-1$ which gives us the $\lambda$-deformed classical $A$-polynomial
 \bea
 &&A^{SO}({\bf 4_1};x,y;\lambda)\cr
 &=&x^4 \lambda^3 (-1 + x \lambda^2)^3y^3\cr
&&+(1 - x)\lambda (-1 + x\lambda^2)^2 (1 - x - x\lambda - 
   x^2\lambda - x^3\lambda - 2 x^2\lambda^2 + 
   x^3\lambda^2 + x^4\lambda^2 + 2 x^3\lambda^3  \cr
   &&\quad+
   5 x^4\lambda^3 + 2 x^5\lambda^3 + x^4\lambda^4 + 
   x^5\lambda^4 - 2 x^6\lambda^4 - x^5\lambda^5 - 
   x^6\lambda^5 - x^7\lambda^5 - x^7\lambda^6 + 
   x^8\lambda^6)y^2\cr
 &&+(1 - x)^2 (-1 + x\lambda^2) (1 - x\lambda - x\lambda^2 - 
   2 x^2\lambda^2 - x^2\lambda^3 + 2 x^3\lambda^3 + 
   x^3\lambda^4 + x^4\lambda^4 - x^3\lambda^5  \cr
&&\quad+   5 x^4\lambda^5 - x^5\lambda^5 + x^4\lambda^6 + 
   x^5\lambda^6 + 2 x^5\lambda^7 - x^6\lambda^7 - 
   2 x^6\lambda^8 - x^7\lambda^8 - x^7\lambda^9 + 
   x^8\lambda^{10})y\cr
   &&+(1 - x)^3 x^4\lambda^6~.
 \eea
In the sequel, one can confirm that the $\lambda$-deformed classical $A$-polynomial $A^{SO}({\bf 4_1};x,y;\lambda)$ obeys the quantizability condition, which can be seen in Figure \ref{fig:Newton-fig8} and Figure \ref{fig:Face-fig8}.

At $\lambda = 1$, it reduces to the  $A$-polynomial associated to the $PSL(2,\bC)$ character variety of the figure-eight
\bea
&& A^{SO}({\bf 4_1};x,y;\lambda=1)\cr
 &=&(x-1)^4(y-1)^2\cr
 &&\times(x^4 - y + 2 x y + 3 x^2 y - 2 x^3 y - 6 x^4 y - 2 x^5 y + 3 x^6 y +  2 x^7 y - x^8 y + x^4 y^2)~.~~~
\eea
 The third factor matches with the non-abelian branch of the $PSL(2,\bC)$ character variety of the figure-eight. (See Example 1 in \S 6.0.12. of \cite{Champanerkar:2003}.) In fact,  the $SL(2,\bC)$ character variety of the figure-eight is given by
\bea
A^{SL(2,\bC)}({\bf 4_1};x,y)=(1-y)(x^4y^2+(x^8-x^6-2x^4-x^{2}+1)y+x^4)~.
\eea
 Therefore, $A^{SO}({\bf 4_1};x,y;\lambda=1)$ is equal to $A^{SL(2,\bC)}({\bf 4_1};x^{1/2},y^{1/2})A^{SL(2,\bC)}({\bf 4_1};x^{1/2},-y^{1/2})$ up to the trivial factors as stated in \eqref{PSL-SL}. In addition, we can obtain the $q$-difference equation for the colored $\frakso(3)$ quantum invariants shown in Table \ref{tab:quantum-a-fig8}.


\section{3d/3d correspondence}\label{3d}
One of the most remarkable developments in recent years has been the program of studying a duality arising from the compactification of the 6d (2,0) superconformal field theory on a certain manifold. Particularly, the partially twisted compactification of the 6d (2,0) theory with Lie algebra $\frakg$ on a 3-manifold $M$ leads to a 3d $\cN=2$ supersymmetric gauge theory $T_{\frak g}[M]$, yielding deep relations between the geometry of the 3-manifold $M$ and the properties of the theory $T_{\frak g}[M]$ \cite{Dimofte:2010tz,Terashima:2011qi,Dimofte:2011ju,Dimofte:2011py}. The relation can be recapitulated by the statement that the partition function of $T_{\frak g}[M]$ on the squashed sphere $S^3_b$ or the superconformal index of $T_{\frak g}[M]$ is equal to the partition function of $G_\bC$ Chern-Simons theory on $M$. As a consequence, the moduli space of flat $G_\bC$ connections on $M$ is identified with the moduli space of supersymmetric vacua in the dual 3d gauge theory  $T_{\frak g}[M]$. This correspondence has been lately placed on a rigorous footing by the localization technique \cite{Yagi:2013fda,Lee:2013ida,Cordova:2013cea}. 

The explicit IR descriptions of $T_{\fraksl(2)}[M]$ for a large class of 3-manifolds $M$ have been established by abelian $\cN=2$ gauge theories, with possibly non-perturbative superpotentials that preserve $U(1)_R$ \cite{Dimofte:2011ju,Dimofte:2011py}. In particular, the cases in which a 3-manifold $M$ is a knot complement $S^3\backslash K$ have been intensively investigated \cite{Dimofte:2011ju,Dimofte:2011py,Beem:2012mb,Dimofte:2013lba,Fuji:2012pi}. Moreover, in this setting, the relation between  ``holomorphic blocks'' $B^\a(x;q)$ and colored Jones polynomials has been investigated in \cite{Beem:2012mb}. Roughly speaking, a holomorphic block $B^\a(x;q)$ is the partition function on  $S^1\times\bR$, labelled by a choice of vacuum $\alpha$ at the asymptotic boundary of spatial $\bR^2$ whose form is 
\be \label{BPSintro}
B^\a(x;q)\,\sim\, \Tr_{\cH(\bR^2;\alpha)} (-1)^R e^{-\beta H} q^{-J+\frac{R}{2}}x^e\,,
\ee
with fugacity $x$ for the $U(1)$ flavor symmetry. Note that the fugacity $x$ is identified with the holonomy eigenvalue $x$ of the gauge connection along the meridian of the tubular neighborhood of $K$ in $SL(2,\bC)$ Chern-Simons theory. Given an abelian $\cN=2$ gauge theory, the holomorphic block can be schematically expressed as 
\be \label{BIintro}
B^\a(x;q)\;\sim\; \int_{\Gamma^\alpha} \frac{dz}{z}\,\big[\theta(w;q)...\big]B_\Delta(w_1;q)\cdots B_\Delta(w_N;q)\,,
\ee
where we choose the appropriate integration cycle $\Gamma^\alpha \subset (\bC^\times)^s$, with $s$ being the number of the $U(1)$ gauge groups. The variables $z\in (\bC^*)^s$ are complexified scalars in the gauge multiplets, and each chiral multiplet contributes a single block 
\be \label{BDintro}
B_\Delta(w;q) = \sum_{n=0}^\infty \frac{w^{-n}}{(q^{-1};q^{-1})_n}\,,
\ee
to the integrand. Note that the parameter $w_i=w_i(z,x)$ depends on the scalars $z$ and the fugacity $x$. Here, the theta-functions $\theta(w;q)$ encode contributions of Chern-Simons and Fayet-Iliopoulos (FI) terms. Then, it was observed in \cite{Beem:2012mb} that the holomorphic block $B(x=q^{2r};q)$ of $T_{\fraksl(2)}[S^3\backslash K]$ at $x=q^{2r}$ coincides with the stable limit of the colored Jones polynomial $J_{[r]}(K;q)$. Nevertheless, the holomorphic block $B(x=q^{2r};q)$ of $T_{\fraksl(2)}[S^3\backslash K]$  does not reproduce the colored Jones polynomial exactly. This appears to be related to the fact that the way the theory $T_{\fraksl(2)}[S^3\backslash K]$ is constructed captures only the non-abelian branch, but not the abelian branch of the $SL(2,\bC)$ character variety. Equivalently, this can be rephrased that the large color limit of colored Jones polynomials provides only the non-abelian branch.\footnote{What is written in this paragraph was explained to S.N. by Sergei Gukov. S.N. is grateful to him.}

On the other hand, as we have seen in \S \ref{sec:super-a-polynomials}, a super-$A$-polynomial encodes  much richer information than an ordinary $A$-polynomial. The most important fact is that it intrinsically encompasses the abelian branch. Thus, it seems more appropriate to consider the 3d/3d correspondence in this setting.  As discussed in \cite{Fuji:2012nx,Fuji:2012pi}, the parameters $a$ or $\lambda$, and $t$ can be interpreted as fugacities in the index for certain global symmetries  $U(1)_{\text{bulk}}$ and $U(1)_F$ in the context of $\cN=2$ gauge theory. Thus, super-$A$-polynomials carry important information about $\cN=2$ gauge theories with those symmetries. Taking into account these features, it would certainly be interesting to elucidate the relation between the ``refined'' holomorphic blocks and the Poincar\'e polynomials of knot homology \cite{Chung:2013}.

As in the $\fraksl(2)$ case, the IR descriptions of $T_{\frakso(3)}[S^3\backslash K]$ can be constructed by abelian $\cN=2$ gauge theories. In fact, the Kaluza-Klein reduction on $S^1$ with $q\to1$ brings \eqref{BIintro} to the form given in \eqref{asymptotics-trefoil} and \eqref{asymptotics-fig8}. In the leading contribution, a single block $B_\Delta$ yields a dilogarithm ($\Li_2$) to the twisted superpotential. Hence, each dilogarithm ($\Li_2$) term in the twisted superpotential $\widetilde{\cW}$ expresses the contribution from a chiral field $\phi_i$. In \eqref{asymptotics-trefoil} and \eqref{asymptotics-fig8},  the parameters $w$, $v$ and $z$ can be interpreted as scalars in $U(1)$ gauge multiplets. Therefore, if the $j$-th dilogarithm term in $\widetilde{\cW}$ is $\pm \Li_2\left(x^{n_1}(-t)^{n_2}a^{n_3}z_{i}^{n_4}\right)$, then the chiral field $\phi_j$ will have charges $\pm n_1$, $\pm n_2$, $\pm n_3$, $\pm n_4$ respectively under the  $U(1)_x$, $U(1)_F$, $U(1)_{\text{bulk}}$ global symmetries and $U(1)_{z_i}$ gauge groups. In addition, the theta-functions reduces to the term $- \log x_i \cdot \log z_j$ for the FI coupling and the term $\frac{k_{ij}}{2} \, \log z_i \cdot \log z_j$ for the supersymmetric Chern-Simons coupling $\frac{k_{ij}}{4 \pi} \int A_i \wedge d A_j$. Using this dictionary, one can read off the theory $T_{\frakso(3)}[S^3\backslash {\bf 3_1}]$ and $T_{\frakso(3)}[S^3\backslash {\bf 4_1}]$ where the $U(1)$ charges of matter contents are depicted in Table \ref{tab:U(1)-trefoil} and  Table \ref{tab:U(1)-fig8}.

 \begin{table}[H]
\begin{equation*}
\begin{array}{|l|c|c|c|c|c|c|c|c|c|c|c|c|}
\hline
& \phi_1 & \phi_2 & \phi_3 & \phi_4 & \phi_5 & \phi_6 & \phi_7 & \phi_8 & \phi_9 & \phi_{10} & \phi_{11} &\text{Parameter} \\
\hline
U(1)_{\rm gauge} & 0 & 0 & 1 & 0 & -1 & 0 & 0 & 0 & 1 & 0 & 0 & z \\
U(1)_{\rm gauge} & 0 & 0 & 0 & 1 & -1 & 0 & 0 & 0 & 0 & 0 & 1 & w \\
U(1)_{\rm gauge} & 0 & 1 & -1 & 0 & 0 & 0 & -1 & 0 & 0 & 0 & 0 & v \\
U(1)_{\rm global}& -1 & 0 & 0 & 0 & 1 & 1 & -1 & 0 & -1 & 0 & 0 & x \\
U(1)_{\rm bulk}& 0 & 0 & 0 & 0 & 0 & 2 & -2 & 0 & 0 & 1 & -1 & \lambda \\
U(1)_{\rm F}& 0 & 0 & 0 & 0 & 0 & 3 & -3 & 1 & -1 & 1 & -1 & t \\
\hline
\end{array}
\end{equation*}
\caption{$U(1)$ charges of the chiral fields corresponding to the twisted superpotential $\widetilde{\cW}^{SO}({\bf 3_1};x,w,v,z)$.}\label{tab:U(1)-trefoil}
\end{table}

 \begin{table}[H]
\begin{equation*}
\begin{array}{|l|c|c|c|c|c|c|c|c|c|c|c|c|c|c|c|c|c|c|}
\hline
& \phi_1 & \phi_2 & \phi_3 & \phi_4 & \phi_5 & \phi_6 & \phi_7 & \phi_8 & \phi_9 & \phi_{10} & \phi_{11} & \phi_{12} & \phi_{13} & \phi_{14} & \phi_{15} &\text{Parameter} \\
\hline
U(1)_{\rm gauge} & 0 & 0 & -1 & 0 & 1 & -1 & 1 & 1 & 0 & 1 & 0 & 0 & 1 & -1 & 2 & z \\
U(1)_{\rm gauge} & 0 & 1 & -1 & 0 & 0 & -1 & 1 & 0 & 0 & 0 & 0 & 0 & 1 & 0 & 1 & w \\
U(1)_{\rm gauge} & 0 & 0 & 0 & 1 & -1 & 0 & -1 & -1 & 0 & 0 & 0 & 0 & 0 & 1 & -1 & v \\
U(1)_{\rm global}& -1 & 0 & 1 & 0 & 0 & 1 & -1 & -2 & 0 & -1 & 1 & 0 & -1 & 1 & -2 & x \\
U(1)_{\rm bulk}& 0 & 0 & 0 & 0 & 0 & 2 & -2 & -2 & 0 & 0 & 2 & 1 & -1 & 1 & -1 & \lambda \\
U(1)_{\rm F}& 0 & 0 & 0 & 0 & 0 & 1 & -1 & -3 & 1 & -1 & 3 & 1 & -1 & 3 & -3 & t \\
\hline
\end{array}
\end{equation*}
\caption{$U(1)$ charges of the chiral fields corresponding to the twisted superpotential $\widetilde{\cW}^{SO}({\bf 4_1};x,w,v,z)$.}\label{tab:U(1)-fig8}
\end{table}


\section{Future directions}\label{sec:discussions}
The study on the structure of the $[r]$-colored Kauffman homology that we have implemented raises several questions. Although we focused on thin knots of the simplest class so far, it would be desirable to study whether these structural properties hold for the Kauffman homology of thick knots. However, the size of colored Kauffman homology for thick knots would be too large to study with our current technique. Nevertheless, it is worth making the following comments. Given the uncolored Kauffman homology of the $(3,4)$-torus knot  $T_{3,4}$ (Appendix B in \cite{Gukov:2011ry}), its uncolored $\frakso(4)$ homology should be isomorphic to the homology with respect to  the $d_4$ differential. However, if we use the differential whose $(\lambda,q,t_r,t_c)$-degree is $\deg d_4=(-1,3,-1,-1)$ as in \eqref{dN-degree}, the homology $ H_*(\scH^\Kauffman_{\yng(1)}(T_{3,4}),d_4)$ would be 41-dimensional while the uncolored $\frakso(4)$ homology of $T_{3,4}$  should be 25-dimensional, which follows from the 5-dimensional uncolored $\fraksl(2)$ homology of $T_{3,4}$. This indicates that  one has to take into account the spectral sequence \cite{Rasmussen:2006} for the $d_4$ differential in the case of thick knots.  

Another important problem concerns the interpretation of the differentials inherent in colored Kauffman homology.
In \cite{Gukov:2005qp}, the differentials in uncolored Kauffman homology are elucidated by the deformations of  the Landau-Ginzburg potential in B-model. One way to relate the colored Kauffman homology to the Landau-Ginzburg  B-model is the algebraic model proposed in \cite{Gorsky:2013jxa}. In this model, the colored HOMFLY homology of a torus knot is identified with a space of differential forms on a certain reduced moduli space. This description is certainly suitable for the interpretation in terms of the Landau-Ginzburg model. Hence, the route via the algebraic model may connect to the explanation of the differentials  in terms of the  Landau-Ginzburg  B-model .

Furthermore, it was observed in \cite{Gorsky:2013jxa} that the colored differentials present in the HOMFLY homology can be associated with the representations of the Lie superalgebra $\fraksl(n|m)$. Therefore, it would be intriguing to consider the relation to Lie superalgebra and its physical meaning in the context of the colored Kauffman homology.

It is also worthwhile to find a new way to look at the $\lambda$-deformed $A$-polynomials of $SO$-type. 
It is conjectured in \cite{Aganagic:2012jb} that the $a$-deformed $A$-polynomials of $SU$-type can be identified with the augmentation polynomials of knot contact homology~\cite{Ng:2010,Ekholm:2010}. This conjecture has been verified in the cases of the $(2,2p+1)$-torus knots \cite{Fuji:2012nx} and the twist knots \cite{Fuji:2012pi,Nawata:2012pg}. It is natural to ask whether or not one can construct augmentation polynomials in knot contact homology analogous to the $\lambda$-deformed $A$-polynomials of $SO$-type.  

Finally, one may be interested in extracting information about quantum $6j$-symbols for $U_q(\frakso(N))$ from the expressions for the invariants of the trefoil and the figure-eight obtained in \S \ref{sec:homologi} as done in \cite{Nawata:2013ppa} for $U_q(\fraksl(N))$. For this purpose, the relations from representation theory will be also useful.  In fact, quantum $6j$-symbols for $U_q(\frakso(N))$  would enable us to evaluate colored $\frakso(N)$ quantum invariants for any knot. The feasibility of this approach is currently under investigation.


\section*{Acknowledgement}
S.N. is indebted to S. Gukov, M. Sto$\check{\text{s}}$i$\acute{\text{c}}$ and P. Su{\l}kowski for collaboration on related topics. Without discussion with them, this paper would have been impossible. In addition, the authors would like to thank M. Aganagic, H.-J. Chung, T. Dimofte, S. Garoufalidis, C. Koutschan, M. Mari\~no, H. Murakami, P. Putrov, R. van der Veen, J. Walcher for helpful discussions and comments. Furthermore, the authors appreciate S. Arthamonov, A. Morozov, S. Shakirov and S. Stevan for sharing Mathematica programs, as well as their results of colored HOMFLY and Kauffman polynomials. The authors are grateful to J. Walcher for organizing the workshop ``Physics and Mathematics of Link Homology'', where a part of the work was carried out. S.N. would like to thank IIT Bombay and Summer Simons Workshop 2013 for the warm hospitality where the initial and final stage of the work had been done, respectively.  S.N. is also grateful to ITEP group and T. Takebe for the invitation to the workshop ``Synthesis of integrabilities in the context of duality between the string theory and gauge theories". The work of S.N. is partially supported by the ERC Advanced Grant no.~246974, {\it``Supersymmetry: a window to non-perturbative physics''}.
\clearpage
\appendix

\section{Conventions and notations}\label{conv}
\begin{description}
\item[Polynomial invariants] \hfill \\

\begin{table}[h]
\begin{tabular}{lp{12cm} l}
$J^{\frakg}_R(K;q)$& The reduced $\frakg$ quantum invariant of a knot $K$ colored by a representation $R$ of $\frakg$ \cr
$\overline J^{\frakg}_R(K;q)$& The unreduced $\frakg$  quantum invariant of a knot $K$ colored by a representation $R$ of $\frakg$ \cr
$P_R(K;a,q)$& The reduced HOMFLY polynomial of a knot $K$ colored by a representation $R$\cr
$\overline P_R(K;a,q)$& The unreduced HOMFLY polynomial of a knot $K$ colored by a representation $R$\cr
$\overline P_{(R,S)}(K;a,q)$& The unreduced HOMFLY polynomial of a knot $K$ colored by a composite representation $(R,S)$\cr
$F_R(K;\lambda,q)$&  The reduced Kauffman polynomial of a knot $K$ colored by  a representation $R$\cr
$\overline F_R(K;\lambda,q)$&  The unreduced Kauffman polynomial of a knot $K$ colored by a representation $R$
\end{tabular}
\nonumber
\end{table}

\item[Knot homology]\hfill\\
\begin{table}[h]
\begin{tabular}{lp{10.5cm} l}
$(\scH^{\rm HOMFLY}_R(K))_{i,j,k,\ell}$& The reduced quadruply-graded HOMFLY homology colored by a representation $R$ with $(a,q,t_r,t_c)$-gradings \cr
$(\wt\scH^{\rm HOMFLY}_R(K))_{i,j,k,\ell}$& The reduced quadruply-graded HOMFLY homology colored by a representation $R$ with $(a,Q,t_r,t_c)$-gradings\cr
$H_*(\scH^\HOMFLY_{R}(K),d)_{i,j,k,\ell}$& The homology  with respect to a differential $d$ acting on $(\scH^\HOMFLY_R(K))_{i,j,k,.
\ell}$  with $(a,q,t_r,t_c)$-gradings \cr
$H_*(\wt\scH^\HOMFLY_{R}(K),d)_{i,j,k,\ell}$& The homology  with respect to  a differential $d$ acting on $(\wt\scH^\HOMFLY_R(K))_{i,j,k,.
\ell}$  with $(a,Q,t_r,t_c)$-gradings \cr
$(\scH^\Kauffman_R(K))_{i,j,k,.
\ell}$& The reduced quadruply-graded Kauffman homology  colored by a representation $R$ with $(\lambda,q,t_r,t_c)$-gradings \cr
$(\wt\scH^\Kauffman_R(K))_{i,j,k,\ell}$& The reduced quadruply-graded Kauffman homology  colored by a representation $R$ with $(\lambda,Q,t_r,t_c)$-gradings \cr
$H_*(\scH^\Kauffman_{R}(K),d)_{i,j,k,\ell}$& The homology with respect to a differential $d$ acting on $(\scH^\Kauffman_R(K))_{i,j,k,.
\ell}$  with $(\lambda,q,t_r,t_c)$-gradings \cr
$H_*(\wt\scH^\Kauffman_{R}(K),d)_{i,j,k,\ell}$& The homology  with respect to a differential $d$ acting on $(\wt\scH^\Kauffman_R(K))_{i,j,k,.
\ell}$  with $(\lambda,Q,t_r,t_c)$-gradings 
\end{tabular}
\nonumber
\end{table}

\clearpage
\item[Poincar\'e polynomials]\hfill\\
\begin{eqnarray*}
\scP_R(K;a,q, t_r,t_c)&:=& \sum_{i,j,k,\ell} a^i q^j t_r^k t_c^\ell ~\dim\; (\scH^\HOMFLY_{R}(K))_{i,j,k,\ell}\cr
\wt\scP_R(K;a,Q, t_r,t_c)&:=& \sum_{i,j,k,\ell} a^i Q^j t_r^k t_c^\ell ~\dim\;(\wt\scH^\HOMFLY_{R}(K))_{i,j,k,\ell}\cr
\scP(\scH^\HOMFLY_{R}(K),d)(a,q,t_r,t_c)&:=& \sum_{i,j,k,\ell} a^i q^j t_r^k t_c^\ell ~\dim H_*(\scH^\HOMFLY_{R}(K),d)_{i,j,k,\ell}\cr
\wt\scP(\wt\scH^\HOMFLY_{R}(K),d)(a,Q,t_r,t_c)&:=& \sum_{i,j,k,\ell} a^i Q^j t_r^k t_c^\ell ~\dim H_*(\wt\scH^\HOMFLY_{R}(K),d)_{i,j,k,\ell}\cr
\scF_R(K;\lambda,q, t_r,t_c)&:=& \sum_{i,j,k,\ell} \lambda^i q^j t_r^k t_c^\ell ~\dim\;(\scH^\Kauffman_{R}(K))_{i,j,k,\ell}\cr
\wt\scF_R(K;\lambda,Q, t_r,t_c)&:=& \sum_{i,j,k,\ell} \lambda^i Q^j t_r^k t_c^\ell ~\dim\;(\wt\scH^\Kauffman_{R}(K))_{i,j,k,\ell}\cr
\scF(\scH^\Kauffman_{R}(K),d)(\lambda,q,t_r,t_c)&:=& \sum_{i,j,k,\ell} \lambda^i q^j t_r^k t_c^\ell ~\dim H_*(\scH^\Kauffman_{R}(K),d_N)_{i,j,k,\ell}\cr
\wt\scF(\wt\scH^\Kauffman_{R}(K),d)(\lambda,Q,t_r,t_c)&:=& \sum_{i,j,k,\ell} \lambda^i Q^j t_r^k t_c^\ell ~\dim H_*(\wt\scH^\Kauffman_{R}(K),d)_{i,j,k,\ell}
\end{eqnarray*}
\item[$A$-polynomials]\hfill\\
\begin{table}[h]
\begin{tabular}{lp{11cm} l}
$A^{SL(2,\bC)}(K;x,y)$&  The $SL(2,\bC)$ character variety of the complement of a knot $K$ in $S^3$\\
$A^{PSL(2,\bC)}(K;x,y)$&  The $PSL(2,\bC)$ character variety of the complement of a knot $K$ in $S^3$\\
$\wh A^{PSL(2,\bC)}(K;\hat x,\hat y;q)$& The quantum $PSL(2,\bC)$ character variety of the complement of a knot $K$ in $S^3$\\
$A^{SU} (K;x,y;a)$& The $a$-deformed $A$-polynomial of $SU$-type for a knot $K$\\
$A^{SO} (K;x,y;\lambda)$&  The $\lambda$-deformed $A$-polynomial of $SO$-type for a knot $K$ \\
$\wh A^{SO} (K;\hat x,\hat y;\lambda,q)$& The quantum $\lambda$-deformed $A$-polynomial of $SO$-type for a knot $K$ \\
$\scA^{SU} (K;x,y;a,t)$& The super-$A$-polynomial of $SU$-type for a knot $K$\\
$\scA^{SO} (K;x,y;\lambda,t)$& The super-$A$-polynomial of  $SO$-type for a knot $K$
\end{tabular}
\nonumber
\end{table}
\end{description}


\section{Integrality conjecture for Kauffman polynomials}\label{sec:integrality}
In this appendix, we check the integrality conjecture  proposed by Mari\~no \cite{Marino:2009mw} for Kauffman polynomials using the invariants of the figure-eight. This is the first non-trivial check of the integrality conjecture for a non-torus knot that provides an indirect evidence for geometric transition in the presence of an orientifold \cite{Sinha:2000ap}.

Similar to Ooguri-Vafa reformulated polynomials,  we can determine 
reformulated invariants $g_R(K;\lambda,q)$ with integrality structure using unreduced colored Kauffman polynomials and unreduced  colored HOMFLY polynomials.  Although the Ooguri-Vafa-Labastida-Mari\~no conjecture for $U(N)$ invariants was proven in \cite{Liu:2007kv,Liu:2010zr}, the Mari\~no's conjecture \cite{Marino:2009mw} still remains open. 
The colored Kauffman polynomials for knots  appear in writing the expectation value of the Ooguri-Vafa operator 
\be
Z_{CS}^{SO/Sp}(K;\lambda,q,v)= \sum_R {\overline F}_R(K;\lambda,q) s_R(v)~,
\ee
in the A-model  topological strings on an orientifolded conifold background 
where $s_R(v)$ are the Schur polynomials in variable $v$. As explained in \eqref{observ}, the topological string  partition function in this background receives the contributions from the oriented  and  
unoriented sectors  \cite{Bouchard:2004iu,Bouchard:2004ri,Borhade:2005pw,Marino:2009mw}.
 The oriented topological string amplitude involves colored HOMFLY polynomials
of knots carrying composite representations $(R,S)$ \cite{Marino:2009mw} 
\be
\log \left[\sum_{R,S} {\overline P}_{(R,S)}(K;a,q) s_R(v) s_S(v)\right] =
\sum_{\wt R} \sum_{d=1}^{\infty}  \frac1dh_{\wt R}(a^d,q^d)s_{\wt R}(v^d)~.
\ee 
It is straightforward to obtain the explicit expressions up to double boxes
\bea
h_{\yng(1)}(K;a,q)&=&2\bar{P}_{\yng(1)}(K;a,q)~,\cr
h_{\yng(2)}(K;a,q)&=&2\bar{P}_{\yng(2)}(K;a,q)+\bar{P}_{(\yng(1),\yng(1))}(K;a,q)-2\bar{P}_{\yng(1)}(K;a,q)^2-\bar{P}_{\yng(1)}(K;a^2,q^2)~,\cr
h_{\yng(1,1)}(K;a,q)&=&2\bar{P}_{\yng(1,1)}(K;a,q)+\bar{P}_{(\yng(1),\yng(1))}(K;a,q)-2\bar{P}_{\yng(1)}(K;a,q)^2+\bar{P}_{\yng(1)}(K;a^2,q^2)~.
\eea
The reformulated invariant is  expected to obey the Ooguri-Vafa integrality structure 
\be\label{OV-form}
h_{R}(K;a,q)= \frac{1}{q-q^{-1}} \sum_{\beta,s} N_{R,\beta,s}(K) a^\beta q^s~,
\ee 
where
$N_{R,\beta,s}$ are integers. The check of this conjecture for the figure-eight requires the
 HOMFLY polynomials colored by composite representations.
Fortunately, the colored HOMFLY polynomial of the figure-eight  colored by composite representation  $(\yng(1),\yng(1))$ is given in p.25 of \cite{Hadji:2006}
\bea
&&{\overline P}_{(\yng(1),\yng(1))} ({\bf 4_1})\cr
&=&3-2 z^2-6 z^4-2 z^6+\left(-2-z^2+2 z^4+z^6\right) \left(\tfrac{1}{\lambda^2}+\lambda^2\right)+\left(1+2 z^2+z^4\right) \left(\tfrac{1}{\lambda^4}+\lambda^4\right) \cr
&=&11-\tfrac{2}{q^6}+\tfrac{6}{q^4}-\tfrac{8}{q^2}-8 q^2+6 q^4-2 q^6+\lambda^{-4}\left(3+\tfrac{1}{q^4}-\tfrac{2}{q^2}-2 q^2+q^4\right)\cr
&&+\lambda^4 \left(3+\tfrac{1}{q^4}-\tfrac{2}{q^2}-2 q^2+q^4\right)+\lambda^{-2}\left(-8+\tfrac{1}{q^6}-\tfrac{4}{q^4}+\tfrac{6}{q^2}+6 q^2-4 q^4+q^6\right)\cr
&&+\lambda^2 \left(-8+\tfrac{1}{q^6}-\tfrac{4}{q^4}+\tfrac{6}{q^2}+6 q^2-4 q^4+q^6\right),
\eea
where $z=q-q^{-1}$. 
Using this data, we can verify the expected  form \eqref{OV-form}
for the figure-eight  upto two boxes:
\bea
\label{h-twohor}
h_{\yng(2)}({\bf 4_1};a,q)&=&-\frac{1}{a^6 q^7(q-q^{-1})}(-a^4+2 a^6-a^8-2 a^2 q^2+5 a^4 q^2-4 a^6 q^2+a^8 q^2+3 q^4\cr
&&-6 a^2 q^4+2 a^4 q^4+2 a^6 q^4-2 a^{10} q^4+a^{12} q^4-q^6+2 a^2 q^6+2 a^4 q^6-5 a^6 q^6\cr
&&-2 a^8 q^6+4 a^{10} q^6-a^{12} q^6+q^8-4 a^2 q^8+2 a^4 q^8+5 a^6 q^8-2 a^8 q^8-2 a^{10} q^8\cr
&&+a^{12} q^8-q^{10}+2 a^2 q^{10}-2 a^6 q^{10}-2 a^8 q^{10}+6 a^{10} q^{10}-3 a^{12} q^{10}-a^4 q^{12}\cr
&&+4 a^6 q^{12}-5 a^8 q^{12}+2 a^{10} q^{12}+a^4 q^{14}-2 a^6 q^{14}+a^8 q^{14})~, 
\eea
\bea
\label{h-twover}
h_{\yng(1,1)}({\bf 4_1};a,q)&=&-\frac{1}{a^6 q^7(q-q^{-1})}(-a^4+2 a^6-a^8+a^4 q^2-4 a^6 q^2+5 a^8 q^2-2 a^{10} q^2+q^4\cr
&&-2 a^2 q^4+2 a^6 q^4+2 a^8 q^4-6 a^{10} q^4+3 a^{12} q^4-q^6+4 a^2 q^6-2 a^4 q^6-5 a^6 q^6\cr
&&+2 a^8 q^6+2 a^{10} q^6-a^{12} q^6+q^8-2 a^2 q^8-2 a^4 q^8+5 a^6 q^8+2 a^8 q^8-4 a^{10} q^8\cr
&&+a^{12} q^8-3 q^{10}+6 a^2 q^{10}-2 a^4 q^{10}-2 a^6 q^{10}+2 a^{10} q^{10}-a^{12} q^{10}\cr
&&+2 a^2 q^{12}-5 a^4 q^{12}+4 a^6 q^{12}-a^8 q^{12}+a^4 q^{14}-2 a^6 q^{14}+a^8 q^{14})~.
\eea 
Subtracting the oriented contribution from the free energy $\log Z_{CS}^{SO/Sp}(K)$, one can write the contribution from the unoriented sector 
invoving the  reformulated polynomials $g_R(K;\lambda,q)$ 
\be
\log Z_{CS}^{SO/Sp}(K;\lambda,q,v) - \half \log \left[\sum_{R,S} {\overline P}_{(R,S)}(K;\lambda,q) s_R(v) s_S(v)\right]=
\sum_{\wt R}\sum_{d:{\rm odd}} {1 \over d} g_{\wt R}(K;\lambda^d,q^d) s_{\wt R}(v^d)~,
\ee
where the variable $\lambda$ is substituted for the variable $a$ in the colored HOMFLY polynomials.
For representations up to two boxes, the  above equation  gives
\bea
g_{\yng(1)}(K;\lambda,q)&=&{\overline F}_{\yng(1)}(K;\lambda,q)-{\overline P}_{\yng(1)}(K;\lambda,q)~,\cr
g_{\yng(2)}(K;\lambda,q)&=&{\overline F}_{\yng(2)}(K;\lambda,q)-\tfrac12{\overline F}_{\yng(1)}(K;\lambda,q)^2 -{\overline P}_{\yng(2)} (K;\lambda,q)+{\overline P}_{\yng(1)}(K;\lambda,q)^2
-\tfrac12{\overline P}_{(\yng(1),\yng(1))}(K;\lambda,q)~, \cr
g_{\yng(1,1)}(K;\lambda,q)&=&{\overline F}_{\yng(1,1)}(K;\lambda,q)- \tfrac12{\overline F}_{\yng(1)}(K;\lambda,q)^2 -{\overline P}_{\yng(1,1)}(K;\lambda,q) +
{\overline P}_{\yng(1)}(K;\lambda,q)^2-\tfrac12{\overline P}_{(\yng(1),\yng(1))}(K;\lambda,q)~.\cr
&&
\eea
The reformulated invariants $g_R(K;\lambda,q)$  have been obtained for many framed torus knots/links \cite{Marino:2009mw,Paul:2010qu,Stevan:2010jh}. Nevertheless, no calculation has been implemented for the reformulated invariants of non-torus knots/links beyond a single box $\yng(1)$ because it is difficult at present to evaluate colored Kauffman polynomials and colored HOMFLY polynomials with composite representations. 

With the colored Kauffman polynomial \eqref{triple-Kauffman-fig8} of the figure-eight,
we can calculate the reformulated polynomials $g_{\yng(2)}({\bf 4_1};\lambda,q)$ and $g_{\yng(1,1)}({\bf 4_1};\lambda,q)$:
\bea
\label{g-twobox}
g_{\yng(2)}({\bf 4_1};\lambda,q)&=&\tfrac{1}{\lambda^6 q^{12}}(-1+\lambda) (1+\lambda) (\lambda-q) (\lambda+q) (-1+\lambda q) (1+\lambda q) \cr
&&\times(-1+\lambda q+\lambda^2 q^4-\lambda^3 q^5-q^6+\lambda q^7+q^8-2 \lambda^2 q^8+\lambda^4 q^8-\lambda^6 q^8+\lambda^3 q^9\cr
&&+2 \lambda^2 q^{10}-2 \lambda^4 q^{10}-\lambda^3 q^{11}+q^{12}-\lambda^2 q^{12}+2 \lambda^4 q^{12}-\lambda^6 q^{12}-\lambda^5 q^{13}+\lambda^6 q^{14}\cr
&&+\lambda^3 q^{15}-\lambda^4 q^{16}-\lambda^5 q^{19}+\lambda^6 q^{20})~,\cr
g_{\yng(1,1)}({\bf 4_1};\lambda,q)&=&\tfrac{1}{q^{12}\lambda ^6}(q-\lambda )(-1+\lambda )(1+\lambda ) (q+\lambda ) (-1+q \lambda ) (1+q \lambda )\cr
&&(-q^8 g-q^{12}+q^{14}+q^{20}+q^{13} \lambda +q^{19} \lambda +q^8 \lambda ^2-2 q^{10} \lambda ^2+2 q^{12} \lambda ^2\cr
&&-q^{16} \lambda ^2+q^5 \lambda ^3-q^9 \lambda ^3+q^{11} \lambda ^3-q^{15} \lambda ^3+q^4 \lambda ^4-2 q^8 \lambda ^4+2 q^{10} \lambda ^4\cr
&&-q^{12} \lambda ^4-q \lambda ^5-q^7 \lambda ^5-\lambda ^6-q^6 \lambda ^6+q^8 \lambda ^6+q^{12} \lambda ^6)~.
\eea
The above two reformulated polynomials agrees with the expected integrality structure  $g_R(K;\lambda,q) \propto \sum_{\beta,s}  N_{R,\beta,s} \lambda^\beta(K) q^s$. 

To see the connection with the configurations of the M2-branes explained in \S\ref{sec:string-theory-a}, one can define the partition $\mu$ from the winding number $w =(w_1, \cdots, w_h)$ as follows: the $i$-th entry of $\mu$ is the number
of $w_j$s that take the value $i$. Then, one can relate the winding number $w$ basis to  the representation $R$ basis by
\bea
\Tr_R(U)=\sum_\mu  \frac{\chi_R(C_\mu)}{z_{\mu}} \prod_{j=1}^{\infty} (\Tr ~U^j)^{\mu_j}~,
\eea
where $z_{\mu}=\prod_i i^{\mu_i} \mu_i! $. Note that  we denote by $\chi_R(C_\mu)$ the character of the symmetric group representation $R$ in the conjugacy class $C_{\mu}$  associated to  the partition $\mu$ such that $\sum_j j \mu_j= |R|$. Thus, to obtain BPS integer coefficients, we need to change the basis in such a way that 
\bea
{\hat h}_R(K;a,q)&=&\sum_S (M^{-1})_{RS}~h_S(K;a,q)~,\cr
{\hat g}_R(K;\lambda,q)&=&\sum_S (M^{-1})_{RS}~g_S(K;\lambda,q)~,
\eea
where the matrix $M_{RS}$ is an invertible matrix given by
\be
M_{RS} =\sum_{\mu}{1 \over z_{\mu}} \chi_R(C_ {\mu}) \chi_S(C_{\mu}) {\prod_{i=1}^{\ell ({\mu})} (q^i-q^{-i})^{\mu_i}\over {q-q^{-1}}}~.
\ee
Writing with the variable $z=q-q^{-1}$ instead of the variable $q$, one can extract BPS integer degeneracies $ \wh N^c_{R;g,\beta}(K)$ from the reformulated invariants ${\hat h}_R$ and ${\hat g}_R$ \cite{Marino:2009mw}
\bea\label{Marino-form}
\hat{h}_R(K;\lambda,z)&=&\sum_{g\geq 0}\sum_{ \beta  \in \mathbb Z}\left(\wh N^{c=0}_{R;g,\beta}(K)z^{2g-1}\lambda^\beta \right)~,\cr
 \hat{g}_R(K;\lambda,z)&=&\sum_{g\geq 0}\sum_{\beta  \in \mathbb Z}\left(\wh N^{c=1}_{R;g,\beta}(K)z^{2g}\lambda^\beta +\wh  N^{c=2}_{R;g,\beta}(K)z^{2g+1}\lambda^\beta \right)~.
\eea
Using  the inverse of the matrix $M_{RS}$, we can write
\bea
\hat{h}_{\yng(2)}({\bf 4_1};a,q)&=&\frac{q~h_{\yng(2)}({\bf 4_1};a,q)+q^{-1}h_{\yng(1,1)}({\bf 4_1};a,q)}{q^2-q^{-2}}\cr
&=&\frac{1}{a^6 q^7(q-q^{-1})}(-a^4+2 a^6-a^8-2 a^6 q^2+4 a^8 q^2-2 a^{10} q^2+q^4-4 a^2 q^4\cr
&&+4 a^4 q^4+2 a^8 q^4-6 a^{10} q^4+3 a^{12} q^4+2 q^6-2 a^2 q^6-5 a^6 q^6+6 a^8 q^6\cr
&&-2 a^{10} q^6+q^8-4 a^2 q^8+4 a^4 q^8+2 a^8 q^8-6 a^{10} q^8+3 a^{12} q^8-2 a^6 q^{10}\cr
&&+4 a^8 q^{10}-2 a^{10} q^{10}-a^4 q^{12}+2 a^6 q^{12}-a^8 q^{12})~,
\eea
\bea
\hat{h}_{\yng(1,1)}({\bf 4_1};a,q)&=&-\frac{q^{-1}h_{\yng(2)}({\bf 4_1};a,q)+q~h_{\yng(1,1)}({\bf 4_1};a,q)}{q^2-q^{-2}}\cr
&=&\frac{1}{a^6 q^7(q-q^{-1})}(a^4-2 a^6+a^8+2 a^2 q^2-4 a^4 q^2+2 a^6 q^2-3 q^4+6 a^2 q^4-2 a^4 q^4\cr
&&-4 a^8 q^4+4 a^{10} q^4-a^{12} q^4+2 a^2 q^6-6 a^4 q^6+5 a^6 q^6+2 a^{10} q^6-2 a^{12} q^6-3 q^8\cr
&&+6 a^2 q^8-2 a^4 q^8-4 a^8 q^8+4 a^{10} q^8-a^{12} q^8+2 a^2 q^{10}-4 a^4 q^{10}+2 a^6 q^{10}\cr
&&+a^4 q^{12}-2 a^6 q^{12}+a^8 q^{12})~,
\eea
and
\bea
\hat{g}_{\yng(2)}({\bf 4_1};\lambda,q)&=&\frac{q~g_{\yng(2)}({\bf 4_1};\lambda,q)+q^{-1}g_{\yng(1,1)}({\bf 4_1};\lambda,q)}{q^2-q^{-2}}\cr
&=&-\frac{1}{q^{11} \lambda ^6}(q-\lambda ) (-1+\lambda ) (1+\lambda ) (q+\lambda ) (-1+q \lambda ) (1+q \lambda )\cr
 &&\times(q^2+q^6-q^{12}-q^{16}-q^3 \lambda -q^7 \lambda -q^9 \lambda -q^{11} \lambda -q^{15} \lambda -q^6 \lambda ^2+q^8 \lambda ^2\cr
 &&-q^{10} \lambda ^2+q^{12} \lambda ^2+q^5 \lambda ^3+q^7 \lambda ^3+q^{11} \lambda ^3+q^{13} \lambda ^3+q^4 \lambda ^4-q^8 \lambda ^4+q^{10} \lambda ^4\cr
 &&-q^{14} \lambda ^4-q \lambda ^5-q^5 \lambda ^5-q^7 \lambda ^5-q^9 \lambda ^5-q^{11} \lambda ^5-q^{13} \lambda ^5-q^{17} \lambda ^5-\lambda ^6\cr
 &&-q^4 \lambda ^6-q^6 \lambda ^6+q^{12} \lambda ^6+q^{14} \lambda ^6+q^{18} \lambda ^6)~,\cr
\hat{g}_{\yng(1,1)}({\bf 4_1};\lambda,q)&=&\frac{q^{-1}g_{\yng(2)}({\bf 4_1};\lambda,q)+q~g_{\yng(1,1)}({\bf 4_1};\lambda,q)}{q^2-q^{-2}}\cr
&=&\frac{1}{q^{11} \lambda ^6}(q-\lambda ) (-1+\lambda ) (1+\lambda ) (q+\lambda ) (-1+q \lambda ) (1+q \lambda )\cr
&&\times(-1-q^4-q^6+q^{12}+q^{14}+q^{18}+q \lambda +q^5 \lambda +q^7 \lambda +q^9 \lambda +q^{11} \lambda +q^{13} \lambda \cr 
&&+q^{17} \lambda +q^4 \lambda ^2-q^8 \lambda ^2+q^{10} \lambda ^2-q^{14} \lambda ^2-q^5 \lambda ^3-q^7 \lambda ^3-q^{11} \lambda ^3-q^{13} \lambda ^3-q^6 \lambda ^4\cr
&&+q^8 \lambda ^4-q^{10} \lambda ^4+q^{12} \lambda ^4+q^3 \lambda ^5+q^7 \lambda ^5+q^9 \lambda ^5+q^{11} \lambda ^5+q^{15} \lambda ^5+q^2 \lambda ^6\cr
&&+q^6 \lambda ^6-q^{12} \lambda ^6-q^{16} \lambda ^6)~.
\eea
The reformulated invariants above are indeed subject to the form \eqref{Marino-form} and we tabulate 
the integral BPS degeneracies in the following tables.
\begin{table}[H]
\parbox{.5\linewidth}{
\begin{tabular}{|c|c c c c c c c |}
\hline
$g\setminus \beta$&-6&-4&-2&0&2&4&6\\
\hline
0& 4 & -10 & 6 & -5 & 16 & -18 & 6\\
1& 1 & -4 & -5& 10 & 9& -14 & 3\\
2& 0 & 0 & -6 & 10 & -2 & -2 & 0\\
3& 0 & 0 & -1 & 2 & -1 & 0 & 0\\
\hline
\end{tabular}
\centering
$\wh N^{c=0}_{\yng(2);g,\beta}({\bf 4_1})$
}
\parbox{.45\linewidth}{
\begin{tabular}{|c|c c c c c c c|}
\hline
$g\setminus \beta$&-6&-4&-2&0&2&4&6\\
\hline
0& 6 & -18 & -16 & -5 & 6 & -10 & 4\\
1& 3 & -14 & 9& 10 & -5& -4 & 1\\
2& 0 & -2 & -2 & 10 & -6 & 0 & 0\\
3& 0 & 0 & -1 & 2 & -1 & 0 & 0\\
\hline
\end{tabular}
\centering
$\wh N^{c=0}_{\yng(1,1);g,\beta}({\bf 4_1})$
}
\end{table}

\begin{table}[H]
\parbox{.45\linewidth}{
\begin{tabular}{|c|c c c c c c |}
\hline
$g\setminus \beta$&-5&-3&-1&1&3&5\\
\hline
0& 5& -19& 34& -38& 25& -7\\
1& 10& -40& 75& -99& 75& -21\\
2& 6& -29& 57& -98& 85& -21\\
3& 1& -9& 18& -47& 45& -8\\
4& 0& -1& 2& -11& 11& -1\\
5& 0& 0& 0& -1& 1& 0\\
\hline
\end{tabular}
\centering
$\wh N^{c=1}_{\yng(2);g,\beta}({\bf 4_1})$
}
\parbox{.45\linewidth}{
\begin{tabular}{|c|c c c c c c c|}
\hline
$g\setminus \beta$&-6&-4&-2&0&2&4&6\\
\hline
0& 10& -32& 40& -45& 65& -55& 17\\
1& 15& -56& 65& -75& 145& -130& 36\\
2& 17& -36& 38& -44& 128& -121& 28\\
3& 1& -10& 10& -11& 56& -55& 9\\
4& 0& -1& 1& -1& 12& -12& 1\\
5&0 &0 & 0& 0& 1&-1 &  0\\
\hline
\end{tabular}
\centering
$\wh  N^{c=2}_{\yng(2);g,\beta}({\bf 4_1})$
}
\end{table}

\begin{table}[H]
\parbox{.45\linewidth}{
\begin{tabular}{|c|c c c c c c |}
\hline
$g\setminus \beta$&-5&-3&-1&1&3&5\\
\hline
0&7& -25 & 38 & -34 & 19 & -5\\
1&21& -75 & 99 & -75 & 40 & -10\\
2&21& -85 & 98 & -57 & 29 & -6\\
3&8& -45& 47& -18 & 9 & -1\\
4&1& -11 &11 & -2 & 1 & 0\\
5&0& -1 & 1 & 0 & 0 & 0\\
\hline
\end{tabular}
\centering
$\wh N^{c=1}_{\yng(1,1);g,\beta}({\bf 4_1})$
}
\parbox{.45\linewidth}{
\begin{tabular}{|c|c c c c c c c|}
\hline
$g\setminus \beta$&-6&-4&-2&0&2&4&6\\
\hline
0& 17 & -55 & 65 & -45 & 40 & -32 & 10\\
1& 36 & -130 & 145& -75 & 65& -56 & 15\\
2& 28 & -121 & 128 & -44 & 38 & -36 & 7\\
3& 9 & -55 & 56 & -11 & 10 & -10 & 1\\
4& 1 & -12 & 12 & -1 & 1 & -1 & 0\\
5& 0 & -1 & 1 & 0& 0 & 0 & 0 \\
\hline
\end{tabular}
\centering
$\wh  N^{c=2}_{\yng(1,1);g,\beta}({\bf 4_1})$
}
\end{table}
To see the integrality structure for higher rank representations, it requires the explicit expressions of the colored HOMFLY polynomials carrying composite representations. We relegate it to future investigation.


\section{Tables and figures} \label{sec:tab-fig}
\begin{table}[H]
\begin{center}
\begin{tabular}{|c|p{12.8cm}|}
\hline 
\textbf{Knot}& $\wh{A}^{PSL(2,\bC)}({\bf 4_1};\hat{x},\hat{y};q)$ \tabularnewline
\hline 
\hline 
${\bf 4_1}$ &\tiny{
$q^{-56}\hat{x}^{10}(q-\hat{x}^2)(q^2+\hat{x}^2)^2(q^3+\hat{x}^2)(q^4+\hat{x}^2)(-1+q \hat{x}^2)(q^8-q^6 \hat{x}^2-q^3 \hat{x}^4-q^5 \hat{x}^4-q^2 \hat{x}^6+\hat{x}^8)(q^{30}-2 q^{26} \hat{x}^2-2 q^{28} \hat{x}^2-q^{21} \hat{x}^4+q^{22} \hat{x}^4+4 q^{24} \hat{x}^4+q^{26} \hat{x}^4-q^{27} \hat{x}^4+q^{17} \hat{x}^6-q^{18} \hat{x}^6+3 q^{19} \hat{x}^6-3 q^{20} \hat{x}^6-3 q^{22} \hat{x}^6+3 q^{23} \hat{x}^6-q^{24} \hat{x}^6+q^{25} \hat{x}^6+q^{14} \hat{x}^8-2 q^{15} \hat{x}^8+2 q^{16} \hat{x}^8-2 q^{17} \hat{x}^8+5 q^{18} \hat{x}^8-2 q^{19} \hat{x}^8+2 q^{20} \hat{x}^8-2 q^{21} \hat{x}^8+q^{22} \hat{x}^8-q^{10} \hat{x}^{10}+q^{11} \hat{x}^{10}-2 q^{12} \hat{x}^{10}+2 q^{13} \hat{x}^{10}-5 q^{14} \hat{x}^{10}+2 q^{15} \hat{x}^{10}-5 q^{16} \hat{x}^{10}+2 q^{17} \hat{x}^{10}-2 q^{18} \hat{x}^{10}+q^{19} \hat{x}^{10}-q^{20} \hat{x}^{10}+q^8 \hat{x}^{12}-2 q^9 \hat{x}^{12}+2 q^{10} \hat{x}^{12}-2 q^{11} \hat{x}^{12}+5 q^{12} \hat{x}^{12}-2 q^{13} \hat{x}^{12}+2 q^{14} \hat{x}^{12}-2 q^{15} \hat{x}^{12}+q^{16} \hat{x}^{12}+q^5 \hat{x}^{14}-q^6 \hat{x}^{14}+3 q^7 \hat{x}^{14}-3 q^8 \hat{x}^{14}-3 q^{10} \hat{x}^{14}+3 q^{11} \hat{x}^{14}-q^{12} \hat{x}^{14}+q^{13} \hat{x}^{14}-q^3 \hat{x}^{16}+q^4 \hat{x}^{16}+4 q^6 \hat{x}^{16}+q^8 \hat{x}^{16}-q^9 \hat{x}^{16}-2 q^2 \hat{x}^{18}-2 q^4 \hat{x}^{18}+\hat{x}^{20})\hat{y}^3$}
\\[3pt] \rule{0pt}{2mm}
&\tiny{
$-q^{-16}(\hat{x}^2-1) \hat{x}^2(1+\hat{x}^2)^2(q^2+\hat{x}^2)(-1+q \hat{x}^2)(1+q \hat{x}^2)(-q+q \hat{x}^2+\hat{x}^4+q^2 \hat{x}^4+q \hat{x}^6-q \hat{x}^8)(q^{12}-2 q^{10} \hat{x}^2+q^{11} \hat{x}^2-2 q^{12} \hat{x}^2-2 q^{13} \hat{x}^2-q^{14} \hat{x}^2-q^7 \hat{x}^4+q^8 \hat{x}^4-2 q^9 \hat{x}^4+4 q^{10} \hat{x}^4+2 q^{11} \hat{x}^4+q^{12} \hat{x}^4+2 q^{13} \hat{x}^4+3 q^{14} \hat{x}^4+q^{15} \hat{x}^4-q^{16} \hat{x}^4-q^{17} \hat{x}^4+q^5 \hat{x}^6-2 q^6 \hat{x}^6+4 q^7 \hat{x}^6-q^8 \hat{x}^6+3 q^9 \hat{x}^6-2 q^{11} \hat{x}^6-4 q^{12} \hat{x}^6+q^{14} \hat{x}^6-q^{15} \hat{x}^6+q^{16} \hat{x}^6+4 q^{17} \hat{x}^6+2 q^{18} \hat{x}^6+2 q^4 \hat{x}^8-3 q^5 \hat{x}^8+3 q^6 \hat{x}^8-2 q^7 \hat{x}^8-q^8 \hat{x}^8-4 q^9 \hat{x}^8-q^{11} \hat{x}^8+2 q^{12} \hat{x}^8+4 q^{13} \hat{x}^8-3 q^{14} \hat{x}^8-4 q^{15} \hat{x}^8-q^{17} \hat{x}^8-4 q^{18} \hat{x}^8+q^{20} \hat{x}^8+q^{21} \hat{x}^8-q^2 \hat{x}^{10}+2 q^3 \hat{x}^{10}-4 q^4 \hat{x}^{10}-3 q^6 \hat{x}^{10}+q^7 \hat{x}^{10}-3 q^8 \hat{x}^{10}+q^9 \hat{x}^{10}+q^{10} \hat{x}^{10}-3 q^{11} \hat{x}^{10}-3 q^{12} \hat{x}^{10}-6 q^{13} \hat{x}^{10}-5 q^{14} \hat{x}^{10}+3 q^{16} \hat{x}^{10}-5 q^{17} \hat{x}^{10}-q^{18} \hat{x}^{10}-q^{19} \hat{x}^{10}-q^{20} \hat{x}^{10}-3 q^{21} \hat{x}^{10}-q^{22} \hat{x}^{10}-q \hat{x}^{12}+2 q^2 \hat{x}^{12}-2 q^3 \hat{x}^{12}+q^4 \hat{x}^{12}-q^5 \hat{x}^{12}+5 q^6 \hat{x}^{12}-q^8 \hat{x}^{12}+2 q^9 \hat{x}^{12}+q^{11} \hat{x}^{12}+3 q^{12} \hat{x}^{12}+7 q^{13} \hat{x}^{12}+3 q^{14} \hat{x}^{12}+7 q^{15} \hat{x}^{12}-4 q^{16} \hat{x}^{12}+q^{17} \hat{x}^{12}+2 q^{18} \hat{x}^{12}+4 q^{19} \hat{x}^{12}-2 q^{20} \hat{x}^{12}+q^{22} \hat{x}^{12}-q^{24} \hat{x}^{12}+2 q \hat{x}^{14}-q^2 \hat{x}^{14}+2 q^3 \hat{x}^{14}+3 q^5 \hat{x}^{14}+q^6 \hat{x}^{14}-q^7 \hat{x}^{14}+3 q^8 \hat{x}^{14}+4 q^9 \hat{x}^{14}+5 q^{10} \hat{x}^{14}+5 q^{12} \hat{x}^{14}-4 q^{13} \hat{x}^{14}+5 q^{14} \hat{x}^{14}-3 q^{15} \hat{x}^{14}+5 q^{16} \hat{x}^{14}+3 q^{17} \hat{x}^{14}+7 q^{18} \hat{x}^{14}-2 q^{19} \hat{x}^{14}+q^{20} \hat{x}^{14}+3 q^{21} \hat{x}^{14}+3 q^{22} \hat{x}^{14}-q^{23} \hat{x}^{14}+q^{24} \hat{x}^{14}+q^{25} \hat{x}^{14}+\hat{x}^{16}-2 q \hat{x}^{16}+q^4 \hat{x}^{16}-q^5 \hat{x}^{16}-q^6 \hat{x}^{16}+q^7 \hat{x}^{16}+2 q^9 \hat{x}^{16}-6 q^{10} \hat{x}^{16}+3 q^{11} \hat{x}^{16}-3 q^{12} \hat{x}^{16}+6 q^{13} \hat{x}^{16}-4 q^{14} \hat{x}^{16}+3 q^{15} \hat{x}^{16}-3 q^{16} \hat{x}^{16}+3 q^{17} \hat{x}^{16}-5 q^{18} \hat{x}^{16}+q^{20} \hat{x}^{16}+3 q^{21} \hat{x}^{16}-3 q^{22} \hat{x}^{16}-q^{23} \hat{x}^{16}+2 q^{24} \hat{x}^{16}-\hat{x}^{18}-q \hat{x}^{18}+q^3 \hat{x}^{18}-3 q^4 \hat{x}^{18}-4 q^5 \hat{x}^{18}-4 q^7 \hat{x}^{18}-q^8 \hat{x}^{18}-8 q^9 \hat{x}^{18}+q^{10} \hat{x}^{18}-7 q^{11} \hat{x}^{18}-q^{12} \hat{x}^{18}-12 q^{13} \hat{x}^{18}-q^{14} \hat{x}^{18}-7 q^{15} \hat{x}^{18}+q^{16} \hat{x}^{18}-8 q^{17} \hat{x}^{18}-q^{18} \hat{x}^{18}-4 q^{19} \hat{x}^{18}-4 q^{21} \hat{x}^{18}-3 q^{22} \hat{x}^{18}+q^{23} \hat{x}^{18}-q^{25} \hat{x}^{18}-q^{26} \hat{x}^{18}+2 q^2 \hat{x}^{20}-q^3 \hat{x}^{20}-3 q^4 \hat{x}^{20}+3 q^5 \hat{x}^{20}+q^6 \hat{x}^{20}-5 q^8 \hat{x}^{20}+3 q^9 \hat{x}^{20}-3 q^{10} \hat{x}^{20}+3 q^{11} \hat{x}^{20}-4 q^{12} \hat{x}^{20}+6 q^{13} \hat{x}^{20}-3 q^{14} \hat{x}^{20}+3 q^{15} \hat{x}^{20}-6 q^{16} \hat{x}^{20}+2 q^{17} \hat{x}^{20}+q^{19} \hat{x}^{20}-q^{20} \hat{x}^{20}-q^{21} \hat{x}^{20}+q^{22} \hat{x}^{20}-2 q^{25} \hat{x}^{20}+q^{26} \hat{x}^{20}+q \hat{x}^{22}+q^2 \hat{x}^{22}-q^3 \hat{x}^{22}+3 q^4 \hat{x}^{22}+3 q^5 \hat{x}^{22}+q^6 \hat{x}^{22}-2 q^7 \hat{x}^{22}+7 q^8 \hat{x}^{22}+3 q^9 \hat{x}^{22}+5 q^{10} \hat{x}^{22}-3 q^{11} \hat{x}^{22}+5 q^{12} \hat{x}^{22}-4 q^{13} \hat{x}^{22}+5 q^{14} \hat{x}^{22}+5 q^{16} \hat{x}^{22}+4 q^{17} \hat{x}^{22}+3 q^{18} \hat{x}^{22}-q^{19} \hat{x}^{22}+q^{20} \hat{x}^{22}+3 q^{21} \hat{x}^{22}+2 q^{23} \hat{x}^{22}-q^{24} \hat{x}^{22}+2 q^{25} \hat{x}^{22}-q^2 \hat{x}^{24}+q^4 \hat{x}^{24}-2 q^6 \hat{x}^{24}+4 q^7 \hat{x}^{24}+2 q^8 \hat{x}^{24}+q^9 \hat{x}^{24}-4 q^{10} \hat{x}^{24}+7 q^{11} \hat{x}^{24}+3 q^{12} \hat{x}^{24}+7 q^{13} \hat{x}^{24}+3 q^{14} \hat{x}^{24}+q^{15} \hat{x}^{24}+2 q^{17} \hat{x}^{24}-q^{18} \hat{x}^{24}+5 q^{20} \hat{x}^{24}-q^{21} \hat{x}^{24}+q^{22} \hat{x}^{24}-2 q^{23} \hat{x}^{24}+2 q^{24} \hat{x}^{24}-q^{25} \hat{x}^{24}-q^4 \hat{x}^{26}-3 q^5 \hat{x}^{26}-q^6 \hat{x}^{26}-q^7 \hat{x}^{26}-q^8 \hat{x}^{26}-5 q^9 \hat{x}^{26}+3 q^{10} \hat{x}^{26}-5 q^{12} \hat{x}^{26}-6 q^{13} \hat{x}^{26}-3 q^{14} \hat{x}^{26}-3 q^{15} \hat{x}^{26}+q^{16} \hat{x}^{26}+q^{17} \hat{x}^{26}-3 q^{18} \hat{x}^{26}+q^{19} \hat{x}^{26}-3 q^{20} \hat{x}^{26}-4 q^{22} \hat{x}^{26}+2 q^{23} \hat{x}^{26}-q^{24} \hat{x}^{26}+q^5 \hat{x}^{28}+q^6 \hat{x}^{28}-4 q^8 \hat{x}^{28}-q^9 \hat{x}^{28}-4 q^{11} \hat{x}^{28}-3 q^{12} \hat{x}^{28}+4 q^{13} \hat{x}^{28}+2 q^{14} \hat{x}^{28}-q^{15} \hat{x}^{28}-4 q^{17} \hat{x}^{28}-q^{18} \hat{x}^{28}-2 q^{19} \hat{x}^{28}+3 q^{20} \hat{x}^{28}-3 q^{21} \hat{x}^{28}+2 q^{22} \hat{x}^{28}+2 q^8 \hat{x}^{30}+4 q^9 \hat{x}^{30}+q^{10} \hat{x}^{30}-q^{11} \hat{x}^{30}+q^{12} \hat{x}^{30}-4 q^{14} \hat{x}^{30}-2 q^{15} \hat{x}^{30}+3 q^{17} \hat{x}^{30}-q^{18} \hat{x}^{30}+4 q^{19} \hat{x}^{30}-2 q^{20} \hat{x}^{30}+q^{21} \hat{x}^{30}-q^9 \hat{x}^{32}-q^{10} \hat{x}^{32}+q^{11} \hat{x}^{32}+3 q^{12} \hat{x}^{32}+2 q^{13} \hat{x}^{32}+q^{14} \hat{x}^{32}+2 q^{15} \hat{x}^{32}+4 q^{16} \hat{x}^{32}-2 q^{17} \hat{x}^{32}+q^{18} \hat{x}^{32}-q^{19} \hat{x}^{32}-q^{12} \hat{x}^{34}-2 q^{13} \hat{x}^{34}-2 q^{14} \hat{x}^{34}+q^{15} \hat{x}^{34}-2 q^{16} \hat{x}^{34}+q^{14} \hat{x}^{36}) \hat{y}^2$}
\\[3pt] \rule{0pt}{2mm}
&\tiny{$+q^2 \hat{x}^2 (-1+q \hat{x})(1+q \hat{x})(-1+q \hat{x}^2)(1+q \hat{x}^2)(1+q^2 \hat{x}^2)^2 (1+q^4 \hat{x}^2) (1-q^2 \hat{x}^2-q^3 \hat{x}^4-q^5 \hat{x}^4-q^6 \hat{x}^6+q^8 \hat{x}^8)(-q+q \hat{x}^2+2 q^2 \hat{x}^2+2 q^3 \hat{x}^2-q^4 \hat{x}^2+2 q^5 \hat{x}^2+\hat{x}^4+q \hat{x}^4-q^2 \hat{x}^4-3 q^3 \hat{x}^4-2 q^4 \hat{x}^4-q^5 \hat{x}^4-2 q^6 \hat{x}^4-4 q^7 \hat{x}^4+2 q^8 \hat{x}^4-q^9 \hat{x}^4+q^{10} \hat{x}^4-2 q \hat{x}^6-4 q^2 \hat{x}^6-q^3 \hat{x}^6+q^4 \hat{x}^6-q^5 \hat{x}^6+4 q^7 \hat{x}^6+2 q^8 \hat{x}^6-3 q^{10} \hat{x}^6+q^{11} \hat{x}^6-4 q^{12} \hat{x}^6+2 q^{13} \hat{x}^6-q^{14} \hat{x}^6-\hat{x}^8-q \hat{x}^8+4 q^3 \hat{x}^8+q^4 \hat{x}^8+4 q^6 \hat{x}^8+3 q^7 \hat{x}^8-4 q^8 \hat{x}^8-2 q^9 \hat{x}^8+q^{10} \hat{x}^8+4 q^{12} \hat{x}^8+q^{13} \hat{x}^8+2 q^{14} \hat{x}^8-3 q^{15} \hat{x}^8+3 q^{16} \hat{x}^8-2 q^{17} \hat{x}^8+q \hat{x}^{10}+3 q^2 \hat{x}^{10}+q^3 \hat{x}^{10}+q^4 \hat{x}^{10}+q^5 \hat{x}^{10}+5 q^6 \hat{x}^{10}-3 q^7 \hat{x}^{10}+5 q^9 \hat{x}^{10}+6 q^{10} \hat{x}^{10}+3 q^{11} \hat{x}^{10}+3 q^{12} \hat{x}^{10}-q^{13} \hat{x}^{10}-q^{14} \hat{x}^{10}+3 q^{15} \hat{x}^{10}-q^{16} \hat{x}^{10}+3 q^{17} \hat{x}^{10}+4 q^{19} \hat{x}^{10}-2 q^{20} \hat{x}^{10}+q^{21} \hat{x}^{10}+q \hat{x}^{12}-q^3 \hat{x}^{12}+2 q^5 \hat{x}^{12}-4 q^6 \hat{x}^{12}-2 q^7 \hat{x}^{12}-q^8 \hat{x}^{12}+4 q^9 \hat{x}^{12}-7 q^{10} \hat{x}^{12}-3 q^{11} \hat{x}^{12}-7 q^{12} \hat{x}^{12}-3 q^{13} \hat{x}^{12}-q^{14} \hat{x}^{12}-2 q^{16} \hat{x}^{12}+q^{17} \hat{x}^{12}-5 q^{19} \hat{x}^{12}+q^{20} \hat{x}^{12}-q^{21} \hat{x}^{12}+2 q^{22} \hat{x}^{12}-2 q^{23} \hat{x}^{12}+q^{24} \hat{x}^{12}-q^2 \hat{x}^{14}-q^3 \hat{x}^{14}+q^4 \hat{x}^{14}-3 q^5 \hat{x}^{14}-3 q^6 \hat{x}^{14}-q^7 \hat{x}^{14}+2 q^8 \hat{x}^{14}-7 q^9 \hat{x}^{14}-3 q^{10} \hat{x}^{14}-5 q^{11} \hat{x}^{14}+3 q^{12} \hat{x}^{14}-5 q^{13} \hat{x}^{14}+4 q^{14} \hat{x}^{14}-5 q^{15} \hat{x}^{14}-5 q^{17} \hat{x}^{14}-4 q^{18} \hat{x}^{14}-3 q^{19} \hat{x}^{14}+q^{20} \hat{x}^{14}-q^{21} \hat{x}^{14}-3 q^{22} \hat{x}^{14}-2 q^{24} \hat{x}^{14}+q^{25} \hat{x}^{14}-2 q^{26} \hat{x}^{14}-2 q^5 \hat{x}^{16}+q^6 \hat{x}^{16}+3 q^7 \hat{x}^{16}-3 q^8 \hat{x}^{16}-q^9 \hat{x}^{16}+5 q^{11} \hat{x}^{16}-3 q^{12} \hat{x}^{16}+3 q^{13} \hat{x}^{16}-3 q^{14} \hat{x}^{16}+4 q^{15} \hat{x}^{16}-6 q^{16} \hat{x}^{16}+3 q^{17} \hat{x}^{16}-3 q^{18} \hat{x}^{16}+6 q^{19} \hat{x}^{16}-2 q^{20} \hat{x}^{16}-q^{22} \hat{x}^{16}+q^{23} \hat{x}^{16}+q^{24} \hat{x}^{16}-q^{25} \hat{x}^{16}+2 q^{28} \hat{x}^{16}-q^{29} \hat{x}^{16}+q^5 \hat{x}^{18}+q^6 \hat{x}^{18}-q^8 \hat{x}^{18}+3 q^9 \hat{x}^{18}+4 q^{10} \hat{x}^{18}+4 q^{12} \hat{x}^{18}+q^{13} \hat{x}^{18}+8 q^{14} \hat{x}^{18}-q^{15} \hat{x}^{18}+7 q^{16} \hat{x}^{18}+q^{17} \hat{x}^{18}+12 q^{18} \hat{x}^{18}+q^{19} \hat{x}^{18}+7 q^{20} \hat{x}^{18}-q^{21} \hat{x}^{18}+8 q^{22} \hat{x}^{18}+q^{23} \hat{x}^{18}+4 q^{24} \hat{x}^{18}+4 q^{26} \hat{x}^{18}+3 q^{27} \hat{x}^{18}-q^{28} \hat{x}^{18}+q^{30} \hat{x}^{18}+q^{31} \hat{x}^{18}-q^7 \hat{x}^{20}+2 q^8 \hat{x}^{20}-q^{11} \hat{x}^{20}+q^{12} \hat{x}^{20}+q^{13} \hat{x}^{20}-q^{14} \hat{x}^{20}-2 q^{16} \hat{x}^{20}+6 q^{17} \hat{x}^{20}-3 q^{18} \hat{x}^{20}+3 q^{19} \hat{x}^{20}-6 q^{20} \hat{x}^{20}+4 q^{21} \hat{x}^{20}-3 q^{22} \hat{x}^{20}+3 q^{23} \hat{x}^{20}-3 q^{24} \hat{x}^{20}+5 q^{25} \hat{x}^{20}-q^{27} \hat{x}^{20}-3 q^{28} \hat{x}^{20}+3 q^{29} \hat{x}^{20}+q^{30} \hat{x}^{20}-2 q^{31} \hat{x}^{20}-2 q^{10} \hat{x}^{22}+q^{11} \hat{x}^{22}-2 q^{12} \hat{x}^{22}-3 q^{14} \hat{x}^{22}-q^{15} \hat{x}^{22}+q^{16} \hat{x}^{22}-3 q^{17} \hat{x}^{22}-4 q^{18} \hat{x}^{22}-5 q^{19} \hat{x}^{22}-5 q^{21} \hat{x}^{22}+4 q^{22} \hat{x}^{22}-5 q^{23} \hat{x}^{22}+3 q^{24} \hat{x}^{22}-5 q^{25} \hat{x}^{22}-3 q^{26} \hat{x}^{22}-7 q^{27} \hat{x}^{22}+2 q^{28} \hat{x}^{22}-q^{29} \hat{x}^{22}-3 q^{30} \hat{x}^{22}-3 q^{31} \hat{x}^{22}+q^{32} \hat{x}^{22}-q^{33} \hat{x}^{22}-q^{34} \hat{x}^{22}+q^{12} \hat{x}^{24}-2 q^{13} \hat{x}^{24}+2 q^{14} \hat{x}^{24}-q^{15} \hat{x}^{24}+q^{16} \hat{x}^{24}-5 q^{17} \hat{x}^{24}+q^{19} \hat{x}^{24}-2 q^{20} \hat{x}^{24}-q^{22} \hat{x}^{24}-3 q^{23} \hat{x}^{24}-7 q^{24} \hat{x}^{24}-3 q^{25} \hat{x}^{24}-7 q^{26} \hat{x}^{24}+4 q^{27} \hat{x}^{24}-q^{28} \hat{x}^{24}-2 q^{29} \hat{x}^{24}-4 q^{30} \hat{x}^{24}+2 q^{31} \hat{x}^{24}-q^{33} \hat{x}^{24}+q^{35} \hat{x}^{24}+q^{15} \hat{x}^{26}-2 q^{16} \hat{x}^{26}+4 q^{17} \hat{x}^{26}+3 q^{19} \hat{x}^{26}-q^{20} \hat{x}^{26}+3 q^{21} \hat{x}^{26}-q^{22} \hat{x}^{26}-q^{23} \hat{x}^{26}+3 q^{24} \hat{x}^{26}+3 q^{25} \hat{x}^{26}+6 q^{26} \hat{x}^{26}+5 q^{27} \hat{x}^{26}-3 q^{29} \hat{x}^{26}+5 q^{30} \hat{x}^{26}+q^{31} \hat{x}^{26}+q^{32} \hat{x}^{26}+q^{33} \hat{x}^{26}+3 q^{34} \hat{x}^{26}+q^{35} \hat{x}^{26}-2 q^{19} \hat{x}^{28}+3 q^{20} \hat{x}^{28}-3 q^{21} \hat{x}^{28}+2 q^{22} \hat{x}^{28}+q^{23} \hat{x}^{28}+4 q^{24} \hat{x}^{28}+q^{26} \hat{x}^{28}-2 q^{27} \hat{x}^{28}-4 q^{28} \hat{x}^{28}+3 q^{29} \hat{x}^{28}+4 q^{30} \hat{x}^{28}+q^{32} \hat{x}^{28}+4 q^{33} \hat{x}^{28}-q^{35} \hat{x}^{28}-q^{36} \hat{x}^{28}-q^{22} \hat{x}^{30}+2 q^{23} \hat{x}^{30}-4 q^{24} \hat{x}^{30}+q^{25} \hat{x}^{30}-3 q^{26} \hat{x}^{30}+2 q^{28} \hat{x}^{30}+4 q^{29} \hat{x}^{30}-q^{31} \hat{x}^{30}+q^{32} \hat{x}^{30}-q^{33} \hat{x}^{30}-4 q^{34} \hat{x}^{30}-2 q^{35} \hat{x}^{30}+q^{26} \hat{x}^{32}-q^{27} \hat{x}^{32}+2 q^{28} \hat{x}^{32}-4 q^{29} \hat{x}^{32}-2 q^{30} \hat{x}^{32}-q^{31} \hat{x}^{32}-2 q^{32} \hat{x}^{32}-3 q^{33} \hat{x}^{32}-q^{34} \hat{x}^{32}+q^{35} \hat{x}^{32}+q^{36} \hat{x}^{32}+2 q^{31} \hat{x}^{34}-q^{32} \hat{x}^{34}+2 q^{33} \hat{x}^{34}+2 q^{34} \hat{x}^{34}+q^{35} \hat{x}^{34}-q^{35} \hat{x}^{36}) \hat{y}$}
\\[3pt] \rule{0pt}{2mm}
&\tiny{$+q^{15} \hat{x}^{10}(-1+q^2 \hat{x})(1+q^2 \hat{x})(-1+q \hat{x}^2)(1+q^4 \hat{x}^2)^2 (1+q^5 \hat{x}^2)(1+q^6 \hat{x}^2)(1-q^4 \hat{x}^2-q^7 \hat{x}^4-q^9 \hat{x}^4-q^{12} \hat{x}^6+q^{16} \hat{x}^8)(1-2 q^4 \hat{x}^2-2 q^6 \hat{x}^2-q^7 \hat{x}^4+q^8 \hat{x}^4+4 q^{10} \hat{x}^4+q^{12} \hat{x}^4-q^{13} \hat{x}^4+q^{11} \hat{x}^6-q^{12} \hat{x}^6+3 q^{13} \hat{x}^6-3 q^{14} \hat{x}^6-3 q^{16} \hat{x}^6+3 q^{17} \hat{x}^6-q^{18} \hat{x}^6+q^{19} \hat{x}^6+q^{16} \hat{x}^8-2 q^{17} \hat{x}^8+2 q^{18} \hat{x}^8-2 q^{19} \hat{x}^8+5 q^{20} \hat{x}^8-2 q^{21} \hat{x}^8+2 q^{22} \hat{x}^8-2 q^{23} \hat{x}^8+q^{24} \hat{x}^8-q^{20} \hat{x}^{10}+q^{21} \hat{x}^{10}-2 q^{22} \hat{x}^{10}+2 q^{23} \hat{x}^{10}-5 q^{24} \hat{x}^{10}+2 q^{25} \hat{x}^{10}-5 q^{26} \hat{x}^{10}+2 q^{27} \hat{x}^{10}-2 q^{28} \hat{x}^{10}+q^{29} \hat{x}^{10}-q^{30} \hat{x}^{10}+q^{26} \hat{x}^{12}-2 q^{27} \hat{x}^{12}+2 q^{28} \hat{x}^{12}-2 q^{29} \hat{x}^{12}+5 q^{30} \hat{x}^{12}-2 q^{31} \hat{x}^{12}+2 q^{32} \hat{x}^{12}-2 q^{33} \hat{x}^{12}+q^{34} \hat{x}^{12}+q^{31} \hat{x}^{14}-q^{32} \hat{x}^{14}+3 q^{33} \hat{x}^{14}-3 q^{34} \hat{x}^{14}-3 q^{36} \hat{x}^{14}+3 q^{37} \hat{x}^{14}-q^{38} \hat{x}^{14}+q^{39} \hat{x}^{14}-q^{37} \hat{x}^{16}+q^{38} \hat{x}^{16}+4 q^{40} \hat{x}^{16}+q^{42} \hat{x}^{16}-q^{43} \hat{x}^{16}-2 q^{44} \hat{x}^{18}-2 q^{46} \hat{x}^{18}+q^{50} \hat{x}^{20})+q^2 \hat{x}^2(-1+q \hat{x})(1+q \hat{x})(-1+q \hat{x}^2)(1+q \hat{x}^2)(1+q^2 \hat{x}^2)^2 (1+q^4 \hat{x}^2) (1-q^2 \hat{x}^2-q^3 \hat{x}^4-q^5 \hat{x}^4-q^6 \hat{x}^6+q^8 \hat{x}^8)$}
\tabularnewline
\hline 
 \end{tabular}
\caption{Quantum $PSL(2,\bC)$ character variety of the figure-eight}
\label{tab:quantum-a-fig8}
\end{center}
\end{table}

\begin{landscape}
\begin{figure}
 \centering
    \includegraphics[width=22cm]{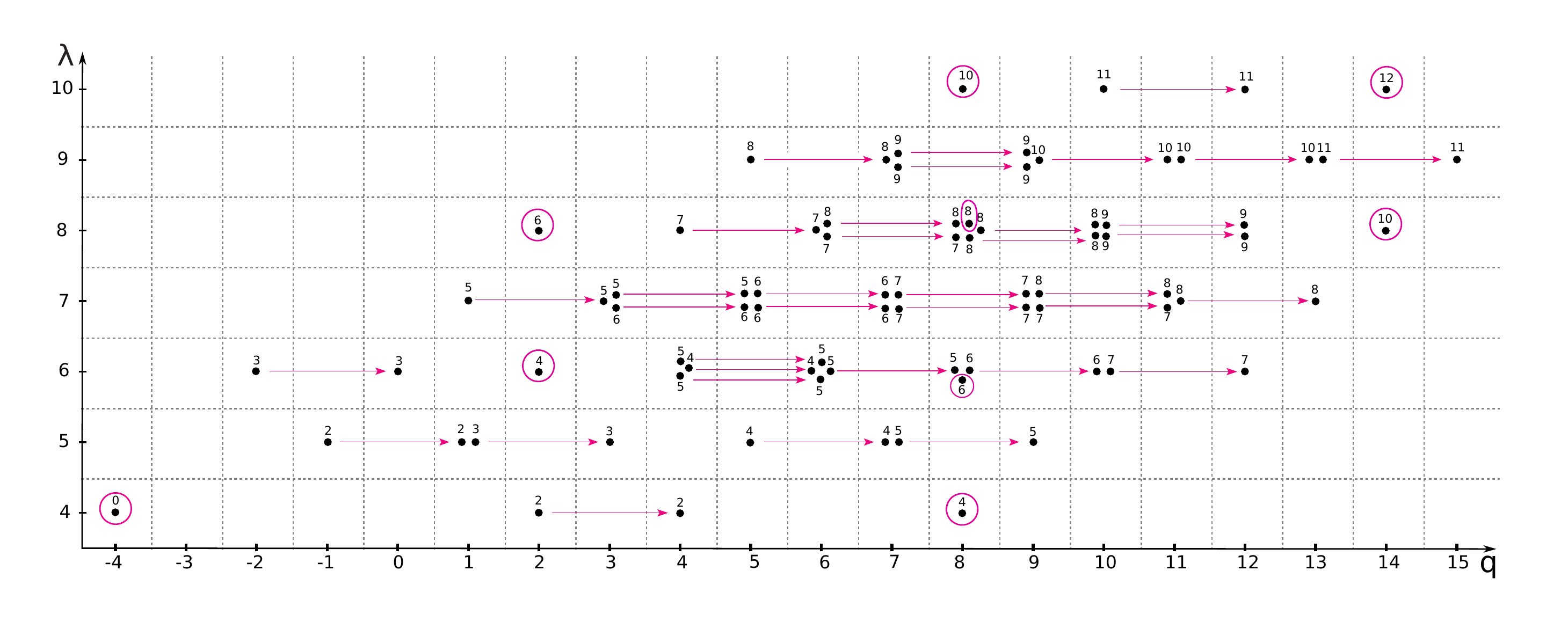}
    \caption{The universal colored differential $d^\leftarrow$  acts on the $[2]$-colored Kauffman homology of the trefoil. The dots enclosed by red circles represent the elements in $H_*(\scH^{\Kauffman}_{[2]},d^\leftarrow)$.}\label{fig:Kauffman-trefoil-2to1}
\end{figure}
\begin{footnotesize}
\bea\label{triple-Kauffman-trefoil-2}
\scF_{\yng(2)}({\bf 3_1};\lambda,q,t,1)&=&\lambda ^4\left(\tfrac{1}{q^4}+q^2 t^2+q^4 t^2+q^8 t^4\right) +\lambda^5\left(\tfrac{t^2}{q}+q t^2+q t^3+q^3 t^3+q^5 t^4+q^9 t^5+q^7 \left(t^4+t^5\right)\right) \cr
&&+ \lambda ^6\left(t^3+\tfrac{t^3}{q^2}+q^2 t^4+q^{12} t^7+q^4 \left(t^4+2 t^5\right)+q^6 \left(t^4+3 t^5\right)+q^8 \left(t^5+2 t^6\right)+q^{10} \left(t^6+t^7\right)\right)\cr
&&+ \lambda ^7\left(q t^5+q^{13} t^8+q^3 \left(2 t^5+t^6\right)+q^5 \left(t^5+3 t^6\right)+2 q^7 \left(t^6+t^7\right)+q^9 \left(3 t^7+t^8\right)+q^{11} \left(t^7+2 t^8\right)\right)\cr
&&+ \lambda ^8\left(q^2 t^6+q^4 t^7+2 q^{12} t^9+q^{14} t^{10}+q^6 \left(2 t^7+t^8\right)+q^8 \left(t^7+4 t^8\right)+q^{10} \left(2 t^8+2 t^9\right)\right)\cr
&&+\lambda ^9\left(q^5 t^8+2 q^9 t^9+q^9 t^{10}+2 q^{11} t^{10}+q^{13} t^{10}+q^{13} t^{11}+q^{15} t^{11}+q^7 \left(t^8+2 t^9\right)\right) +\lambda ^{10}\left(q^8 t^{10}+q^{10} t^{11}+q^{12} t^{11}+q^{14} t^{12}\right) ~.\\
\scF(\scH^{\Kauffman}_{[2]},d^\leftarrow)&=&\lambda^4(q^{-4}+q^{8}t^4)+\lambda^6(q^2t^4+q^8t^6)+\lambda^8(q^2t^6+q^8t^8+q^{14}t^{10})+\lambda^{10}(q^8t^{10}+q^{14}t^{12})=\scF_{[1]}(\lambda^2 ,q^2,t^2 q^2)
\eea
\end{footnotesize}
\end{landscape}

\begin{landscape}
\begin{figure}
 \centering
    \includegraphics[width=23cm]{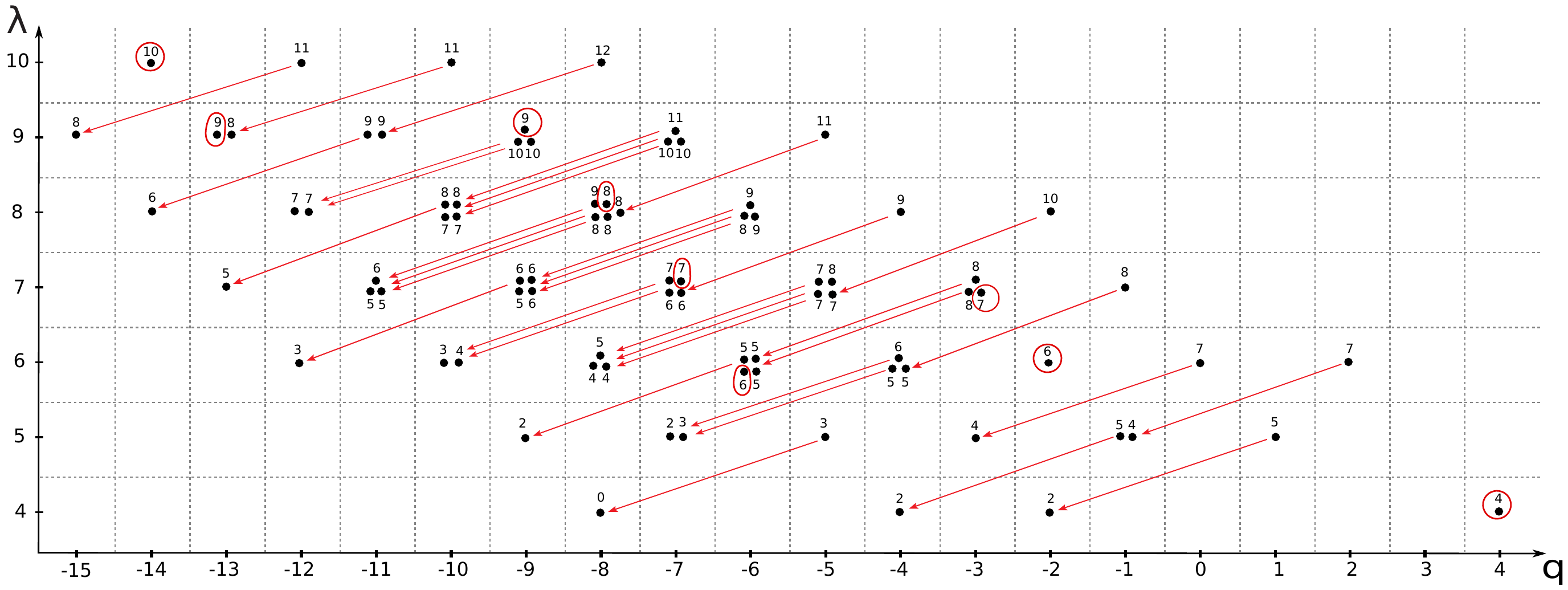}
    \caption{The $[1^2]$-colored Kauffman homology of the trefoil. The dots enclosed by red circles represent the elements in $H_*(\scH^{\Kauffman}_{[1^2]},d_{-2})$.}\label{fig:Kauffman-trefoil-11}
\end{figure}
\begin{tiny}
\bea
\scF_{\yng(1,1)}({\bf 3_1};\lambda,q,t_r=1,t_c=t)&=&\lambda ^4\left(\tfrac{1}{q^8}+\tfrac{t^2}{q^4}+\tfrac{t^2}{q^2}+q^4 t^4\right) + \lambda ^5 \left(\tfrac{t^2}{q^9}+\tfrac{t^2}{q^7}+\tfrac{t^3}{q^7}+\tfrac{t^3}{q^5}+\tfrac{t^4}{q^3}+\tfrac{t^4}{q}+\tfrac{t^5}{q}+q t^5\right) +\lambda ^6\left(\tfrac{t^3}{q^{12}}+\tfrac{t^3}{q^{10}}+\tfrac{t^4}{q^{10}}+\tfrac{2 t^4}{q^8}+\tfrac{t^5}{q^8}+\tfrac{3 t^5}{q^6}+\tfrac{2 t^5}{q^4}+\tfrac{t^6}{q^6}+\tfrac{t^6}{q^4}+\tfrac{t^6}{q^2}+t^7+q^2 t^7\right) \cr
&&+ \lambda ^7\left(\tfrac{t^5}{q^{13}}+\tfrac{2 t^5}{q^{11}}+\tfrac{t^5}{q^9}+\tfrac{t^6}{q^{11}}+\tfrac{3 t^6}{q^9}+\tfrac{2 t^6}{q^7}+\tfrac{2 t^7}{q^7}+\tfrac{3 t^7}{q^5}+\tfrac{t^7}{q^3}+\tfrac{t^8}{q^5}+\tfrac{2 t^8}{q^3}+\tfrac{t^8}{q}\right)+ \lambda ^8 \left(\tfrac{t^6}{q^{14}}+\tfrac{2 t^7}{q^{12}}+\tfrac{2 t^7}{q^{10}}+\tfrac{2 t^8}{q^{10}}+\tfrac{4 t^8}{q^8}+\tfrac{t^8}{q^6}+\tfrac{t^9}{q^8}+\tfrac{2 t^9}{q^6}+\tfrac{t^9}{q^4}+\tfrac{t^{10}}{q^2}\right)\cr
&&+ \lambda ^9\left(\tfrac{t^8}{q^{15}}+\tfrac{t^8}{q^{13}}+\tfrac{t^9}{q^{13}}+\tfrac{2 t^9}{q^{11}}+\tfrac{t^9}{q^9}+\tfrac{2 t^{10}}{q^9}+\tfrac{2 t^{10}}{q^7}+\tfrac{t^{11}}{q^7}+\tfrac{t^{11}}{q^5}\right)+\lambda ^{10}\left(\tfrac{t^{10}}{q^{14}}+\tfrac{t^{11}}{q^{12}}+\tfrac{t^{11}}{q^{10}}+\tfrac{t^{12}}{q^8}\right) ~.
\eea
\end{tiny}
\bea\label{triple-Kauffman-trefoil-sp2}
\scF(\scH^{\Kauffman}_{[1^2]},d_{-2})(\lambda,q,1,t)=\lambda^4q^4t^4+\lambda^6(q^{-6}+q^{-2})t^6+\lambda^7(q^{-7}+q^{-3})t^7+\lambda^8q^{-8}t^8+\lambda^9(q^{-13}+q^{-9})t^9+\lambda^{10}q^{-14}t^{10}~.
\eea
\end{landscape}

\begin{figure}
 \centering
    \includegraphics[width=15cm]{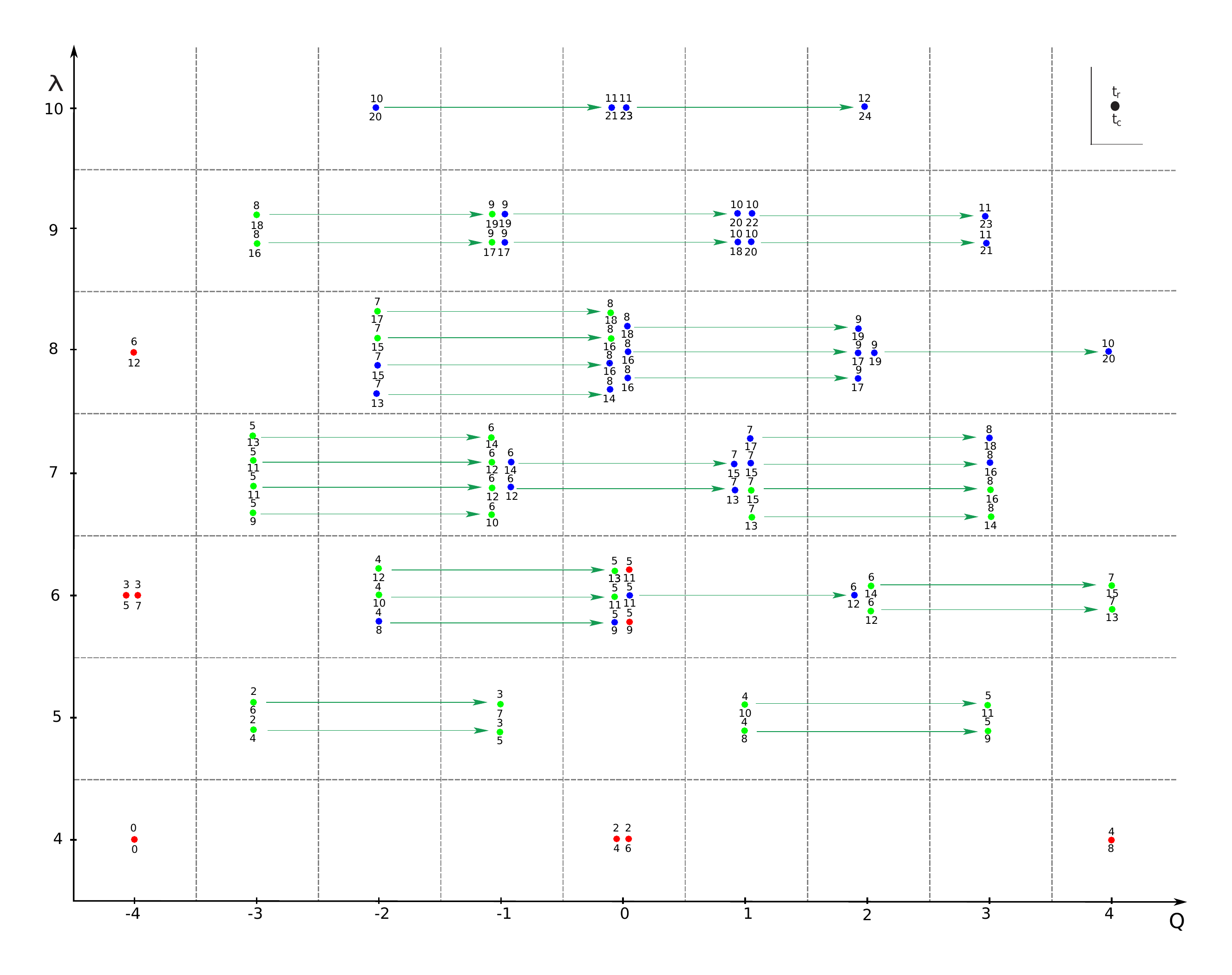}
    \caption{The universal differential $d^{\rm univ}_{\to}$ acts on the $[2]$-colored quadruply-graded Kauffman homology of the trefoil. The red dots represent the elements in $H_*(\wt\scH^\Kauffman_{[2]}({\bf 3_1}),d^{\rm univ}_{\to})$. The red, green and blue dots consist of the terms corresponding to $k=2,1$ and $0$ in the first summation of \eqref{quad-Kauffman-trefoil} respectively.}\label{fig:quad-Kauffman-trefoil-2-universal}
\end{figure}
\begin{footnotesize}
\bea\label{quad-Kauffman-trefoil-2-color}
&&\wt\scF_{\yng(2)}({\bf 3_1};\lambda,Q,t_r,t_c)\cr
&=&{\color{red}\lambda^4 \left(\tfrac{1}{Q^4}+t_c^4 t_r^2+t_c^6 t_r^2+Q^4 t_c^8 t_r^4\right)+\lambda^6 \left(\tfrac{t_c^5 t_r^3}{Q^4}+\tfrac{t_c^7 t_r^3}{Q^4}+t_c^9 t_r^5+t_c^{11} t_r^5\right)+\tfrac{\lambda^8 t_c^{12} t_r^6}{Q^4}}\cr
&&{\color{forestgreen}+\lambda^5 \left(\tfrac{t_c^4 t_r^2}{Q^3}+\tfrac{t_c^6 t_r^2}{Q^3}+\tfrac{t_c^5 t_r^3}{Q}+\tfrac{t_c^7 t_r^3}{Q}+Q t_c^8 t_r^4+Q t_c^{10} t_r^4+Q^3 t_c^9 t_r^5+Q^3 t_c^{11} t_r^5\right)}\cr
&& {\color{forestgreen}+\lambda^6 \left(\tfrac{t_c^{10} t_r^4}{Q^2}+\tfrac{t_c^{12} t_r^4}{Q^2}+t_c^{11} t_r^5+t_c^{13} t_r^5+Q^2 t_c^{12} t_r^6+Q^2 t_c^{14} t_r^6+Q^4 t_c^{13} t_r^7+Q^4 t_c^{15} t_r^7\right)}\cr
&&{\color{forestgreen}+\lambda^7 \left(\tfrac{t_c^9 t_r^5}{Q^3}+\tfrac{2 t_c^{11} t_r^5}{Q^3}+\tfrac{t_c^{13} t_r^5}{Q^3}+\tfrac{t_c^{10} t_r^6}{Q}+\tfrac{2 t_c^{12} t_r^6}{Q}+\tfrac{t_c^{14} t_r^6}{Q}+Q t_c^{13} t_r^7+Q t_c^{15} t_r^7+Q^3 t_c^{14} t_r^8+Q^3 t_c^{16} t_r^8\right)}\cr
&&{\color{forestgreen}+\lambda^8 \left(\tfrac{t_c^{15} t_r^7}{Q^2}+\tfrac{t_c^{17} t_r^7}{Q^2}+t_c^{16} t_r^8+t_c^{18} t_r^8\right)+\lambda^9 \left(\tfrac{t_c^{16} t_r^8}{Q^3}+\tfrac{t_c^{18} t_r^8}{Q^3}+\tfrac{t_c^{17} t_r^9}{Q}+\tfrac{t_c^{19} t_r^9}{Q}\right)}\cr
&&{\color{blue}+\lambda^6 \left(\tfrac{t_c^8 t_r^4}{Q^2}+t_c^9 t_r^5+t_c^{11} t_r^5+Q^2 t_c^{12} t_r^6\right)+\lambda^{10} \left(\tfrac{t_c^{20} t_r^{10}}{Q^2}+t_c^{21} t_r^{11}+t_c^{23} t_r^{11}+Q^2 t_c^{24} t_r^{12}\right)}\cr
&&{\color{blue}+\lambda^7 \left(\tfrac{t_c^{12} t_r^6}{Q}+\tfrac{t_c^{14} t_r^6}{Q}+Q t_c^{13} t_r^7+2 Q t_c^{15} t_r^7+Q t_c^{17} t_r^7+Q^3 t_c^{16} t_r^8+Q^3 t_c^{18} t_r^8\right)}\cr
&&{\color{blue}+\lambda^8 \left(\tfrac{t_c^{13} t_r^7}{Q^2}+\tfrac{t_c^{15} t_r^7}{Q^2}+t_c^{14} t_r^8+3 t_c^{16} t_r^8+t_c^{18} t_r^8+2 Q^2 t_c^{17} t_r^9+2 Q^2 t_c^{19} t_r^9+Q^4 t_c^{20} t_r^{10}\right)}\cr
&&{\color{blue}+\lambda^9 \left(\tfrac{t_c^{17} t_r^9}{Q}+\tfrac{t_c^{19} t_r^9}{Q}+Q t_c^{18} t_r^{10}+2 Q t_c^{20} t_r^{10}+Q t_c^{22} t_r^{10}+Q^3 t_c^{21} t_r^{11}+Q^3 t_c^{23} t_r^{11}\right)}
\eea
\end{footnotesize}
\clearpage

\begin{figure}
 \centering
    \includegraphics[width=15cm]{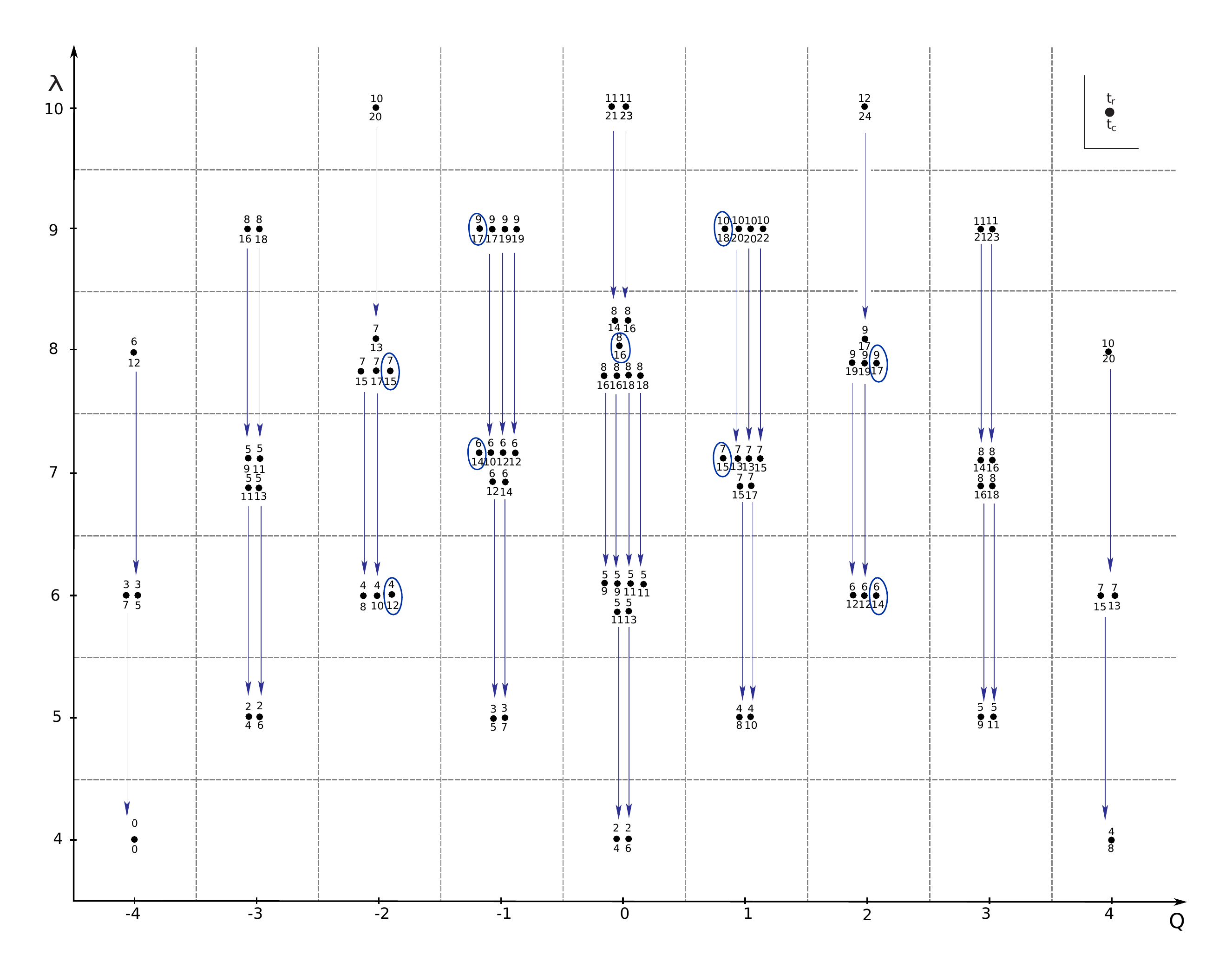}
    \caption{The colored differential $d_{[2]\to[1]}^-$ acts on the $[2]$-colored quadruply-graded Kauffman homology of the trefoil. The dots enclosed by blue circles represent the elements in $H_*(\wt\scH^{\Kauffman}_{[2]}({\bf 3_1}),d_{[2]\to[1]}^-)$. One can convince oneself that the homology $H_*(\wt\scH^{\Kauffman}_{[2]}({\bf 3_1}),d_{[2]\to[1]}^-)$ is isomorphic to the uncolored homology $\wt\scH^{\Kauffman}_{[1]}({\bf 3_1})$ shown in Figure \ref{fig:Kauffman-uncolor-trefoil}.}\label{fig:quad-Kauffman-trefoil-2to1-}
\end{figure}
\begin{footnotesize}
\bea\label{quad-Kauffman-trefoil-2}
&&\wt\scF_{[2]}({\bf 3_1};\lambda,Q,t_r,t_c)\cr
&=& \lambda ^4\left(\tfrac{1}{Q^4}+t_c^4 t_r^2+t_c^6 t_r^2+Q^4 t_c^8 t_r^4\right)\cr
&&+\lambda ^5\left(\tfrac{t_c^4 t_r^2}{Q^3}+\tfrac{t_c^6 t_r^2}{Q^3}+\tfrac{t_c^5 t_r^3}{Q}+\tfrac{t_c^7 t_r^3}{Q}+Q t_c^8 t_r^4+Q t_c^{10} t_r^4+Q^3 t_c^9 t_r^5+Q^3 t_c^{11} t_r^5\right) \cr
&&+\lambda ^6\left(\tfrac{t_c^5 t_r^3}{Q^4}+\tfrac{t_c^7 t_r^3}{Q^4}+\tfrac{t_c^8 t_r^4}{Q^2}+\tfrac{t_c^{10} t_r^4}{Q^2}+\tfrac{t_c^{12} t_r^4}{Q^2}+2 t_c^9 t_r^5+3 t_c^{11} t_r^5+t_c^{13} t_r^5+2 Q^2 t_c^{12} t_r^6+Q^2 t_c^{14} t_r^6+Q^4 t_c^{13} t_r^7+Q^4 t_c^{15} t_r^7\right) \cr
&&+ \lambda ^7\left(\tfrac{t_c^9 t_r^5}{Q^3}+\tfrac{2 t_c^{11} t_r^5}{Q^3}+\tfrac{t_c^{13} t_r^5}{Q^3}+\tfrac{t_c^{10} t_r^6}{Q}+\tfrac{3 t_c^{12} t_r^6}{Q}+\tfrac{2 t_c^{14} t_r^6}{Q}+2 Q t_c^{13} t_r^7+3 Q t_c^{15} t_r^7+Q t_c^{17} t_r^7+Q^3 t_c^{14} t_r^8+2 Q^3 t_c^{16} t_r^8+Q^3 t_c^{18} t_r^8\right)\cr
&&+ \lambda ^8\left(\tfrac{t_c^{12} t_r^6}{Q^4}+\tfrac{t_c^{13} t_r^7}{Q^2}+\tfrac{2 t_c^{15} t_r^7}{Q^2}+\tfrac{t_c^{17} t_r^7}{Q^2}+t_c^{14} t_r^8+4 t_c^{16} t_r^8+2 t_c^{18} t_r^8+2 Q^2 t_c^{17} t_r^9+2 Q^2 t_c^{19} t_r^9+Q^4 t_c^{20} t_r^{10}\right)\cr
&&+\lambda ^9\left(\tfrac{t_c^{16} t_r^8}{Q^3}+\tfrac{t_c^{18} t_r^8}{Q^3}+\tfrac{2 t_c^{17} t_r^9}{Q}+\tfrac{2 t_c^{19} t_r^9}{Q}+Q t_c^{18} t_r^{10}+2 Q t_c^{20} t_r^{10}+Q t_c^{22} t_r^{10}+Q^3 t_c^{21} t_r^{11}+Q^3 t_c^{23} t_r^{11}\right) \cr
&&+ \lambda ^{10} \left(\tfrac{t_c^{20} t_r^{10}}{Q^2}+t_c^{21} t_r^{11}+t_c^{23} t_r^{11}+Q^2 t_c^{24} t_r^{12}\right)
\eea
\bea
&&\wt\scF(\scH^\Kauffman_{[2]}({\bf 3_1}),d_{[2]\to[1]}^- )(\lambda,Q,t_r,t_c)=\lambda^4 t_r^4t_c^{12} \wt\scF_{[1]}({\bf 3_1};\lambda,Q,t_r,t_c)\cr
&=&\lambda^6(Q^{-2}  t_c^{12} t_r^4 + Q^2 t_c^{14} t_r^6)  + \lambda^7(
Q^{-1} t_c^{14} t_r^6  + Q t_c^{15} t_r^7 ) +\lambda^8 (Q^{-2} t_c^{15} t_r^7  + t_c^{16} t_r^8  +  Q^2 t_c^{17} t_r^9) \cr
 && +\lambda^9 (Q^{-1}t_c^{17} t_r^9  +  Q t_c^{18} t_r^{10} )
\eea
\end{footnotesize}

\begin{figure}
 \centering
    \includegraphics[width=15cm]{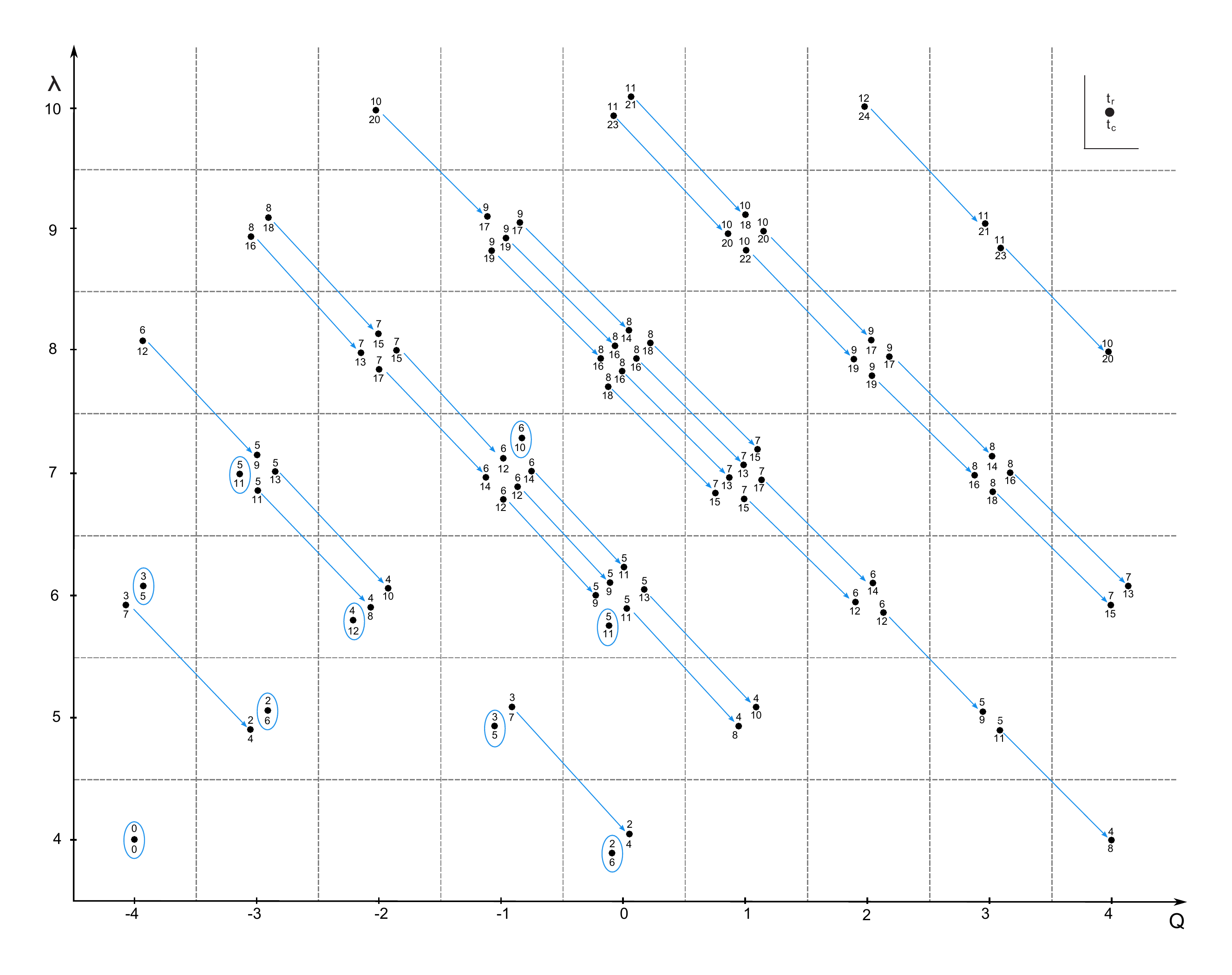}
    \caption{The colored differential $d_{[2]\to[1]}^+$ acts on the $[2]$-colored quadruply-graded Kauffman homology of the trefoil. The dots enclosed by light blue circles represent the elements in $H_*(\wt\scH^{\Kauffman}_{[2]}({\bf 3_1}),d_{[2]\to[1]}^-)$. One can convince oneself that the homology $H_*(\wt\scH^{\Kauffman}_{[2]}({\bf 3_1}),d_{[2]\to[1]}^+)$ is isomorphic to the uncolored homology $\wt\scH^{\Kauffman}_{[1]}({\bf 3_1})$ shown in Figure \ref{fig:Kauffman-uncolor-trefoil}.}\label{fig:quad-Kauffman-trefoil-2to1+}
\end{figure}
\begin{footnotesize}
\bea
&&\wt\scF(\scH^\Kauffman_{[2]}({\bf 3_1}),d_{[2]\to[1]}^+ )(\lambda,Q,t_r,t_c)\cr
&=&\lambda^4(Q^{-4} + t_c^{6}t_r^2)  + \lambda^5(Q^{-3}t_c^6t_r^2 + Q^{-1}t_c^5t_r^3) +\lambda^6 (Q^{-4} t_c^{5} t_r^3  + Q^{-2}t_c^{12} t_r^4  +  t_c^{11} t_r^5) \cr
 && +\lambda^7 (Q^{-3}t_c^{11} t_r^5  +  Q^{-1} t_c^{10} t_r^{6} )
\eea
\end{footnotesize}
\clearpage

\begin{figure}
 \centering
    \includegraphics[width=15cm]{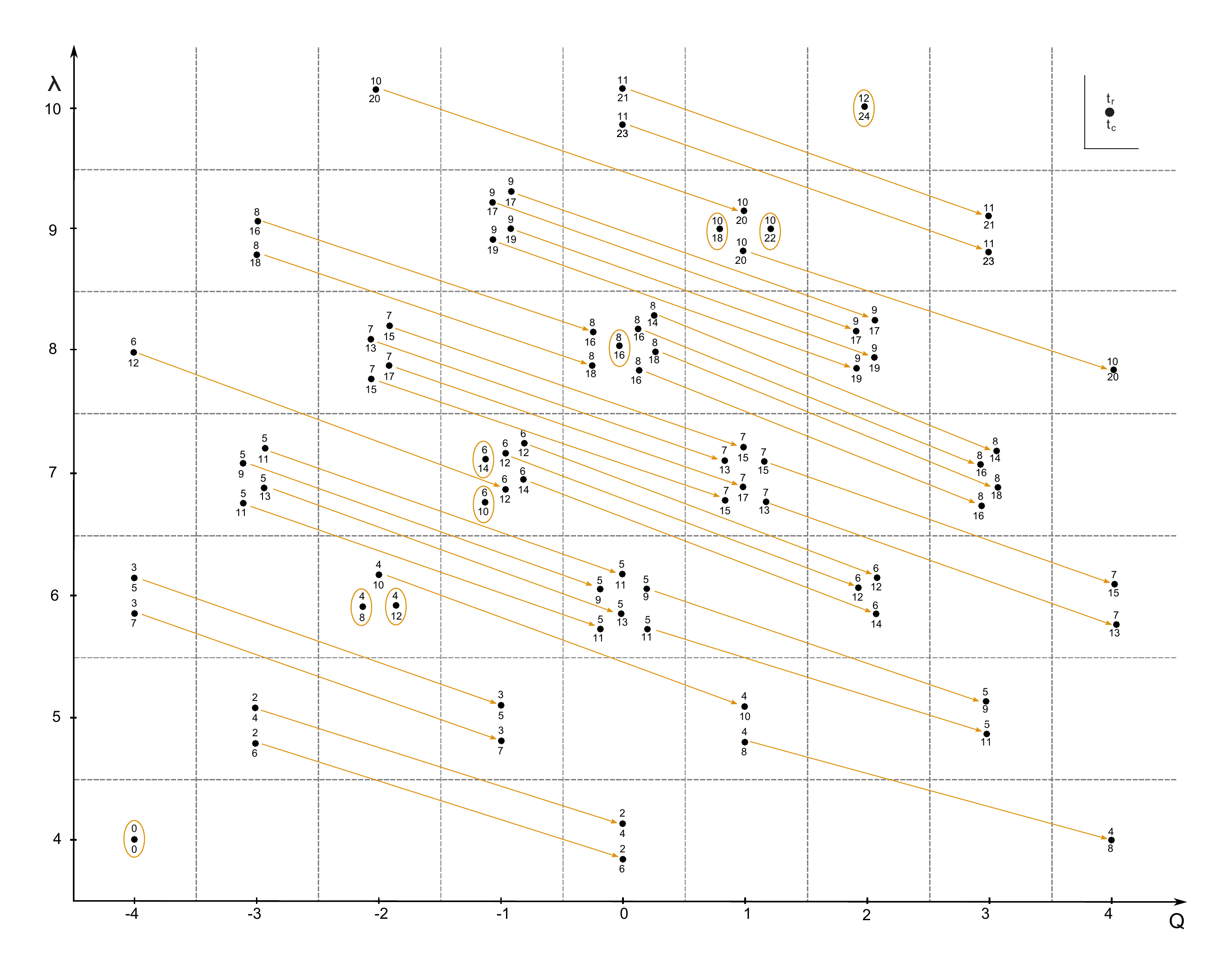}
    \caption{The diagonal differential $d_{\rm diag}^+$ acts on the $[2]$-colored quadruply-graded Kauffman homology of the trefoil. The dots enclosed by light brown circles represent the elements in $H_*(\wt\scH^{\Kauffman}_{[2]}({\bf 3_1}),d_{\rm diag}^+)$. It is easy to see that they are aligned along the diagonal line $\lambda(x)-Q(x)=8$.  The homology $H_*(\wt\scH^{\Kauffman}_{[2]}({\bf 3_1}),d_{\rm diag}^+)$ is isomorphic to the uncolored homology $\wt\scH^{\HOMFLY}_{[2]}({\bf 3_1})$.}\label{fig:quad-Kauffman-trefoil-diagonal+}
\end{figure}
\begin{footnotesize}
\bea
&&\wt\scF(\scH^\Kauffman_{[2]}({\bf 3_1}),d_{\rm diag}^+ )(\lambda,Q,t_r,t_c)\cr
&=&\lambda^4Q^{-4}  + \lambda^6(Q^{-2}t_c^8t_r^4 + Q^{-2}t_c^{12}t_r^4) +\lambda^7 (Q^{-1} t_c^{10} t_r^6  + Q^{-1}t_c^{14} t_r^6 ) \cr
 && +\lambda^8 t_c^{16}t_r^8 + \lambda^9 (Q t_c^{18} t_r^{10}  +  Q t_c^{22} t_r^{10} )+\lambda^{10} Q^2 t_c^{24}t_r^{12}
\eea
\end{footnotesize}
\clearpage

\begin{figure}
 \centering
    \includegraphics[width=15cm]{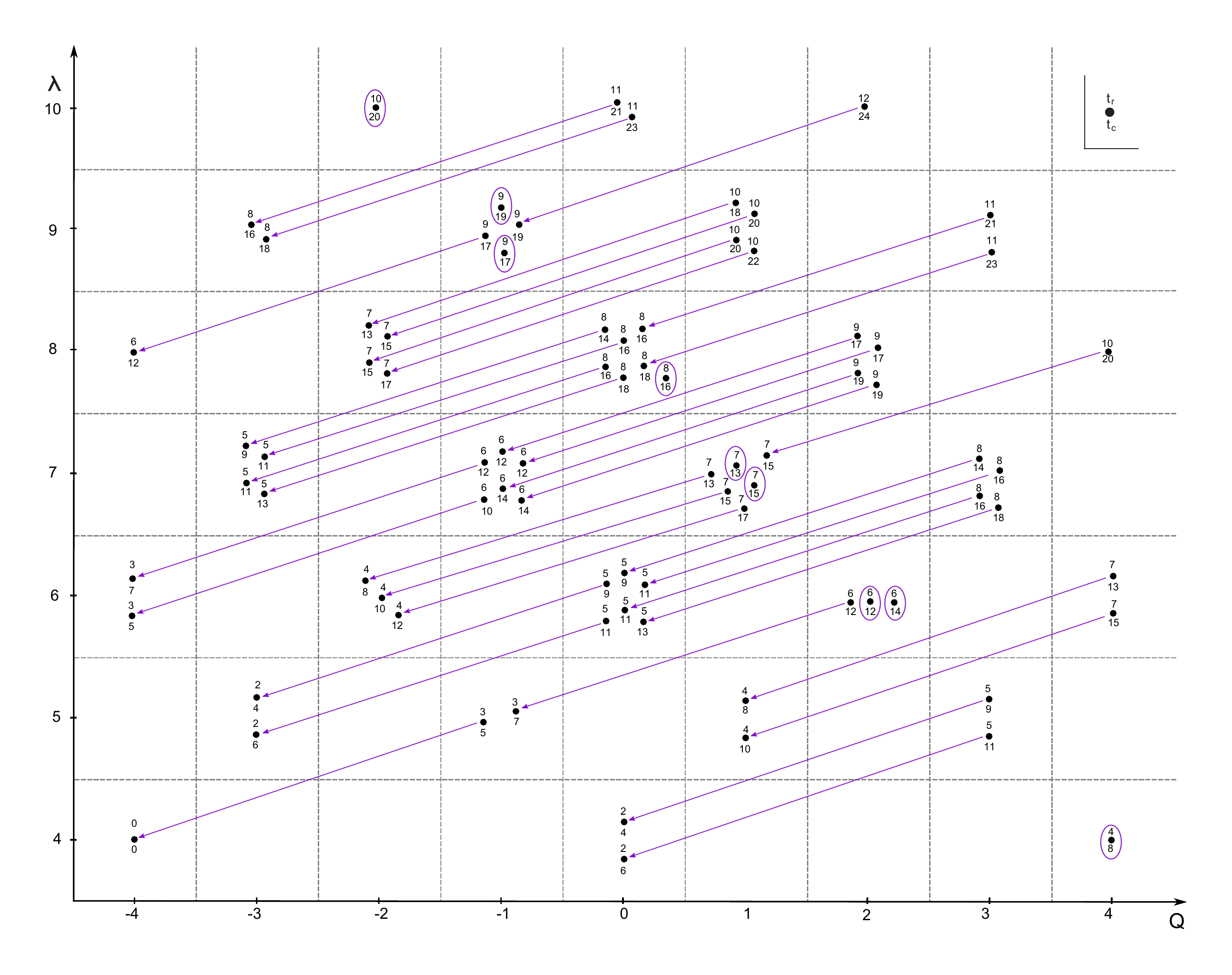}
    \caption{The diagonal differential $d_{\rm diag}^-$ acts on the $[2]$-colored quadruply-graded Kauffman homology of the trefoil. The dots enclosed by violet circles represent the elements in $H_*(\wt\scH^{\Kauffman}_{[2]}({\bf 3_1}),d_{\rm diag}^-)$. It is easy to see that they are aligned along the diagonal line $\lambda(x)+Q(x)=8$.  The homology $H_*(\wt\scH^{\Kauffman}_{[2]}({\bf 3_1}),d_{\rm diag}^-)$ is isomorphic to the uncolored homology $\wt\scH^{\HOMFLY}_{[2]}({\bf 3_1})$.}\label{fig:quad-Kauffman-trefoil-diagonal-}
\end{figure}
\begin{footnotesize}
\bea
&&\wt\scF(\scH^\Kauffman_{[2]}({\bf 3_1}),d_{\rm diag}^- )(\lambda,Q,t_r,t_c)\cr
&=&\lambda^4Q^4t_c^8t_r^4  + \lambda^6(Q^2t_c^{12}t_r^6 + Q^2t_c^{14}t_r^6) +\lambda^7 (Q t_c^{13} t_r^7  + Q t_c^{15} t_r^7)+  \lambda^8 t_c^{16} t_r^8 \cr
 && +\lambda^9 (Q^{-1}t_c^{17} t_r^9  +  Q^{-1} t_c^{19} t_r^{9} )+\lambda^{10}Q^{-2}t_c^{20}t_r^{10}
\eea
\end{footnotesize}
\clearpage

\begin{figure}
 \centering
    \includegraphics[width=15cm]{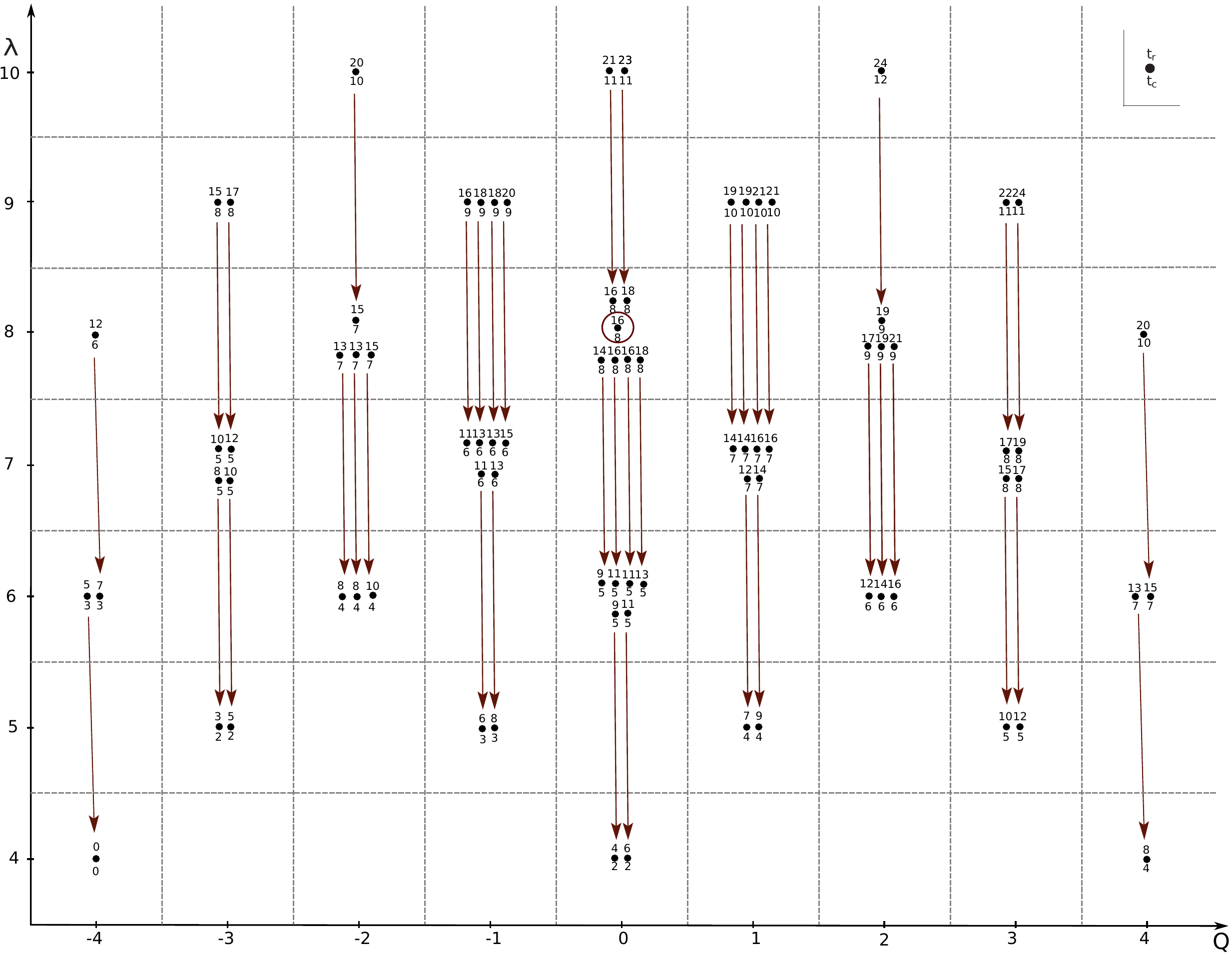}
    \caption{The canceling differential $d_{[1^r]\to[0]}^+$ acts on the $[1^2]$-colored quadruply-graded Kauffman homology of the trefoil.}\label{fig:quad-Kauffman-trefoil-11-cancel}
\end{figure}

\begin{footnotesize}
\bea
&&\wt\scF_{\yng(1,1)}({\bf 3_1};\lambda,Q,t_r,t_c)=\wt\scF_{\yng(2)}({\bf 3_1};\lambda,Q^{-1}t_c^{-1}t_r^{-2},t_c,t_r)\cr
&=&\lambda ^4\left(\tfrac{1}{Q^4}+t_c^2 t_r^4+t_c^2 t_r^6+Q^4 t_c^4 t_r^8\right) \cr
&&+\lambda ^5\left(\tfrac{t_c^2 t_r^3}{Q^3}+\tfrac{t_c^2 t_r^5}{Q^3}+\tfrac{t_c^3 t_r^6}{Q}+Q t_c^4 t_r^7+\tfrac{t_c^3 t_r^8}{Q}+Q t_c^4 t_r^9+Q^3 t_c^5 t_r^{10}+Q^3 t_c^5 t_r^{12}\right) \cr
&&+\lambda ^6\left(\tfrac{t_c^3 t_r^5}{Q^4}+\tfrac{t_c^3 t_r^7}{Q^4}+\tfrac{2 t_c^4 t_r^8}{Q^2}+2 t_c^5 t_r^9+\tfrac{t_c^4 t_r^{10}}{Q^2}+3 t_c^5 t_r^{11}+Q^2 t_c^6 t_r^{12}+t_c^5 t_r^{13}+Q^4 t_c^7 t_r^{13}+Q^2 t_c^6 t_r^{14}+Q^4 t_c^7 t_r^{15}+Q^2 t_c^6 t_r^{16}\right)\cr
&&+\lambda ^7\left(\tfrac{t_c^5 t_r^8}{Q^3}+\tfrac{2 t_c^5 t_r^{10}}{Q^3}+\tfrac{2 t_c^6 t_r^{11}}{Q}+\tfrac{t_c^5 t_r^{12}}{Q^3}+Q t_c^7 t_r^{12}+\tfrac{3 t_c^6 t_r^{13}}{Q}+3 Q t_c^7 t_r^{14}+\tfrac{t_c^6 t_r^{15}}{Q}+Q^3 t_c^8 t_r^{15}+2 Q t_c^7 t_r^{16}+2 Q^3 t_c^8 t_r^{17}+Q^3 t_c^8 t_r^{19}\right) \cr
&&+\lambda ^8\left(\tfrac{t_c^6 t_r^{12}}{Q^4}+\tfrac{2 t_c^7 t_r^{13}}{Q^2}+t_c^8 t_r^{14}+\tfrac{2 t_c^7 t_r^{15}}{Q^2}+4 t_c^8 t_r^{16}+Q^2 t_c^9 t_r^{17}+2 t_c^8 t_r^{18}+2 Q^2 t_c^9 t_r^{19}+Q^4 t_c^{10} t_r^{20}+Q^2 t_c^9 t_r^{21}\right)\cr
&&+\lambda ^{9}\left(\tfrac{t_c^8 t_r^{15}}{Q^3}+\tfrac{t_c^9 t_r^{16}}{Q}+\tfrac{t_c^8 t_r^{17}}{Q^3}+\tfrac{2 t_c^9 t_r^{18}}{Q}+2 Q t_c^{10} t_r^{19}+\tfrac{t_c^9 t_r^{20}}{Q}+2 Q t_c^{10} t_r^{21}+Q^3 t_c^{11} t_r^{22}+Q^3 t_c^{11} t_r^{24}\right) \cr
&&+\lambda ^{10}\left(\tfrac{t_c^{10} t_r^{20}}{Q^2}+t_c^{11} t_r^{21}+t_c^{11} t_r^{23}+Q^2 t_c^{12} t_r^{24}\right) 
\eea
\end{footnotesize}
\newpage

\begin{figure} 
 \centering
    \includegraphics[width=16cm]{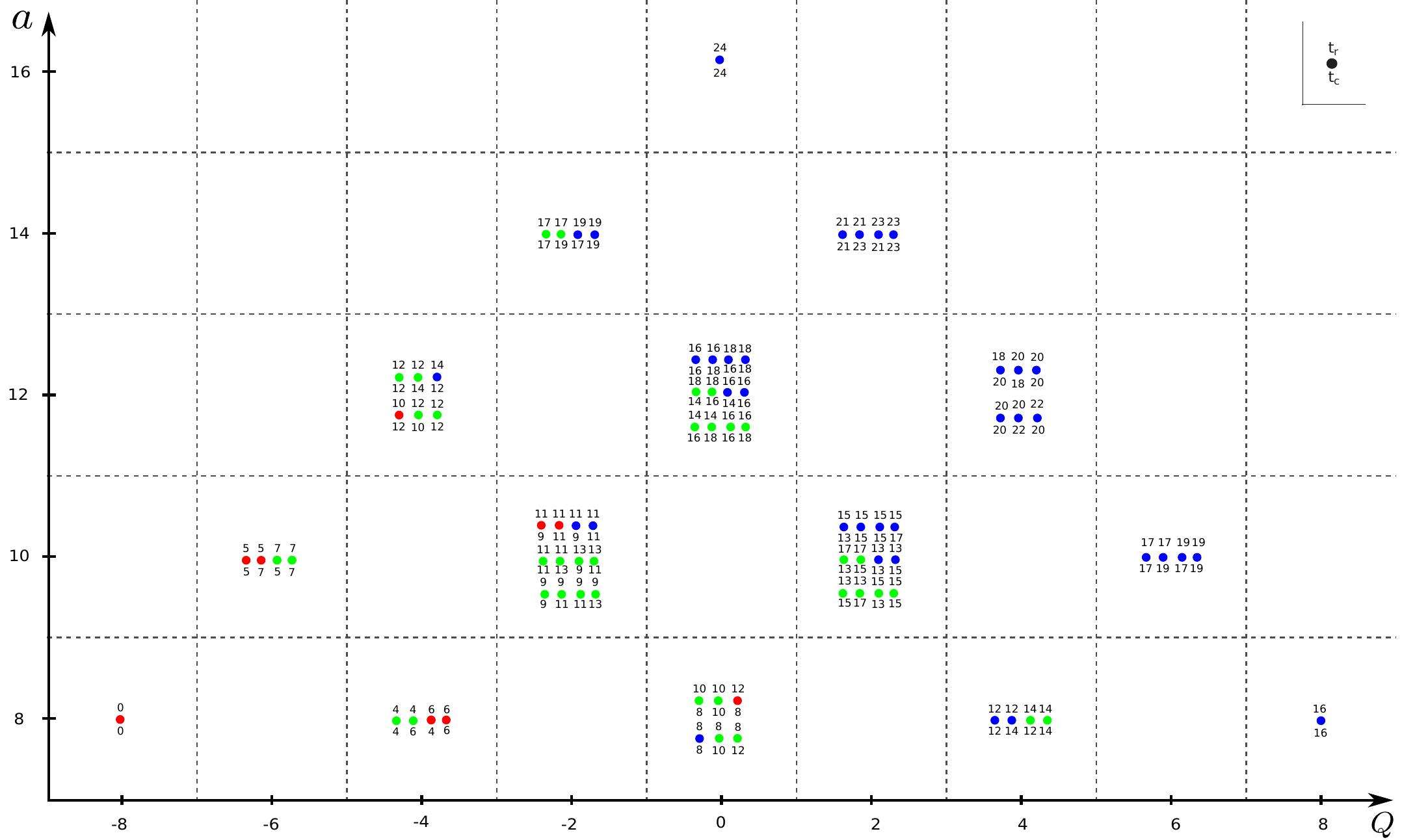}
    \caption{The $[2,2]$-colored HOMFLY homology of the figure-eight. The red dots represent the elements in $H_*(\wt\scH^\HOMFLY_{[2,2]}({\bf 3_1}),d^+_{[2,2]\to[2]})$. The red, green and blue dots consist of the terms corresponding to $k=2,1$ and $0$ in the first summation of \eqref{quad-HOMFLY-2row-trefoil} respectively.}
    \label{fig:HOMFLY-trefoil-22}
\end{figure}
\begin{footnotesize}
\bea
&&\wt\scP_{\yng(2,2)}({\bf 3_1};a,Q,t_r,t_c)\cr
&&={\color{red}a^8 \left(\tfrac{1}{Q^8}+\tfrac{t_c^4 t_r^6}{Q^4}+\tfrac{t_c^6 t_r^6}{Q^4}+t_c^8 t_r^{12}\right)+a^{10} \left(\tfrac{t_c^5 t_r^5}{Q^6}+\tfrac{t_c^7 t_r^5}{Q^6}+\tfrac{t_c^9 t_r^{11}}{Q^2}+\tfrac{t_c^{11} t_r^{11}}{Q^2}\right)+\tfrac{a^{12} t_c^{12} t_r^{10}}{Q^4} }\cr
&&{\color{forestgreen}+ a^8 \left(\tfrac{t_c^4 t_r^4}{Q^4}+\tfrac{t_c^6 t_r^4}{Q^4}+t_c^{10} t_r^8+t_c^{12} t_r^8+t_c^8 t_r^{10}+t_c^{10} t_r^{10}+Q^4 t_c^{12} t_r^{14}+Q^4 t_c^{14} t_r^{14}\right)}\cr
&& {\color{forestgreen}+a^{10} \Big(\tfrac{t_c^5 t_r^7}{Q^6}+\tfrac{t_c^7 t_r^7}{Q^6}+\tfrac{t_c^9 t_r^9}{Q^2}+\tfrac{2 t_c^{11} t_r^9}{Q^2}+\tfrac{t_c^{13} t_r^9}{Q^2}+\tfrac{t_c^{11} t_r^{11}}{Q^2}+\tfrac{t_c^{13} t_r^{11}}{Q^2}+\tfrac{t_c^9 t_r^{13}}{Q^2}+\tfrac{t_c^{11} t_r^{13}}{Q^2}}\cr
&& {\color{forestgreen}\hspace{5cm}+Q^2 t_c^{15} t_r^{13}+Q^2 t_c^{17} t_r^{13}+Q^2 t_c^{13} t_r^{15}+Q^2 t_c^{15} t_r^{15}+Q^2 t_c^{13} t_r^{17}+Q^2 t_c^{15} t_r^{17}\Big)}\cr
&&{\color{forestgreen}+a^{12} \left(\tfrac{t_c^{10} t_r^{12}}{Q^4}+\tfrac{2 t_c^{12} t_r^{12}}{Q^4}+\tfrac{t_c^{14} t_r^{12}}{Q^4}+t_c^{16} t_r^{14}+t_c^{18} t_r^{14}+t_c^{16} t_r^{16}+t_c^{18} t_r^{16}+t_c^{14} t_r^{18}+t_c^{16} t_r^{18}\right)+a^{14} \left(\tfrac{t_c^{17} t_r^{17}}{Q^2}+\tfrac{t_c^{19} t_r^{17}}{Q^2}\right)}\cr
&&{\color{blue}+a^8 \left(t_c^8 t_r^8+Q^4 t_c^{12} t_r^{12}+Q^4 t_c^{14} t_r^{12}+Q^8 t_c^{16} t_r^{16}\right)}\cr
&&{\color{blue}+a^{10} \Big(\tfrac{t_c^9 t_r^{11}}{Q^2}+\tfrac{t_c^{11} t_r^{11}}{Q^2}+Q^2 t_c^{13} t_r^{13}+Q^2 t_c^{15} t_r^{13}+Q^2 t_c^{13} t_r^{15}+2 Q^2 t_c^{15} t_r^{15}+Q^2 t_c^{17} t_r^{15}}\cr
&& {\color{blue}\hspace{6cm}+Q^6 t_c^{17} t_r^{17}+Q^6 t_c^{19} t_r^{17}+Q^6 t_c^{17} t_r^{19}+Q^6 t_c^{19} t_r^{19}\Big)}\cr
&&{\color{blue}+a^{12} \left(\tfrac{t_c^{12} t_r^{14}}{Q^4}+t_c^{14} t_r^{16}+2 t_c^{16} t_r^{16}+t_c^{18} t_r^{16}+t_c^{16} t_r^{18}+t_c^{18} t_r^{18}+Q^4 t_c^{20} t_r^{18}+Q^4 t_c^{18} t_r^{20}+2 Q^4 t_c^{20} t_r^{20}+Q^4 t_c^{22} t_r^{20}+Q^4 t_c^{20} t_r^{22}\right)}\cr
&&{\color{blue}+a^{14} \left(\tfrac{t_c^{17} t_r^{19}}{Q^2}+\tfrac{t_c^{19} t_r^{19}}{Q^2}+Q^2 t_c^{21} t_r^{21}+Q^2 t_c^{23} t_r^{21}+Q^2 t_c^{21} t_r^{23}+Q^2 t_c^{23} t_r^{23}\right)+a^{16} t_c^{24} t_r^{24}}
\eea
\end{footnotesize}

\clearpage

\bibliography{CS}{}
\bibliographystyle{JHEP}
\end{document}